\documentclass{pasj00}

\begin{document}
\SetRunningHead{Tanaka et al.}{A Group of Quiescent Early-type Galaxies at $z=1.6$}

\title{An X-ray Detected Group of Quiescent Early-type Galaxies at $z=1.6$\\in the Chandra Deep Field South}

\author{
Masayuki~\textsc{Tanaka}\altaffilmark{1},
Alexis~\textsc{Finoguenov}\altaffilmark{2,3},
Mohammad~\textsc{Mirkazemi}\altaffilmark{2},
David~J.~\textsc{Wilman}\altaffilmark{2},
John~S.~\textsc{Mulchaey}\altaffilmark{4},
Yoshihiro~\textsc{Ueda}\altaffilmark{5},
Yongquan~\textsc{Xue}\altaffilmark{6,7,8},
William~N.~\textsc{Brandt}\altaffilmark{6,7},
Nico~\textsc{Cappelluti}\altaffilmark{9}
}
\altaffiltext{1}{Institute for the Physics and Mathematics of the Universe, The University of Tokyo,  5-1-5 Kashiwanoha, Kashiwa-shi, Chiba 277-8583, Japan}
\altaffiltext{2}{Max-Planck Institut f\"{u}r extraterrestrische Physik, Giessenbachstrasse, D-85748 Garching bei M\"{u}nchen, Germany}
\altaffiltext{3}{University of Maryland, Baltimore County, 1000 Hilltop Circle,  Baltimore, MD 21250, USA}
\altaffiltext{4}{The Observatories of the Carnegie Institution of Science, 813 Santa Barbara Street, Pasadena, CA 91101, USA}
\altaffiltext{5}{Department of Astronomy, Kyoto University, Kyoto 606-8502, Japan}
\altaffiltext{6}{Department of Astronomy and Astrophysics, Pennsylvania State University, University Park, PA 16802, USA}
\altaffiltext{7}{Institute for Gravitation and the Cosmos, Pennsylvania State University, University Park, PA 16802, USA}
\altaffiltext{8}{Key Laboratory for Research in Galaxies and Cosmology, Department of Astronomy, University of Science and Technology of China, Chinese Academy of Sciences, Hefei, Anhui 230026, China}
\altaffiltext{9}{INAF-Osservatorio Astronomico di Bologna, Via Ranzani 1, 40127 Bologna, Italy}


\maketitle

\begin{abstract}
We report the discovery of an X-ray group of galaxies located at a
high redshift of $z=1.61$ in the Chandra Deep Field South.  Based on
the 4Msec Chandra data, the group is first identified as an extended
X-ray source.  We use a wealth of deep multi-wavelength data to
identify the optical counterpart -- our red sequence finder detects a
significant over-density of galaxies at $z\sim1.6$.  The brightest group galaxy is
spectroscopically confirmed at $z=1.61$ based on published
spectroscopic redshifts.  Using this as a central redshift of the
group, we measure an X-ray luminosity of $L_{0.1-2.4\rm
  keV}=(1.8\pm0.6)\times10^{43}\rm\ erg\ s^{-1}$, which then translates
into a group mass of $(3.2\pm0.8)\times10^{13}\rm\ M_\odot$.  This is
the lowest mass group ever confirmed at $z>1.5$.  The deep
optical-nearIR images from CANDELS reveal that the group exhibits a
surprisingly prominent red sequence and most of the galaxies are
consistent with a formation redshift of $z_f=3$.  A detailed analysis
of the spectral energy distributions of the group member candidates
confirms that most of them are indeed passive galaxies.  Furthermore,
their structural parameters measured from the near-IR CANDELS images
show that they are morphologically early-type.
The newly identified group at $z=1.61$ is dominated by quiescent
early-type galaxies and the group appears similar to those
in the local Universe.  One possible difference is the high
fraction of AGN --- $38^{+23}_{-20}$\% of the bright group member
candidates are AGN, which might indicate a role for AGN in quenching
of star formation.  However, a statistical sample of high-$z$ groups
is needed to draw a general picture of groups at this redshift.  Such
a sample will hopefully be available in near future surveys.
\end{abstract}

\section{Introduction}

Modern observing facilities can reach galaxy groups and clusters at
$z=1.5$ and beyond.  The number of such high-$z$ systems has been
increasing in recent years and we are now in the process of building a
statistical sample which will give us a detailed picture of
group/cluster evolution at high redshifts.  It is a widely accepted
fact that galaxy clusters today are dominated by quiescent early-type
galaxies, while a significant fraction of field galaxies are star
forming late-type galaxies.  This clearly shows that the formation and
evolution of cosmic large-scale structure affects galaxy evolution.
However, the interplay between galaxies and their surrounding structure
remains unclear.  One observational approach to this question is to
study groups and clusters over a range of redshift to directly trace
their evolution and to identify the epoch when quiescent early-type
galaxies become the dominant population.  High-$z$ groups are statistically
likely progenitors of the present-day clusters and they are thus key objects
to improve our understanding of the origin of the environmental
dependence observed in the local universe.

In this respect, the current frontier of high-$z$ systems is at
$z\sim1.5$ -- a number of such systems have been reported to date.
\citet{mullis05} reported on the discovery of a rich $z=1.39$ cluster.
It is very massive \citep{jee09} and exhibits a tight red sequence
\citep{lidman08}.  Star forming galaxies are absent in the cluster core
\citep{bauer11}.
\citet{stanford05} discovered a cluster at $z=1.41$ based on
infrared photometry from Spitzer.  \citet{brodwin11} carried out
a spectroscopic follow-up observation of the cluster and confirmed
more than 10 members.
They also presented a spectroscopically confirmed cluster at $z=1.49$
selected from the Spitzer photometry.
\citet{stanford06} presented a cluster located at
$z=1.45$.  Follow-up spectroscopic observations and Chandra
observations are reported in Hilton et al. (2007,
2010)\nocite{hilton07,hilton10}.  Although exhibiting a prominent red
sequence \citep{bielby10}, this cluster also hosts a significant number
of emission line objects in the core \citep{hayashi10}.  In the last
few years, the redshift barrier of $z=1.5$ has been broken.
\citet{papovich10} and \citet{tanaka10a} independently confirmed a
group at $z=1.62$ in the Subaru/XMM-Newton Deep Field (SXDF).
\citet{pierre11} performed a Chandra observation of the group.
Although the exposure time was rather short and the detection
was marginal, they reported a consistent flux with that of \citet{tanaka10a}.
The XMM-Newton Deep Cluster Project reported the discovery of a few X-ray
bright clusters at $z\sim1.5$ \citep{fassbender11,nastasi11,santos11}.
Even higher redshift systems are now detected.
\citet{henry10} reported on a possible X-ray group at $z=1.75$ and
\citet{stanford12} recently presented a spectroscopically confirmed
X-ray cluster at $z=1.75$.
\citet{gobat11} presented a color-selected group at $z=2.07$.
There are a few more photo-$z$ selected group candidates at $z\sim2$
\citep{andreon09,spitler11}.
However, some of these very high-$z$ systems still require convincing
confirmation with spectroscopic redshifts.

At $z\gtrsim2$, there are few securely confirmed, gravitationally
bound systems but a number of authors have reported the discovery of
so-called proto-clusters.  The definition of a proto-cluster is often
ambiguous, but here we define it as a system that exhibits a
significant over-density of galaxies but is yet to be gravitationally
bound.  Such proto-clusters have often been identified by tracing
emission line galaxies or color-selected galaxies (e.g.,
\cite{venemans02,kurk04,matsuda04,venemans07}).  Distant radio
galaxies have been frequent targets for such observations (see
\cite{miley08} for a review).  Some of the reported proto-clusters may
collapse and evolve into a cluster at later times.  These
proto-clusters tend to show a significant fraction of star forming
galaxies.  This is at least partly due to the way they are identified,
but a large population of star forming galaxies at high redshift is
not unexpected.  Galaxy formation occurs at density peaks -- thus
proto-clusters are an obvious place to find active star formation in
the early universe.  Clusters change their nature with time from a
place for star forming galaxies to a place for quiescent, passively
evolving galaxies.  To identify this key, cradle-to-grave transition
epoch remains an important, but as yet unfulfilled goal.

In order to address these questions, we are conducting a systematic
survey of groups and clusters of galaxies using X-ray and
optical-nearIR data in deep fields such as COSMOS, CFHT Deep Fields,
and SXDF \citep{finoguenov07,bielby10,finoguenov10}.  These group
catalogs include systems over a wide redshift range ($0<z<1.5$) and are
a powerful probe of cosmology (e.g., \cite{finoguenov10}).  At the same
time, they provide an ideal data set to study the evolution of galaxies
across environment and time.  As part of this project, we have
constructed an X-ray group catalog in the Chandra Deep Field South
(CDFS) as presented by Finoguenov et al.  (2012 in prep).  During the
course of this work, we have identified a high redshift group located
at $z=1.6$.  This system is in the large-scale structure at $z=1.6$
discovered by \citet{kurk09} and is the first gravitationally bound,
X-ray detected system in that structure.  It happens to fall within the
area covered by the CANDELS survey \citep{grogin11,koekemoer11}, where
deep, high-quality HST imaging data are available.  The near-IR CANDELS
images neatly probe the rest-frame optical wavelengths of galaxies at
$z=1.6$.  We take this unique opportunity to study the galaxy
population of this $z=1.6$ group with a depth equivalent to that
typically only found at $z=0$.

The layout of this paper is as follows.  We review the data that we
use in this work, including the creation of our X-ray group catalog in
Section 2, followed by analyses of the $z=1.6$ group detection via
extended X-ray emission and color-magnitude diagrams in Section 3.  We
perform a detailed analysis of spectral energy distributions (SEDs) of
the group galaxies in Section 4 and examine morphology of the galaxies
in Section 5.  Section 6 discusses implications of our finding for the
origin of the environmental dependence of galaxy properties and
Section 7 concludes the paper.  Unless otherwise stated, we assume a
flat universe with $\rm H_0=72\ km\ s^{-1}\ km^{-1}$, $\Omega_M=0.26$
and $\Omega_\Lambda=0.74$.  Magnitudes are given in the AB system.  We
use the following abbreviations : AGN for active galactic nucleus,
BGG for brightest group galaxy, CDFS for Chandra Deep Field South,
FWHM for full-width at half maximum, 
IMF for initial mass function,
PDF for probability distribution function,
PSF for point spread function,
SED for spectral energy distribution, 
SFR for star formation rate, and SXDF for Subaru/XMM-Newton Deep Field.

\section{Data and a Catalog of X-ray Groups in CDFS}

We base our analysis on a wealth of public data available in CDFS.
We first summarize the X-ray and optical-IR data that we use.
We then briefly describe our X-ray group catalog.

\subsection{X-ray data from Chandra and XMM-Newton}

The Chandra Deep Field South has been a frequent target of X-ray
observations with both Chandra and XMM-Newton.  After the first 1Ms
Chandra observation (\cite{giacconi02}), the exposure was recently
extended to 2 Ms (\cite{luo08,luo10,rafferty11}) and later to 4Ms
(\cite{xue11}), via a large Director's Discretionary Time project.
CDFS now provides our most sensitive 0.5--8 keV view of distant AGNs,
starburst galaxies, normal galaxies, and galaxy groups.  For the
detection of extended sources, of value is both the 3.3 Ms XMM coverage
\citep{comastri11} and the extended area covered by Chandra
(\cite{lehmer05}).  The group studied in this paper is detected
independently in both Chandra and XMM data with a consistent source
flux in the observed 0.5-2 keV band (see next section).
We briefly outline the reduction of these data sets below.
We note that we combine the Chandra and XMM-Newton data
for the group identification in Finoguenov et al. 2012 (in prep), but
we primarily use the Chandra data in this paper due to its superb
spatial resolution.

In the Chandra analysis, we have applied a conservative event screening and the
modeling of the quiescent background.  We have filtered the 
light-curve events using the {\sc lc\_clean} tool in order to remove normally
undetected particle flares.  The background model maps have been evaluated
with the prescription of \citet{hickox06}.  We estimated the
particle background by using the ACIS stowed position
(http://cxc.cfa.harvard.edu/contrib/maxim/acisbg) observations and rescaling
them by the ratio $\frac{cts_{9.5-12 keV, data}}{ cts_{9.5-12 keV, stowed}}$.
The cosmic background flux has been evaluated by subtracting
the particle background maps from the real data and masking the area
occupied by the detected sources.  Rapidly broadening Chandra PSF with off-axis
angle produces a large gradient in the resolved fraction of the cosmic
background, which is the dominant source of systematics in our background
subtraction.
However, we note that the group that we study in this paper is
only $\sim3$ arcmin away from the field center, where the Chandra PSF
is still sharp.
We combine the data with and without point source subtraction
and analyze them independently in Sect 3.2.

For the XMM-Newton analysis, we have followed the prescription outlined in
\citet{finoguenov07} on data screening and background evaluation, with
updates described in \citet{bielby10}. After cleaning those observations,
the resulting net total observing time with XMM-Newton are 1.946Ms for pn,
2.552Ms for MOS1, 2.530Ms for MOS2.
We carefully remove point sources following \citet{finoguenov09} and \citet{finoguenov10}.
This is done independently from the Chandra data to
allow for AGN variability and difference in the astrometry.
Furthermore, we do not detect individual sources and catalog them,
but instead we directly work with images.
This is important for XMM-Newton because it is confusion limited at the
depths of CDFS (in the soft band) and the source deblending is not trivial.
We are left with extended sources in the combined image.
We describe the source detection in Sect 2.3 and 3.2.

\subsection{Optical and IR data from the literature}

We base our analysis mainly on two public sets of optical-IR photometry drawn from
MUSIC \citep{grazian06,santini09} and CANDLES \citep{grogin11,koekemoer11}.
In addition,
we make extensive use of photometric redshifts based on the deep multi-band
photometry available in this field in order to identify group members.

The MUSIC catalog is a deep, multi-wavelength catalog of the CDFS field
\citep{grazian06,santini09}.  The catalog that we use is from
\citet{santini09}, which is an extended version of the one presented by
\citet{grazian06} and contains 15-band photometry spanning from the
$u$-band to 24$\mu m$.  For details of the catalog construction,
readers are referred to \citet{grazian06} and \citet{santini09}.  But,
in short, the objects are first detected in the ACS $z$-band and
photometry in the other bands is performed by convolving the $z$-band
image to match with the PSF in the other bands and scaling the fluxes.
Objects that are not detected in the $z$-band are detected and measured
in the $K_s$ and IRAC $4.6\mu m$ images with the $z$-band detected objects
masked out.

In addition to the MUSIC catalog, we use the deep ACS and WFC3 imaging
data from CANDELS \citep{grogin11,koekemoer11}.  CANDELS is a public
imaging survey using the ACS and WFC3 camera on board the Hubble Space
Telescope.  The observation is made mainly in three bands: F814W, F125W
and F160W.  We make stacked F814W, F125W, and F160W images by combining
data from the first 4 epochs of deep observations using the weight maps
supplied by the CANDELS team.  The resultant images reach limiting
magnitudes of 28.3, 27.3, and 27.0 within an aperture of size
$4\times$FWHM.  The PSF of the F160W image is $\sim0.2$ arcsec and the
PSFs of the other images are smoothed to this size using the Gaussian
kernel.  Object detection and photometry is performed using Source
Extractor \citep{bertin96}.  Objects are detected in the F160W image
and photometry in the other bands is performed in dual image mode.  We
adopt MAG\_AUTO for total magnitudes and use magnitudes measured in
$4\times$FWHM apertures for colors.  For simplicity, we often denote
F814W, F125W, and F160W as $I,\ J,$ and $H$, respectively, in what
follows.  We do not make an attempt to combine the MUSIC and CANDELS
catalogs because the photometry is performed in different ways and the
MUSIC catalog already contains the $i$, $J$, and $H$-band photometry
measured in a self-consistent manner.  CANDELS goes deeper than MUSIC,
but most of our analyses is not limited by depth of imaging.  For these
reasons, we use the catalogs separately for complimentary analyses.

\citet{rafferty11} computed photometric redshifts (photo-$z$'s) using
a combined photometric catalog collected from the literature
\citep{wolf04,gawiser06,grazian06,wolf08,nonino09,taylor09,damen11}.
They used the public photo-$z$ code ZEBRA \citep{feldmann06}.  We note
that they focused on AGNs in their paper, but they computed
photo-$z$'s for all galaxies within the extended CDFS.  By comparing
the photo-$z$'s with spectroscopic redshifts, they achieved
$\sigma|\Delta z/(1+z_{spec})|=0.03$ with an outlier rate of 4\% for
bright galaxies with $R<24$.  Here outliers are defined as those with
$|\Delta z/ (1+z_{spec})|>0.2$.  For fainter galaxies, the accuracy
degrades to $\sigma=0.06$ and an outlier rate of 14\%, although the
spectroscopic sample is heterogeneous.  We denote their photo-$z$'s as
$z_{phot,rafferty}$.

Later in the paper, we perform the spectral energy distribution (SED)
fitting of group galaxies using the MUSIC catalog. The photometric
redshift is fit as part of this procedure, and we shall make extensive
use of these photo-$z$'s.  We denote photometric redshifts from our SED
fitting as $z_{phot, tanaka}$.  We use $z_{phot,rafferty}$ and
$z_{phot,tanaka}$ for the selection of group members.

In addition to those sources of photometric redshift, we also examine
photo-$z$'s from \citet{cardamone10}, which we denote
$z_{phot,cardamone}$.  Their photo-$z$'s are based on the multi-band
data from the MUSYC survey \citep{gawiser06,cardamone10}.  With 18
medium-band photometry in the optical, they have achieved excellent
photo-$z$'s for bright objects \citep{cardamone10}.  However, their
broad band imaging is shallower (and covers a wider area), and the
catalog is $BVR$-selected. This means that the fainter and redder
candidate group members are not included in their catalog.  For this
reason, we do not use $z_{phot,cardamone}$ for the primary selection of
group members.

As each data set has its own strengths, we use different catalogs for
different purposes.  To be specific, the selection of the group member
candidates is based on $z_{phot, tanaka}$ and $z_{phot,rafferty}$.  We
use the CANDELS data for the color-magnitude diagrams in Section 3
because the high quality CANDELS photometry neatly probes the
wavelengths around the 4000\AA\ break.  The CANDELS data are also used
for structural analysis in Section 5.  We base our analysis of SED
fitting of the group members on the multi-band MUSIC catalog in order
to cover a wide wavelength range in Section 4.  We remind the readers
of which photometric catalog is used at each stage of the analysis.

\subsection{A catalog of X-ray groups in CDFS}

Using the X-ray data and \citet{rafferty11} photo-$z$ catalog described
above, we construct a catalog of X-ray groups in CDFS.  We will give a
full detail of the catalog construction in Finoguenov et al. (2012 in
prep).  But, we briefly outline our algorithm here.

On the mosaic of coadded XMM and Chandra images, group candidates are
first identified as extended sources through a classical wavelet
transform technique with a careful removal of point sources
\citep{finoguenov09}.  In order to identify optical counterparts of the
extended X-ray emission, we apply an efficient red sequence finder as
described by \citet{finoguenov10} and \citet{bielby10}.  The red
sequence finder computes a significance of the red sequence in the
region of an extended X-ray source, assuming a given redshift, by
comparing observed magnitudes and colors of galaxies with a model red
sequence at that redshift constructed using the recipe by
\citet{lidman08}.  We achieve an efficient removal of fore-/background
galaxies from the red sequence in question using the excellent
photometric redshifts \citep{rafferty11}.  Then, for each redshift at
which a significant signal is detected the red sequence is visually
inspected, and the group is assigned a redshift and confidence flags.
We use all the public spectroscopic redshifts from the literature
\citep{cristiani00,croom01,strolger04,szokoly04,vanderwel04,doherty05,lefevre05,mignoli05,ravikumar07,vanzella08,popesso09,balestra10,cooper11}
to help identify each system and obtain its spectroscopic redshift
during this final step.

In total, we have spectroscopically identified 39 systems.  We have
another 8 systems with low quality flags.  The secure groups are
located mostly at $z<1$, but a good fraction of them have higher
redshifts.  Most of the identified systems are low-mass groups with
$<3\times10^{13}\rm M_\odot$.  Such low-mass X-ray groups at relatively
high redshift are interesting, and the unprecedented X-ray depth in the
CDFS field provides an exciting opportunity to study these systems.  We
will discuss properties of these groups and their brightest members in
Wilman et al. (in prep).  The group we focus on in this paper is the
spectroscopically confirmed, highest redshift group in our group
catalog.

\section{A Galaxy Group at $z=1.6$}

\subsection{Overview of the Discovery}

\begin{figure*}
  \begin{center}
    \FigureFile(170mm,80mm){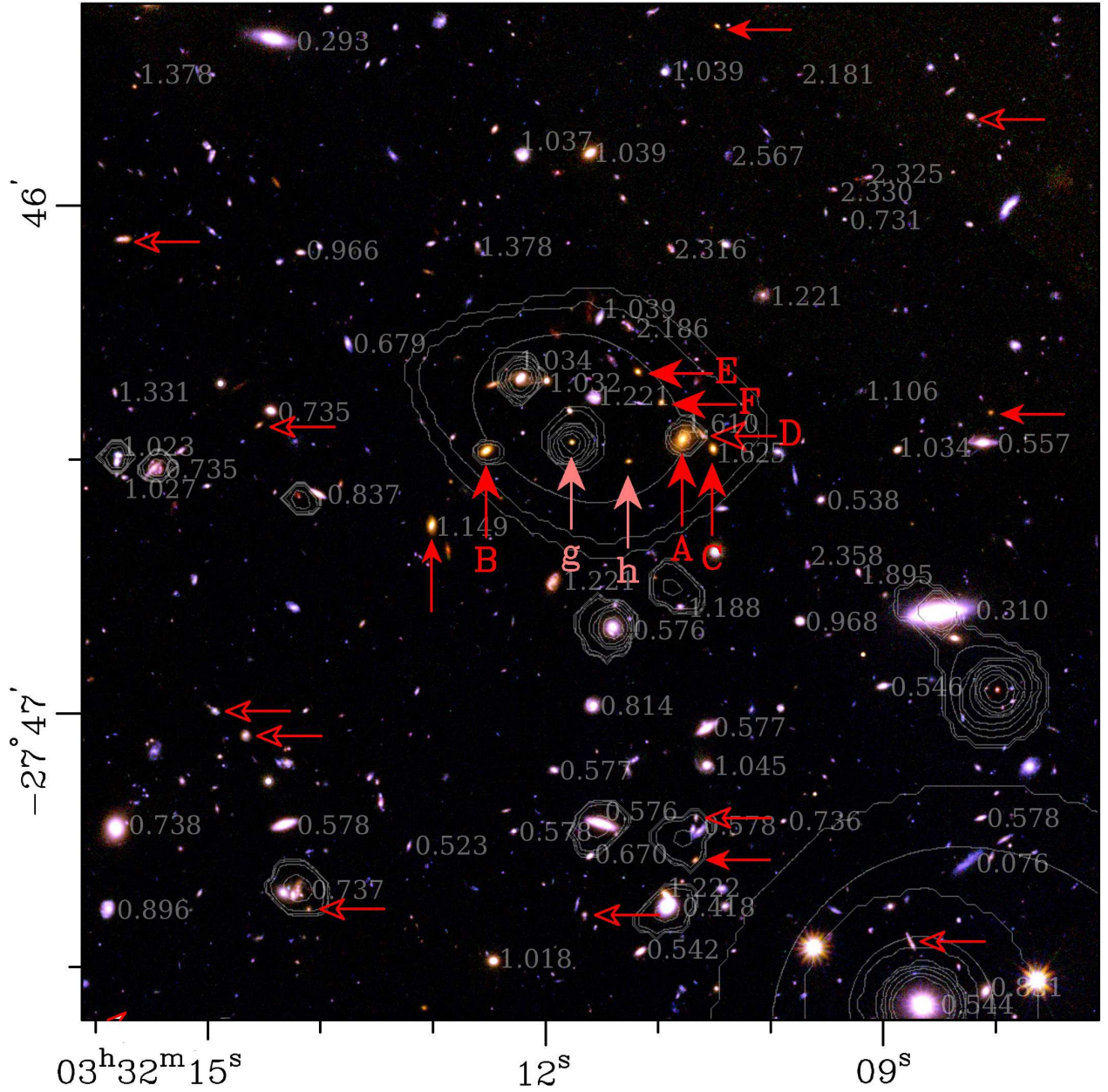}
  \end{center}
  \caption{ The CANDELS $IJH$ pseudo-color image of the group.  The
    contours show the X-ray emission.  The arrows indicate bright
    ($H<24$) galaxies with $P_{gr}\geq0.16$ (see Section 3.3 for details).
    There are two types of arrow in the figure.  The filled arrows
    point to galaxies with red colors and within $\Delta|I-H|<0.5$ from
    the model red sequence formed at $z_f=3$ (see Section 3.2 for
    details).  The open arrows are bluer galaxies.  The six brightest
    galaxies in the core of the group that have $P_{gr}\geq0.16$ both from
    \citet{rafferty11} and ourselves are labeled object-A to F (i.e.,
    {\it good candidates}; see Section 3.2).  {\it Likely candidates}
    are labeled object-g and h.  The numbers show spectroscopic
    redshifts from the literature.  }
  \label{fig:color_pic}
\end{figure*}

During the course of the group identification, we have discovered a
promising high-$z$ group candidate.  We have detected extended X-ray
emission not far from the center of the Chandra field of view.  A red
sequence finder has yielded a significant signal around $z=1.6$ for
this X-ray source.  If we assume a group redshift of $z=1.6$, the group
has a very low X-ray luminosity, which suggests that it is a low-mass
group.   Such a high-$z$, low-mass group is an interesting object
because it is likely a progenitor of a present-day cluster of typical mass.
For this system, a wealth of deep, multi-wavelength data are available,
and furthermore, the group has recently been observed by CANDELS and
the deep optical-nearIR images taken with WFC3 are publicly available.
Using the CANDELS images, we present a color picture of the group in
Fig.  \ref{fig:color_pic}.  In this section, we provide evidence that
this system is a group located at $z=1.61$ with extended X-ray
emission.  First, we make a robust analysis of the X-ray emission and
show that it is extended.  We then show that there is a clear
over-density of galaxies at $z=1.6$ around the extended X-ray emission
and that the brightest group galaxy (BGG) is spectroscopically confirmed at $z=1.61$.
Furthermore, the galaxies form a tight red sequence, which is a
ubiquitous feature of rich groups and clusters, at least at lower
redshift.  All of these results lead us to conclude that this system is
a real group located at $z=1.61$.  Below, we describe each piece of
evidence in detail.

\subsection{Extended X-ray Emission from the Group}

The Chandra observation reveals a number of point sources in and around
the group. 
Together with the point sources, we detect extended, larger-scale emission
around the group based on the wavelet analysis
(outer contours in Fig. \ref{fig:color_pic}).
The reader is referred to Sect. 3 of \citet{finoguenov07} for technical details
of the wavelet analysis.
The system is
only $\sim3$ arcmin away from the center of the Chandra field of view
where the PSF is very sharp and well-behaved ($\sim1.5$ arcsec,
estimated using the profile of nearby point sources). 
As can be seen in Fig.  \ref{fig:color_pic}, we detect the extended X-ray
emission over a scale of 20 arcsec in radius, which is significantly larger than
the PSF size.
The extended flux is only $3\times10^{-16}\rm\ erg\ s^{-1}\ cm^{-2}$
as detailed below, but it is one of the two brightest extended objects
in the Chandra best PSF area of the CDFS field ($\lesssim4'$ from the center).
The other object is a group at $z=0.73$ mentioned by \citet{cimatti02}.
Extended X-ray emission can be caused by the inverse-Compton scattering
of the CMB photons off energetic electrons in radio jets (e.g., \cite{jelic12}),
but there is no radio source around the X-ray emission \citep{miller08}.
Such non-thermal origin of the X-ray emission is thus unlikely.

We perform a simple test to prove  that the extended emission is real.
We carry out a simple
and robust analysis of counting photons around the group by masking
point sources in the Chandra raw photon image.
We first detect
point-like objects in the image with Source Extractor \citep{bertin96}.
The detected objects are then
masked out.  The masking radius is set to be three times the half-width
at half-maximum (HWHM), which visual inspection shows us is large
enough to exclude point sources and their extended PSF wings.  We have
confirmed that our conclusion does not change if we change this
radius to twice or four times the HWHM.  We have also confirmed that
our result remains the same if we use the point source catalog of
\citet{xue11} to mask the point sources.

\begin{figure}[htb]
  \begin{center}
    \FigureFile(80mm,80mm){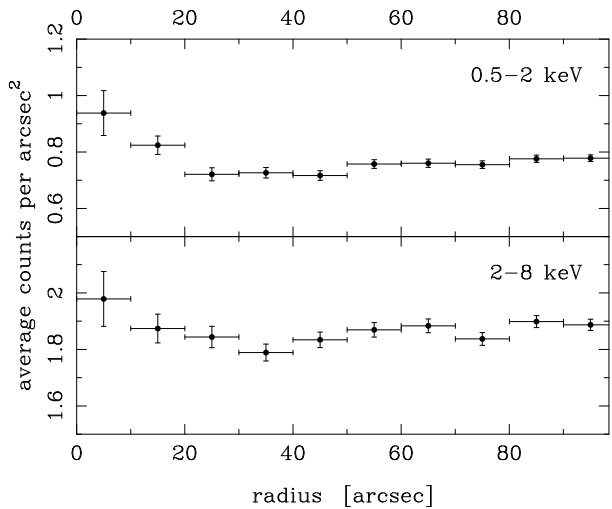}
  \end{center}
  \caption{
    Average photon counts per square arcsec in annuli
    with a width of 10 arcsec running from the center of the system 
    to the outskirts.
    We here take the location of the BGG as the center.
    The top and bottom panels are for the soft and
    hard bands, respectively.
    Point sources are masked out and the vertical error bars show the Poisson noise.
  }
  \label{fig:xray_ext}
\end{figure}

Fig. \ref{fig:xray_ext} shows the average photon counts per unit area
in annuli with a width of 10 arcsec for the 0.5--2 keV and 2--8 keV
bands.
Here, we take the BGG as the center.
In the soft band, the average photon counts are clearly higher
in the center and the X-ray emission is extended to 20 arcsec in
radius, which is consistent with the X-ray extent from the wavelet
analysis.  If we subtract the local background using the average counts
from a radius of 30-60 arcsec, and interpolate over each masked
area assuming the average counts within that annulus, we measure a
total photon count of $146.4\pm 40.7$ within an aperture of 20 arcsec
radius from the center (i.e., $3.6\sigma$ detection).
If we take the center of the wavelet contours in Fig. \ref{fig:color_pic}
as the center, we still obtain a signal of $3.3\sigma$.
We apply the same analysis using the
0.5--2 keV Chandra image independently reduced by \citet{xue11} and
obtained the same result -- there is a clear excess in the photon counts
at the center and we measure a significant photon count from the group
($137.6\pm39.5$).  The photon count is fully consistent with ours.

The group is not detected in the hard band at a significant level as
shown in Fig. \ref{fig:xray_ext}; we measure a photon count of
$81.6\pm59.0$ within 20 arcsec radius after the background subtraction
(i.e., $1.4\sigma$).
The extended X-ray emission is therefore soft.  This is consistent with
the extended emission being due to intra-group medium and it rules out a significant
contamination from unresolved, obscured sources.  Furthermore, the flux
limit for point sources in the 4Ms Chandra data at the location of the
group is about $2\times10^{-17}\rm erg\ s^{-1}\ cm^{-2}$ in the soft
band.  In order to explain the X-ray flux of the group, one needs at
least 15 unresolved point sources, which would correspond to an
over-density of factor 30 on the scale of 15 arcsec.  This is highly
unlikely to happen given the clustering properties of AGNs on these
scales.

\begin{figure*}
  \begin{center}
    \FigureFile(80mm,80mm){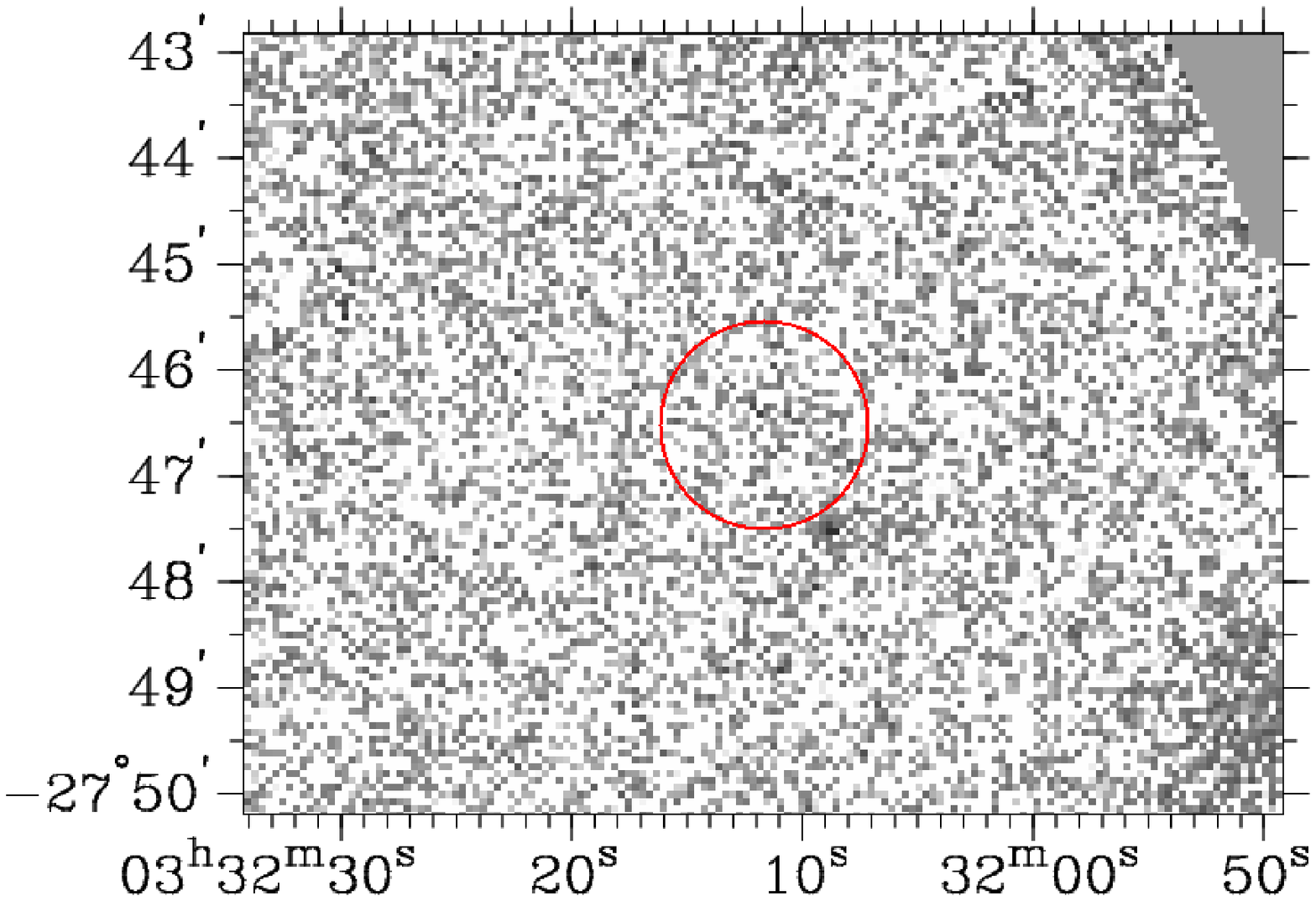}\hspace{0.5cm}
    \FigureFile(80mm,80mm){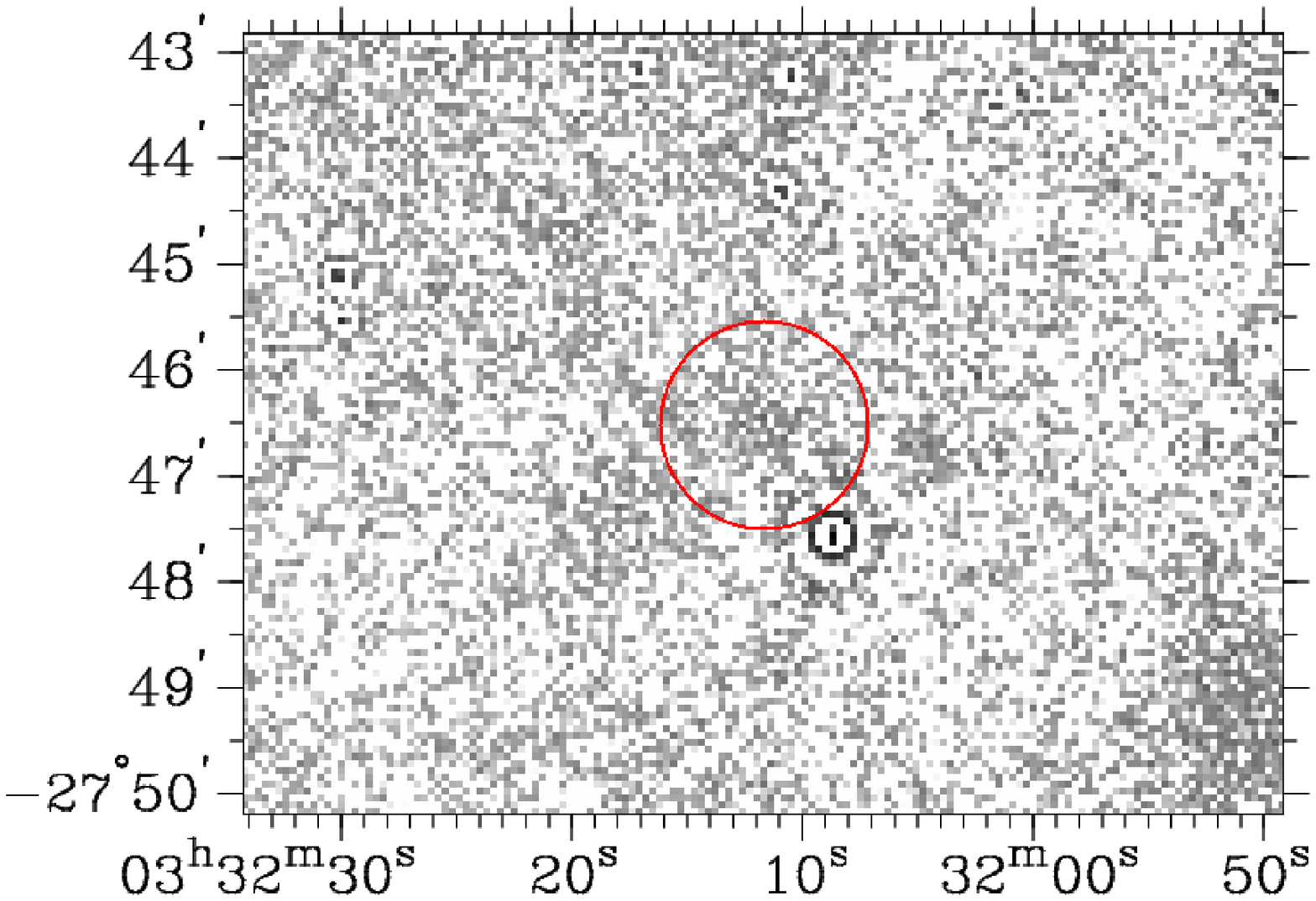}
  \end{center}
  \caption{
    Raw X-ray maps from Chandra (left) and XMM-Newton (right) with
    point sources subtracted. For details of the point source subtraction,
    the reader is refereed to Finoguenov et al. (2009, 2010).
    The circle is 1 arcmin radius around the group.
    The source on the bottom-right of the circle is a residual of
    the point source subtraction.
    The point source there is the 2nd brightest object  in CDFS
    in the soft-band with a flux of $4\times10^{-14}\rm\ erg\ s^{-1}\ cm^{-2}$ \citep{xue11}.
  }
  \label{fig:xray_map}
\end{figure*}

Given the slightly low signal-to-noise ratio of the X-ray detection,
we perform a further analysis by carefully subtracting 
point sources following Finoguenov et al. (2009, 2010).
We present a Chandra and XMM-Newton X-ray maps Fig. \ref{fig:xray_map}.
The X-ray emission is not readily seen by eye in the Chandra map, but
this is not unexpected because the significance of the detection is not very high.
From our experience, only high significance ($>5\sigma$) extended
emission can be easily recognized by eye on a 2D X-ray map.
If we measure the flux following \citet{finoguenov10}, we obtain
$f_{0.5-2.0 keV}=3.1\pm1.0\times10^{-16}\rm\ erg\ s^{-1}\ cm^{-2}$ ($3\sigma$)
from the Chandra image (i.e., $3\sigma$).
On the other hand, the extended emission is relatively easily seen
in the XMM-Newton data thanks to the larger photon collecting area.
The emission is clearly extended over $\sim20''$ in radius, which is consistent
with the wavelet detection in Fig. \ref{fig:color_pic}.
Although we carefully removed point sources in the XMM-Newton data,
one may worry about the point source contamination due to the poor
spatial resolution of XMM-Newton.
We argue that such contamination is likely small.
We measure an X-ray flux of $4.6\pm0.6\times10^{-16}\rm\ erg\ s^{-1}\ cm^{-2}$
at a significance level of $8\sigma$ in the XMM-Newton data.
This flux is consistent with the Chandra flux quoted above.
The overlap in the error bars is not large between the two fluxes
and this might be indicative of a small residual
of point source fluxes in XMM-Newton, but the contamination is small in any case.
These independent detections of the X-ray emission with consistent fluxes
from Chandra and XMM-Newton is strong evidence for
the extended X-ray emission.

Based on all the analyses presented here, we conclude that the detected
X-ray emission is real and it originates from hot thermal plasma
trapped in a deep potential well of the system.
We obtain consistent fluxes between Chandra and XMM-Newton, but
we base our analysis on the Chandra flux to be conservative
in what follows.

\begin{figure}
  \begin{center}
    \FigureFile(80mm,80mm){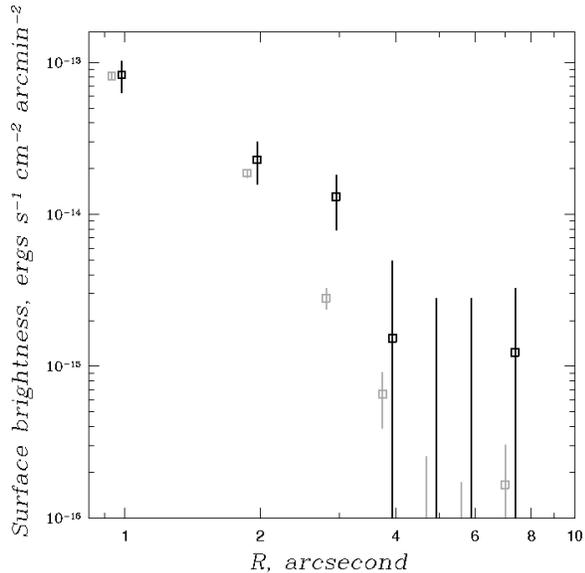}
  \end{center}
  \caption{ The black squares show the X-ray radial profile of the BGG.
    For comparison, the gray squares show the profile of a
    strong AGN separated from the BGG by only $12"$.  At this
    close proximity, the profiles of these two sources are directly
    comparable.  The points are shifted by 5\% in radius for clarity
    and the AGN profile is normalized to that of BGG at the center.  }
  \label{fig:cool_core}
\end{figure}

Finally, we note that we obtain tentative evidence for a cool core in the group.
The compact X-ray emission around the BGG (object-A as defined
later) is slightly more extended than the PSF as shown in Fig.
\ref{fig:cool_core}, although the difference is not large.  If
confirmed, this group would be the highest redshift system with a
cool core.  It may also help explain the galaxy population because a
group with a cool core is unlikely to have recently experienced a
violent merger event (we will elaborate on this point in Section 5).
However, given the large error bars in Fig. \ref{fig:cool_core},
the object may well be a faint AGN.  BGG is not detected in
the 2-8 keV band and the lower limit on the photon index is $1.14$
\citep{xue11}, which does not rule out either an AGN or cool core
origin for the emission.

\subsection{An over-density of galaxies at $z=1.6$}

Having detected the extended X-ray emission, we now look for the
optical counterpart using deep optical-NIR imaging.  For this, we have
run a red sequence finder as described in Section 2.3.  We have
detected a $6.2\sigma$ signal at $z\sim1.6$ after calibrating the model
red sequence with observed color of galaxies using the MUSIC catalog.
No other significant red sequence signal is found at other redshifts.
There are three galaxies at $z_{spec}=1.03$ within the X-ray contours
in Fig. \ref{fig:color_pic}, but we argue that they are not the primary
counterpart of the X-ray emission in Appendix 1.

In order to define group member candidates, we use $z_{phot,rafferty}$ and
$z_{phot,tanaka}$.
Given that photo-$z$'s are sensitive to a change in input photometry and templates,
we use both photo-$z$ estimates in order to reduce any systematic uncertainties
arising from photo-$z$'s.
We define a member candidate as a galaxy with

\begin{equation}
P_{gr}=\int_{z_{gr}-0.06\times (1+z_{gr})}^{z_{gr}+0.06\times (1+z_{gr})} P(z) dz \geq 0.16,
\end{equation}

\noindent
where $z_{gr}$ is the group redshift ($z_{gr}=1.61$ as discussed later)
and $P(z)$ is the photometric redshift probability distribution
function (PDF) from the fitting procedure.  \citet{rafferty11}'s
photo-$z$'s have an accuracy of $\Delta
z_{phot,rafferty}/(1+z_{spec})=0.06$ for faint sources.  Motivated by
this, we integrate the PDF over the interval of $z_{gr}-0.06\times
(1+z_{gr})$ to $z_{gr}+0.06\times (1+z_{gr})$.  If the integrated
probability exceeds 0.16 we define the galaxy as a group member
candidate.  In other words, if a galaxy is consistent with being at
$|z_{phot}-z_{gr}|/(1+z_{gr})<0.06$ within $1\sigma$, it is a candidate
for group membership.

The two photo-$z$'s agree for most objects, but there are some discrepant cases.
We define two categories here:

\begin{itemize}
\item {\it Good candidates} are those with photo$-z$'s consistent with being at $z_{gr}$
(see Eq. 1) both from $z_{phot,rafferty}$ and $z_{phot,tanaka}$.
\item {\it Likely candidates} are those that have $z_{phot,tanaka}$ consistent with
being at $z_{gr}$, while they have lower/higher $z_{phot,rafferty}$.  Note that even if both photo-$z$'s
are consistent, an object can fall in this category if its photo-$z$'s are not
very reliable (e.g., bad $\chi^2$).
\end{itemize}

\noindent
There is obviously a third category of galaxies with $z_{phot,rafferty}$ at $z_{gr}$,
but with lower/higher $z_{phot,tanaka}$.  There is no such galaxy in the core of the group
and we do not include the third category in the following discussions.
We refer to the bright ($H<24$) good/likely candidates as object-X, where
X is an upper/lower case letter, respectively, in the order of the $H$-band brightness.
We do not discuss fainter galaxies in detail
because their photo-$z$'s are uncertain and their field contamination probabilities
are high (see Section 3.4). 
Table \ref{tab:photoz} summarizes photo-$z$'s of the candidates.


\begin{table*}
  \begin{center}
    \caption{ Photo-$z$'s and $P_{gr}$ for good/likely member
      candidates based on the three different photo-$z$ catalogs.
      Object-g has $P_{gr}\geq0.16$ both from $z_{tanaka}$ and
      $z_{rafferty}$, but given the poor SED fit shown in Section 4, we
      classify it only as a likely candidate.  }
    \label{tab:photoz}
    \begin{tabular}{cccrcrcc} 
      ID       & $z_{phot,tanaka}$   　 & $P_{gr,tanaka}$ &$z_{phot,rafferty}$ & $P_{gr,rafferty}$ & $z_{phot,cardamone}$\\\hline
      object-A & $1.62^{+0.08}_{-0.08}$ & $0.97$ & $1.60^{+0.03}_{-0.02}$ & $1.00$ & $1.61^{+0.02}_{-0.02}$ \\
      object-B & $1.64^{+0.06}_{-0.09}$ & $0.94$ & $1.59^{+0.03}_{-0.02}$ & $1.00$ & $1.62^{+0.03}_{-0.02}$ \\
      object-C & $1.64^{+0.06}_{-0.08}$ & $0.95$ & $1.56^{+0.03}_{-0.03}$ & $0.97$ & --- \\
      object-D & $1.59^{+0.26}_{-0.10}$ & $0.63$ & $1.58^{+0.34}_{-0.13}$ & $0.49$ & $1.67^{+0.05}_{-0.04}$\\
      object-E & $1.56^{+0.07}_{-0.09}$ & $0.87$ & $1.41^{+0.04}_{-0.04}$ & $0.16$ & --- \\
      object-F & $1.64^{+0.12}_{-0.13}$ & $0.72$ & $1.67^{+0.05}_{-0.10}$ & $0.86$ & --- \\
      object-g & $1.57^{+0.23}_{-0.25}$ & $0.43$ & $1.59^{+0.07}_{-0.10}$ & $0.91$ & $2.94^{+0.11}_{-0.14}$ \\
      object-h & $1.61^{+0.11}_{-0.10}$ & $0.80$ & $1.18^{+0.06}_{-0.06}$ & $0.07$ & --- \\

    \end{tabular}
  \end{center}
\end{table*}

Now, let us go back to Fig. \ref{fig:color_pic}, in which we show
the spatial distribution of good and likely candidates.   We observe a clear
concentration of galaxies at $z_{phot}\sim1.6$ around the extended
X-ray emission.  Most of them are within the X-ray contours, which
roughly corresponds to $0.5r_{200}$, where $r_{200}$ is the radius
within which the mean interior density is 200 times the critical
density of the universe.  
The BGG is not at the center of the X-ray emission, but is
offset to the West.  This might be partly due to a strong concentration
of the X-ray point sources in the Eastern part of the system causing
low-level contamination.  Based on the 3rd nearest-neighbor density
using galaxies with $H<24$ and $P_{gr}\geq0.16$, we observe a $5\sigma$
over-density.

We emphasize that this is not the first detection of an over-density at
this redshift in the CDFS.  In fact, a large scale structure at this
redshift was already found by \citet{kurk09}, see their Fig. 2.  Our
group is on the edge of the GOODS field and is located about 5 arcmin
(2.5 Mpc) away from the main region of over-density studied by
\citet{kurk09}.  
In our X-ray map, there are a few low-significance sources
around the main \citet{kurk09} region, but we defer detailed discussions on
the region in Finoguenov et al. (in prep).
On the other hand, our group shows the clear extended X-ray emission
as discussed above.  It is likely the first gravitationally bound,
X-ray bright system found in the $z=1.6$ structure.

\begin{figure*}
  \begin{center}
    \FigureFile(160mm,80mm){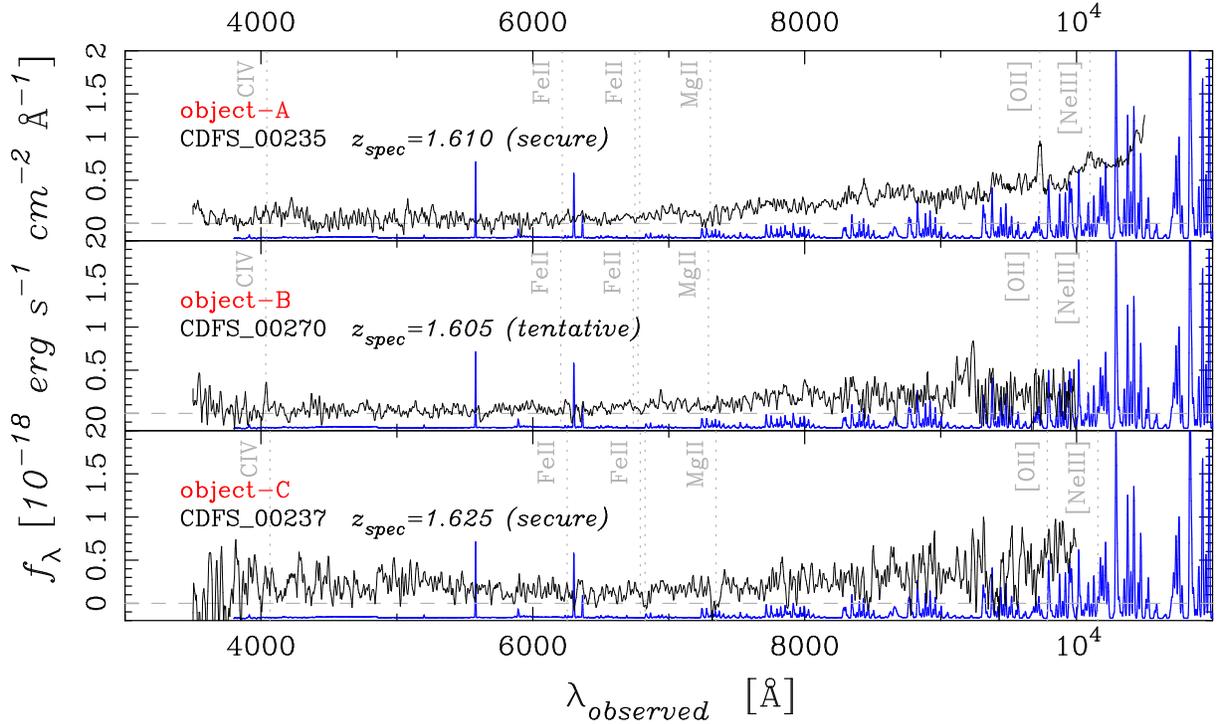}
  \end{center}
  \caption{
    The spectra of three galaxies around the group
    from the K20 survey \citep{mignoli05}.
    The blue lines are arbitrarily scaled sky spectra
    (the spectra do not come with associated noise spectra).
    Prominent spectral features are indicated with labels.
    The spectra are object-A, B, and C from top to bottom, respectively,
    and the spectra are smoothed over $15\rm\AA$ with a top-hat filter.
  }
  \label{fig:spec}
\end{figure*}

As can be seen in Fig. \ref{fig:color_pic}, there are a large number of spectroscopic
redshifts from the literature.  The BGG is confirmed at $z=1.61$,
which gives a strong constraint on the central redshift of the group.
Object-C is also confirmed to be a group member, at $z=1.625$ (see below).
The redshifts indicated in Fig. \ref{fig:color_pic} are secure redshifts only, but
there are a large number of less secure spectroscopic redshifts from the literature.
We summarize secure and less secure redshifts around $z\sim1.6$ in Fig. \ref{fig:spec}.
We briefly comment on each spectrum below.

{\bf Object-A:} This is the BGG of the group.  It exhibits a
strong emission line, which is identified as {\sc [oii]} redshifted by
$z_{spec}=1.61$.  The galaxy has $z_{phot,rafferty}=1.60$ and
$z_{phot,tanaka}=1.62$, consistent with the spectroscopic redshift
(Table \ref{tab:photoz}).  As discussed above, the X-ray emission
around object-A might be a cool core.  If we translate the observed
{\sc [oii]} luminosity into SFR using the formula from
\citet{kennicutt98}, we obtain a SFR of 3.6 $\rm M_\odot\ yr^{-1}$.  We
do not apply a correction for slit loss and this SFR is thus a lower
limit.  As discussed below, we find a SFR of $<2\rm\ M_\odot\ yr^{-1}$
from SED fitting and from the MIPS flux, which is lower than that from
{\sc [oii]}.  This suggests that object-A hosts an AGN that contributes
to the observed {\sc [oii]} emission.
\nocite{tanaka11a,tanaka11b}Tanaka (2011a,b) show that an excess
emission line luminosity that cannot be fully accounted for by star
formation can be efficiently used to identify AGNs.  It is likely that
object-A hosts an AGN.  It remains unclear whether the slightly
extended component is a cool core, but it is possible that the X-ray
emission is a combination of point-like AGN and extended cool core.

{\bf Object-B:} This object is given $z=1.605$ in the K20 survey \citep{mignoli05} with 
a low quality flag.  There is a possible {\sc Civ} line, although
the feature is not convincing.  This galaxy is an X-ray point source
and the possible {\sc Civ} line might be due to the AGN.
We note that the photometric redshifts of the galaxy are $z_{phot,rafferty}=1.59$
and $z_{phot,tanaka}=1.64$, being consistent with the spectroscopic redshift
if the feature is real.

{\bf Object-C:} The galaxy shows relatively clear Fe{\sc ii} and Mg{\sc
  ii} absorption features.  It is given $z=1.615$ in the K20 catalog,
but we slightly tweak the redshift ($\Delta z=+0.01$) so that we can
fit the absorption lines better.  However, we do not argue that our
redshift is more precise than the original one.  Rather, this gives us
a level of uncertainty in the redshift.  The photo-$z$'s of this galaxy
are $z_{phot,rafferty}=1.56$ and $z_{phot,tanaka}=1.64$.  The K20
catalog assigns a low quality flag to this object, but given the
relatively strong absorption features and consistent photo-$z$, we
consider it is a secure redshift.

Overall, we have two secure redshifts from the K20 survey both at
$z\sim1.61$.  The central redshift of the group is likely the redshift
of the BGG, $z=1.61$.  An interesting point worth noting here is
that three of the nine good/likely candidates are AGNs (objects A, B
and g).  Table \ref{tab:physical_props} summarizes their absorption
corrected X-ray luminosities taken from \citet{xue11} together with
other physical parameters derived in Section 4.  Object-g is a
moderately strong AGN with $L_X\sim5\times10^{43}\rm\ erg\ s^{-1}$ at
0.5--8 keV\footnote{ This object is nominally a good candidate given
  the consistent $z_{phot,tanaka}$ and $z_{phot,rafferty}$.  However,
  $\chi^2_\nu$ of $z_{phot,tanaka}$ is bad (see Table \ref{tab:physical_props})
  and also this
  object is given $z_{phot}=2.8$ by \citet{xue11}.  We downgrade the
  object to a likely candidate.  We scale its $L_X$ from $z=2.8$ to
  $z=1.6$ and quote it here.  }.  The other two, one of which is BGG,
are weak AGNs with $L_X\sim4\times10^{42}\rm\ erg\ s^{-1}$.  If
we assume that all the six good candidates are group members, we obtain
an AGN fraction and binomial statistical error of
$0.33^{+0.28}_{-0.21}$.  If we include likely candidates, this fraction
becomes $0.38^{+0.23}_{-0.20}$.  Although the uncertainty is large,
such a high fraction of X-ray AGNs has not been observed in any other
cluster.  \citet{martini07} reported that a fraction of AGNs with
$L_X>10^{42}\rm\ erg\ s^{-1}$ hosted by $M_R<-20$ mag galaxies, which
very roughly matches with our selection of group members in terms of
stellar mass, is about 1\% in local clusters.  Previous studies have
shown that the AGN fraction in clusters seems to increase with
increasing redshift \citep{eastman07,martini09}.  An increasing AGN
fraction is also observed in a spectroscopic study \citep{tanaka11}.
Recently, \citet{fassbender12} reported on a detection of AGNs
in high-z clusters.
This group may extend this increasing AGN fraction to higher redshifts.  
However, a larger sample of groups at this redshift is obviously needed to
establish it.


\begin{table*}
  \begin{center}
    \caption{ Physical properties of galaxies derived from the SED fits
      along with absorption-corrected X-ray luminosities at rest-frame
      0.5--8.0 keV from \citet{xue11} and SFRs from MIPS assuming the
      MIPS to total IR luminosity conversion from \citet{reddy06} and
      to SFRs using \citet{kennicutt98}.  MIPS fluxes are taken from
      the MUSIC catalog \citep{santini09}.  We fit the SEDs of object-A and C
      with redshifts fixed to their spectroscopic redshifts.  Object-g is contaminated by
      AGN and so we fail to reliably measure its stellar mass and SFR
      from the SED fits or from MIPS.  Also, this object was assigned
      $z_{phot}=2.8$ in \citet{xue11}.  We rescale the X-ray luminosity
      to $z=1.6$.
      The quoted uncertainties are statistical uncertainties only.
    }
    \label{tab:physical_props}
    \begin{tabular}{cccccccc} 
      ID       &  $\chi_\nu$ & $z_{phot,tanaka}$ & $\rm M_{stellar}$ [$10^{10}\rm M_\odot$] &  SFR [$\rm M_\odot\ yr^{-1}$] & $L_X \rm [10^{42}\ erg\ s^{-1}]$ & SFR$_{\rm MIPS}\ \rm [M_\odot\ yr^{-1}$]\\\hline
      object-A & $1.6$ & $z_{spec}=1.61$      & $31.6^{+0.1}_{-13.9}$ & $0.79^{+0.03}_{+0.01}$  & 4.3 & $<2.0$\\
      object-B & $0.3$ & $1.64^{+0.06}_{-0.09}$ & $7.9^{+1.1}_{-0.1}$ & $0.06^{+0.16}_{-0.01}$    & 4.7 & $1.5\pm0.6$\\
      object-C & $0.6$ & $z_{spec}=1.62$      & $4.5^{+0.1}_{-0.1}$ & $0.03^{+0.01}_{-0.01}$    & --- & $<1.4$\\
      object-D & $0.7$ & $1.59^{+0.26}_{-0.10}$ & $1.4^{+1.2}_{-0.4}$ & $22.91^{+13.41}_{-4.71}$  &--- &  $25\pm1$\\
      object-E & $0.2$ & $1.56^{+0.07}_{-0.09}$ & $1.8^{+1.1}_{-0.6}$ & $0.05^{+0.08}_{-0.04}$    & --- & $0.6\pm0.6$\\
      object-F & $0.4$ & $1.64^{+0.12}_{-0.13}$ & $1.3^{+0.6}_{-0.4}$ & $0.06^{+0.04}_{-0.03}$    & --- & $<1.9$\\
      object-g & $2.5$ & $1.57^{+0.23}_{-0.25}$ & --- & ---  & 49  & ---\\
      object-h & $0.3$ & $1.61^{+0.11}_{-0.10}$ & $0.6^{+0.4}_{-0.2}$ & $0.08^{+0.32}_{-0.06}$    & --- & $<1.3$\\

    \end{tabular}
  \end{center}
\end{table*}

 Finally, we note that there is a bright red galaxy whose photo-$z$'s are consistent
 with $z=1.6$ ($z_{phot,tanaka}=1.63$, $z_{phot,rafferty}=1.45$, and
 $z_{phot,cardamone}=1.57$), but its secure emission-line redshift from
 the K20 spectrum is $z=1.149$.
 The object is located 
 at $\rm R.A.=03^h32^m13^s.01$, $\rm Dec.=-27^\circ46'37''.9$,
just outside of the group X-ray emission in Fig. \ref{fig:color_pic}.
 We cannot fit the SED of this object at $z=1.149$
 with any reasonable $\chi^2$ and there are also large residuals after
 subtracting the best-fit model galaxy for the morphological analysis
 described in Section 5.  It is possible that there are two overlapping
 galaxies with the brighter galaxy at $z\sim1.6 $ dominating the overall
 flux, and the fainter one at $z=1.15$ responsible for the emission line
 in the spectrum.  Deep near-IR spectroscopy of the object will be
 useful to measure its continuum redshift.  To be conservative, we do
 not include this object in the analysis.

\subsection{Color-magnitude diagrams}

\begin{figure*}[ht]
  \begin{center}
    \FigureFile(160mm,80mm){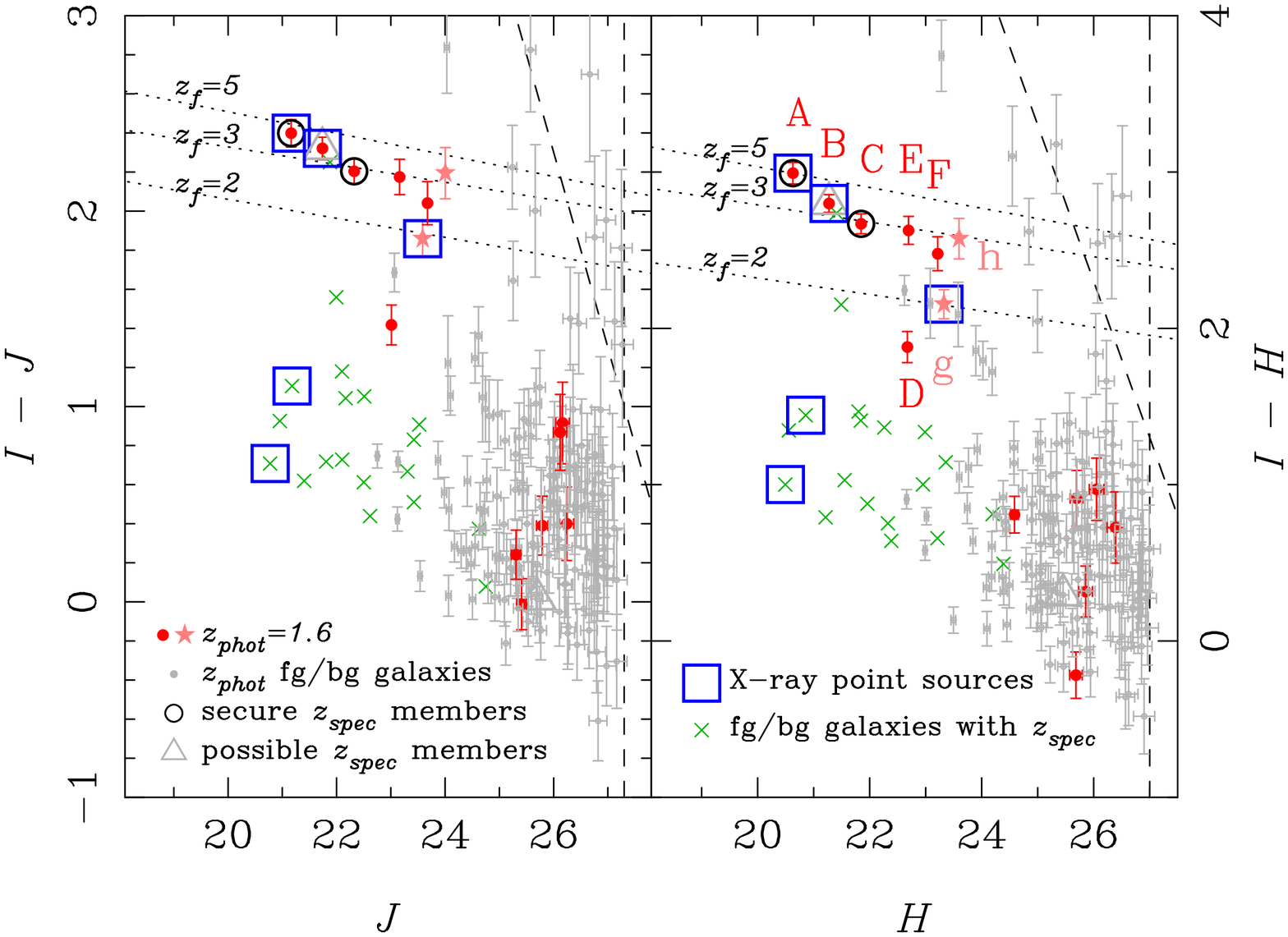}
  \end{center}
  \caption{ Color-magnitude diagrams based on the CANDELS data.  The
    left and right panels respectively show $I-J$ vs. $J$ and $I-H$ vs
    $H$.  All the galaxies within $r_{200}$ of the group center are
    plotted here.  The large points are galaxies with $P_{gr}\geq0.16$.
    Among them, the red/pink symbols with $H<24$ are the good/likely
    candidates selected from photo-$z$ and are labeled as object-A to h
    in the order of the $H$-band magnitude.  The gray points are
    fore-/background galaxies based on photo-$z$.  The dark circles and light triangles
    indicate secure and possible spectroscopic members, respectively.
    The crosses indicate fore-/background galaxies with secure
    spec-$z$'s.
    The blue squares show X-ray sources.  The vertical and slanted
    dashed lines are $5\sigma$ limits.  The dotted lines show model red
    sequence of galaxies formed at $z_f=2,\ 3,$ and 5.}
  \label{fig:cmd1}
\end{figure*}

A ubiquitous feature of rich galaxy groups and clusters in the low
redshift Universe, in addition to over-density of galaxies, is the
tight red sequence.  Red, passively evolving galaxies in groups and
clusters form a sequence of galaxies on the color-magnitude diagram
with a shallow tilt due to the mass-metallicity relation
\citep{kodama97}.  The group fortunately falls within the footprint of
CANDELS, providing publicly available, high quality, deep optical and
near-IR HST images.  Based on the CANDELS data, we present $I-J$ vs $J$
and $I-H$ vs $H$ diagrams in Fig. \ref{fig:cmd1}.

The photo-$z$ selected galaxies with $P_{gr}\geq0.16$ form a surprisingly
prominent red sequence.  There are 2 secure members and 6 membership
candidates with $H<24$ in the group.  Strikingly, a significant
fraction have red colors on both diagrams.  The brightest galaxy (A) is
slightly redder than the other galaxies and is consistent with being
formed at $z_f=5$.  The remaining galaxies are on the model red
sequence formed at $z_f=3$ with two exceptions.  Object-g is located
near the $z_f=2$ sequence, but this object is a powerful X-ray
point-source and the photometry may be affected by the central AGN.
Object-D is clearly bluer, and is forming stars as shown in the next
section.  We find that -- except for these two objects -- all massive
galaxies are classified using a color-color diagram (e.g.,
\cite{williams09,balogh09}) as passive, not as dusty star forming
galaxies.  There are a few faint blue galaxies with $H>24$ and
$P_{gr}\geq0.16$, but the bright members are almost exclusively red.

We argue here that the dominance of bright red galaxies is not
due to biases in the photo-$z$'s.  Photo-$z$'s tend to be more accurate
for quiescent galaxies than for star forming ones due to the prominent
4000\AA\ break of quiescent galaxies.  This bias may result in
an enhanced red fraction, but as shown in Fig. \ref{fig:cmd1},
a significant fraction of bright blue galaxies have secure spec-$z$'s
and none of them is at the group redshift.
This is strong evidence for the bright group members being predominantly red.
We further note that our photo-$z$'s are reasonably good even for
blue galaxies.
We find that, for $z_{phot,rafferty}$, an outlier rate\footnote{
Outliers here are defined as those with $|z_{phot}-z_{spec}|/(1+z_{spec})>0.15$
following the standard definition.
} of red galaxies with
$I-H>1.5$ (which roughly selects red sequence galaxies)
at  $1<z_{spec}<2$ is $8\pm3\%$.  The outlier rate for bluer galaxies is slightly larger;  $12\pm3\%$.
The outlier rates are similar for $z_{phot,tanaka}$;  $7\pm3\%$ and $14\pm3\%$
for red and blue galaxies, respectively.
In both photo-$z$'s, the dispersion is 20\% larger for blue galaxies
than for red galaxies.
Photo-$z$'s are indeed better for red galaxies, but even
for blue galaxies, the outlier rate is only $\sim13\%$.
There are 8 galaxies with $H<24$ and $I-H<1.5$ that do not have spec-$z$'s.
Even if we miss 1 blue galaxy among them,
it is still fair to say that
the bright group galaxies are mostly red.

The dominance of bright red galaxies is surprising in a low mass
group at such a high redshift.  The red sequence extends to
$H\sim23.5$, which corresponds to $\sim m_{H}^*+2$.  There is no
photo-$z$ selected red sequence galaxy at fainter magnitude, suggesting
a truncation of the red sequence.  This truncation magnitude ($m^*+2$)
is similar to that observed in groups at $z=1.2$
\citep{tanaka07}\footnote{ \citet{tanaka07} used the $K_s$-band to
  measure the truncation magnitude, while we use the $H$-band here.
  However, we focus only on passive galaxies, and the truncation
  magnitude does not depend on a pass-band when expressed with respect
  to $m^*$.  }.
However, we should be reminded that photo-$z$'s become less accurate
at fainter magnitudes.  The apparent truncation might be partly
due to increased errors in the photo-$z$'s.

\begin{figure}[ht]
  \begin{center}
    \FigureFile(85mm,80mm){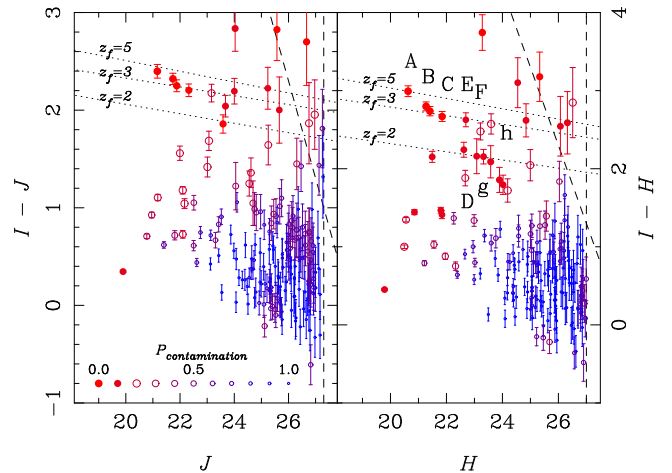}
  \end{center}
  \caption{
    The same color-magnitude diagram as in Fig. \ref{fig:cmd1}, but here we do not
    use photometric redshifts.  We instead perform statistical subtraction of
    fore-/background galaxies.  Galaxies with smaller, bluer symbols are more
    likely to be contamination. 
  }
  \label{fig:cmd2}
\end{figure}

We further check if the observed red sequence is real using statistical
subtraction of fore-/background galaxies without relying on
photo-$z$'s.  Using the recipe described by \citet{tanaka05}, we
compute a probability that a given galaxy is in the fore-/background
(i.e. contamination).  We use the entire CDFS CANDELS field as a
control sample for the field subtraction.  Fig. \ref{fig:cmd2} shows
the contamination probability of each galaxy.  We do not take a
Monte-Carlo approach for the statistical subtraction as done by
\citet{tanaka05}; instead we explicitly show the contamination
probability.  It is clear that the bright galaxies on the red sequence
are unlikely to be contamination.  There is only a small statistical
probability that any of the photo-$z$ selected good or likely members
from Fig. \ref{fig:cmd1} are fore-/background galaxies, and Fig.
\ref{fig:cmd2} shows that the red sequence is clearly real.  There are
many faint, blue galaxies around the group, but most of them are likely
to be fore-/background galaxies as indicated by their large
contamination probabilities. 
There seems to be a few sheets of structure in the foreground of
the group and bright blue galaxies have somewhat small
contamination probabilities.
But, most of them are spectroscopically confirmed foreground galaxies
as shown in Fig. \ref{fig:cmd1}.

To sum up, there is plenty of evidence to support the convincing
detection of an X-ray group at $z=1.6$ in the CDFS.  First, the
extended X-ray emission is detected, which strongly suggests that this
is a physically bound system.  Second, a clear over-density of galaxies
is observed around the X-ray emission at $z_{phot}\sim1.6$.  The BGG
is spectroscopically confirmed at $z=1.61$, which is likely the
central redshift of the group.  Third, the member candidates form a
tight red sequence.  All these results lead us to conclude that this is
a galaxy group located at $z=1.61$.

\subsection{X-ray Properties of the Group}

Using the group redshift of $z=1.61$, we summarize physical properties
of the group measured from X-rays in Table \ref{tab:xray_props}.
From the X-ray luminosity, we estimate the mass of the group using
the calibration against weak-lensing mass from \citet{leauthaud10}:

\begin{equation}
\frac{M_{200} E(z)}{M_0} = A \left(\frac{L_X E(z)^{-1}}{L_{X,0}} \right)^\alpha,
\end{equation}

\noindent
where $E(z)=\sqrt{(1+z)^3\Omega_M+\Omega_\Lambda}$, 
$M_0=10^{13.7}\rm\ M_\odot$, $L_{X,0}=10^{42.7}\ \rm erg\ s^{-1}$,
$\log (A)=0.03\pm0.06$, $\alpha=0.64\pm0.03$.
This group turns out to be a low-mass system with $M_{200}=(3.2\pm0.8)\times10^{13}\rm M_\odot$.
The uncertainty here include X-ray flux uncertainty and
the scatter in the $M_{200}-L_X$ relation.
There is a possible Eddington bias that this group is over-bright
for its mass, but its effect is estimated to be $\lesssim10$\% \citep{leauthaud10}.
The $M_{200}-L_X$ relation used here is measured at $z<1$ and
and we have extrapolated it to $z=1.6$.  This will introduce
a systematic error, amount of which is not straightforward
to estimate because no calibration is available at $z>1$ due to the lack
of statistical sample of groups and clusters there.
At any rate, it is likely the lowest mass system confirmed so far at $z>1.5$
(c.f. $6\times10^{13}\rm\ M_\odot$, the mass of another $z=1.6$ group in SXDF
measured in the same way, \cite{tanaka10a}). Such a high-$z$ low-mass
group is an interesting object to study the origin of the environmental
dependence of galaxy properties and may be an early progenitor which
will grow into a present-day cluster.
Motivated by this, we perform detailed analyses of the properties of
the group members in the following sections.

\begin{table*}
  \begin{center}
    \caption{
      Physical properties of the group measured from X-ray.  The 3rd and 4th columns
      show the X-ray flux in the observed 0.5 -- 2.0 keV band and X-ray luminosity
      at rest-frame 0.1--2.4 keV both within $r_{500}$ (this is a radius,
      within which the mean interior density is 500 times the critical density of the universe).
      For details of the flux estimate, refer to \citet{finoguenov07}.
      The 5th column is $r_{200}$ and the last column shows $M_{200}$, which is mass
      contained within $r_{200}$ calibrated against weak-lensing mass at $z<1$ \citep{leauthaud10}.
      The uncertainties quoted are statistical only and our $M_{200}$ estimate is subject to
      systematic uncertainties (see text for details).
    }
    \label{tab:xray_props}
    \footnotesize
    \begin{tabular}{ccccccc}
      R.A.        & Dec.        & $f_X\ (0.5-2\ \rm keV)$ & $L_X\ (0.1-2.4\ \rm keV)$ & $r_{200}$ & $\rm M_{200}$ \\\hline
      $03^h\ 32^m\ 11^s.7$ & $-27^\circ\ 46'\ 34''$ &  $(3.1\pm1.0)\times10^{-16}\rm\ erg\ s^{-1}\ cm^{-2}$ & $(1.8\pm0.6)\times10^{43}\rm\ erg\ s^{-1}$ & 44'' or 370 kpc & $(3.2\pm0.8)\times10^{13}\rm M_\odot$\\
    \end{tabular}
  \end{center}
\end{table*}

\section{Stellar Populations of the Group Members}

\subsection{Spectral energy distribution fitting}

As shown in Section 3, the group consists of 8 relatively bright
galaxies, most of which have red colors.  
In this section,
we attempt to put tighter constraints on the stellar populations of
the galaxies by using the multi-wavelength data from MUSIC \citep{santini09}.

We fit the observed broad-band photometry with a suite of model
templates generated with an updated version of the \citet{bruzual03}
code, which incorporates an improved treatment of the thermally
pulsating AGB stars.
Given that most of the group galaxies exhibit red
colors, it is reasonable to assume exponentially declining star
formation histories (c.f., \cite{maraston10}).  We assume the Chabrier
IMF \citep{chabrier03} and solar metallicity for all the models.  Model
parameters such as age and metallicity are degenerate and we prefer to
use solar metallicity models only because super-solar or sub-solar
metallicity models introduce larger degeneracies and they do not
reproduce the mass-metallicity relation \citep{tanaka11a}.
We use the attenuation law from \citet{calzetti00}.  We fit the
observed photometry ($u_{VIMOS},\ u_{38},\ u_{50},\ B,\ V,\ R,\ I,\ z,$
$\ J,\ H,\ K_s,\ 3.5\mu m,\ 4.6\mu m,$$\ 5.7\mu m,\ 8.0\mu m$) with the
model templates in the linear flux scale using the standard
$\chi^2$-minimizing technique.

Model templates do not always perfectly match with the observed SEDs of
galaxies and we apply a crude correction to the templates to reduce
the SED mismatches.  We use galaxies with spectroscopic redshifts
drawn from the literature and fit them with redshifts fixed to
their spectroscopic redshifts using the MUSIC photometry.
Because the spectroscopic galaxies span a wide range in redshift,
the observed photometry covers a wide range of rest-frame wavelengths.
By comparing observed fluxes with the best-fitting model fluxes,
we measure systematic flux stretches in the model templates
as a function of rest-frame wavelength.  We also measure a flux dispersion,
which can be used as an uncertainty in the models at a given wavelength.
We apply this template error function to all the templates used in the SED fitting.
For details, the reader is refereed to Appendix 2.
We note that the template error function increases errors on
all the parameters derived from the SED fitting, but we deem that
the increased errors are more realistic estimates.  Note as well
that the template error function does not completely remove the systematics.
There are remaining systematics arising from, e.g., the assumption
of the star formation histories of galaxies, but they are more
difficult to evaluate and are not corrected for in this paper.

For object-g, which is a strong X-ray source, we use a set of composite templates of
AGNs and galaxies.  We use AGN templates from \citet{polletta07} and apply extinction
of $\tau_V=-$0.2, 0, +0.2 and +0.4.  The negative extinction makes sense because
the empirical templates already include extinction.
We restrict galaxy templates to relatively young ages of $<3$ Gyr with $\tau=1$ Gyr.
We then normalize the spectra at 5500\AA\ and merge an AGN template
and a galaxy template with ratios of 1:2, 1:1, 2:1 and 4:1.
We fit object-g using these AGN-galaxy composite templates.

\begin{figure*}
  \begin{center}
    \FigureFile(80mm,80mm){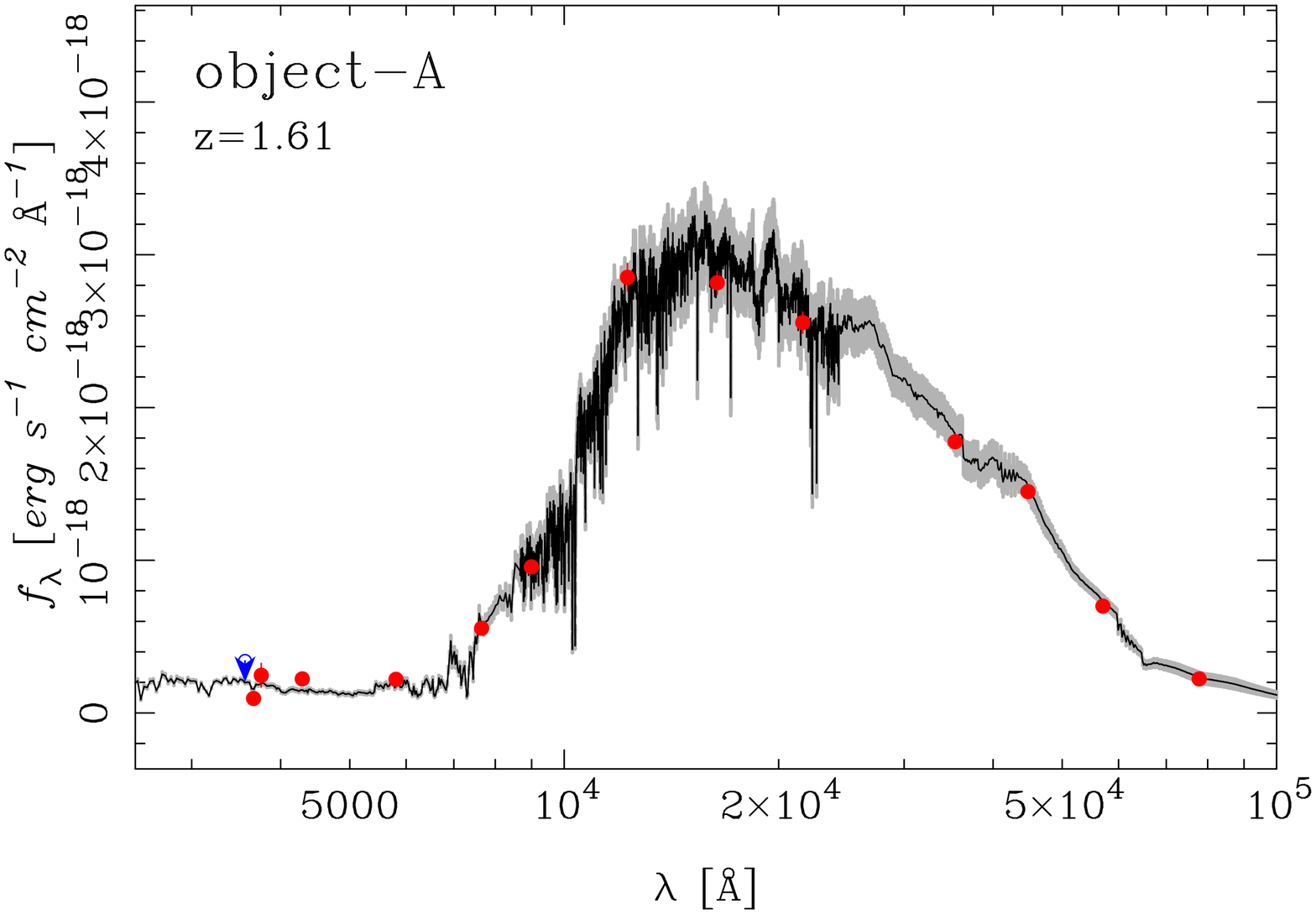}\hspace{0.5cm}
    \FigureFile(80mm,80mm){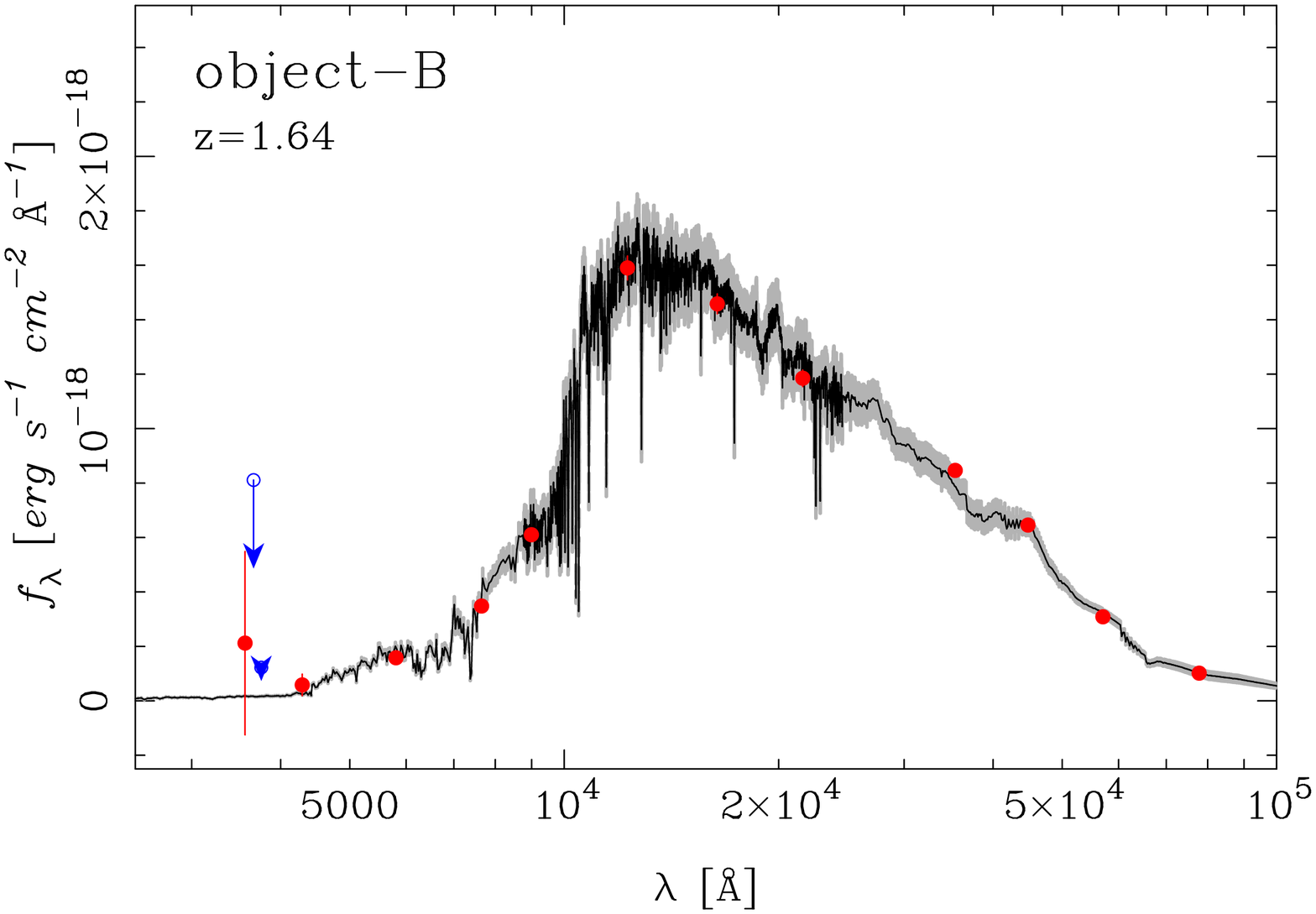}\\
    \FigureFile(80mm,80mm){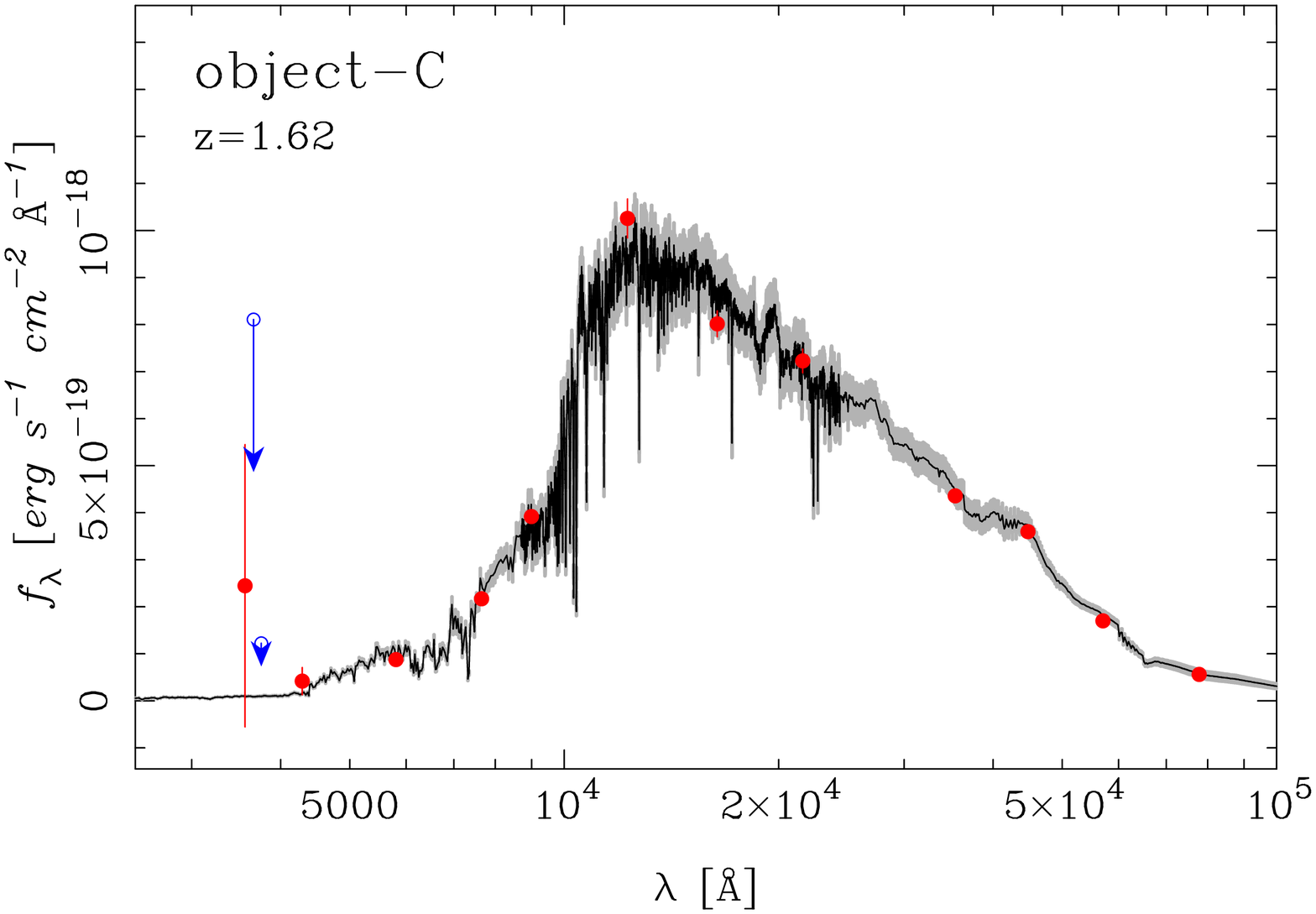}\hspace{0.5cm}
    \FigureFile(80mm,80mm){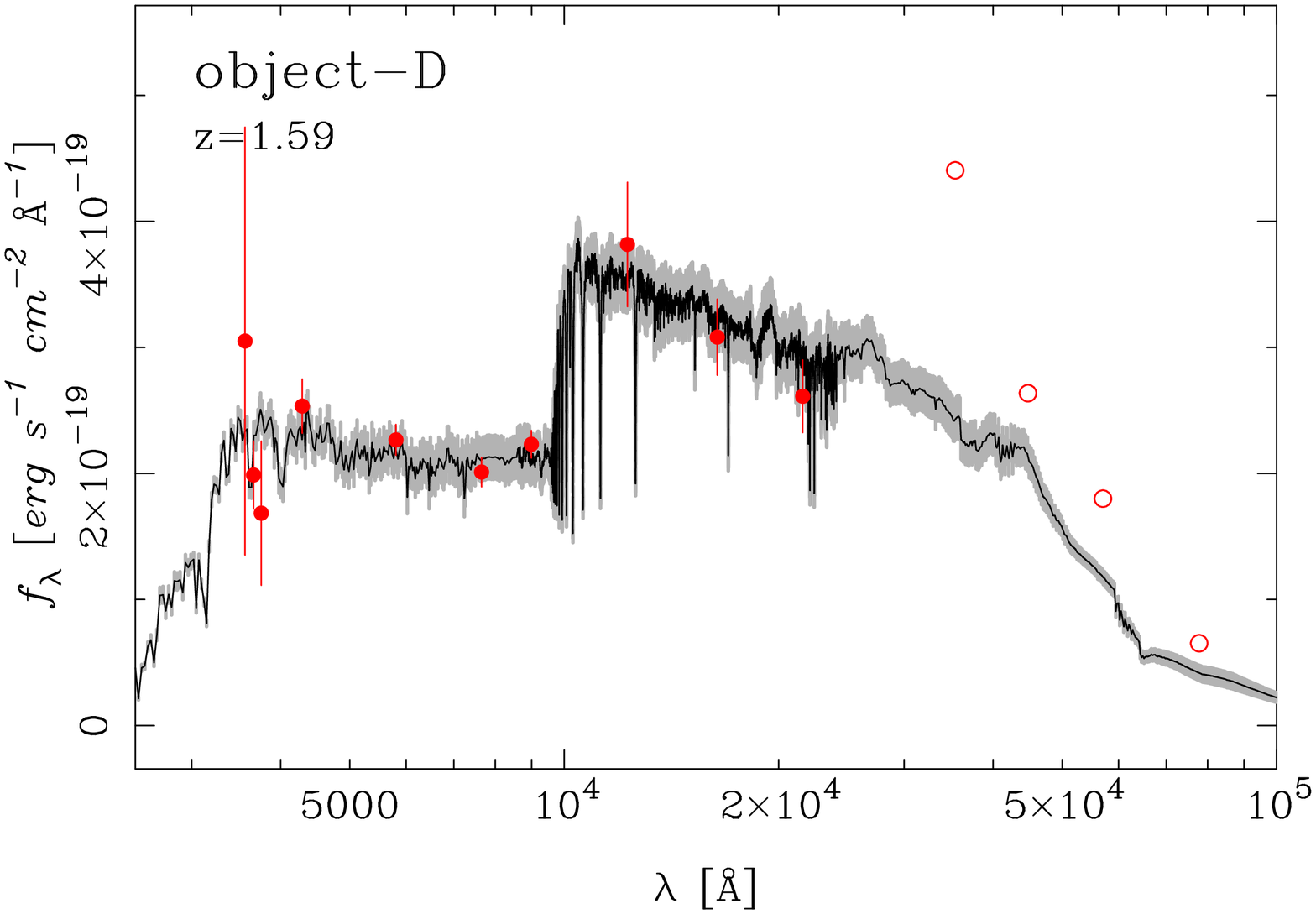}\\
    \FigureFile(80mm,80mm){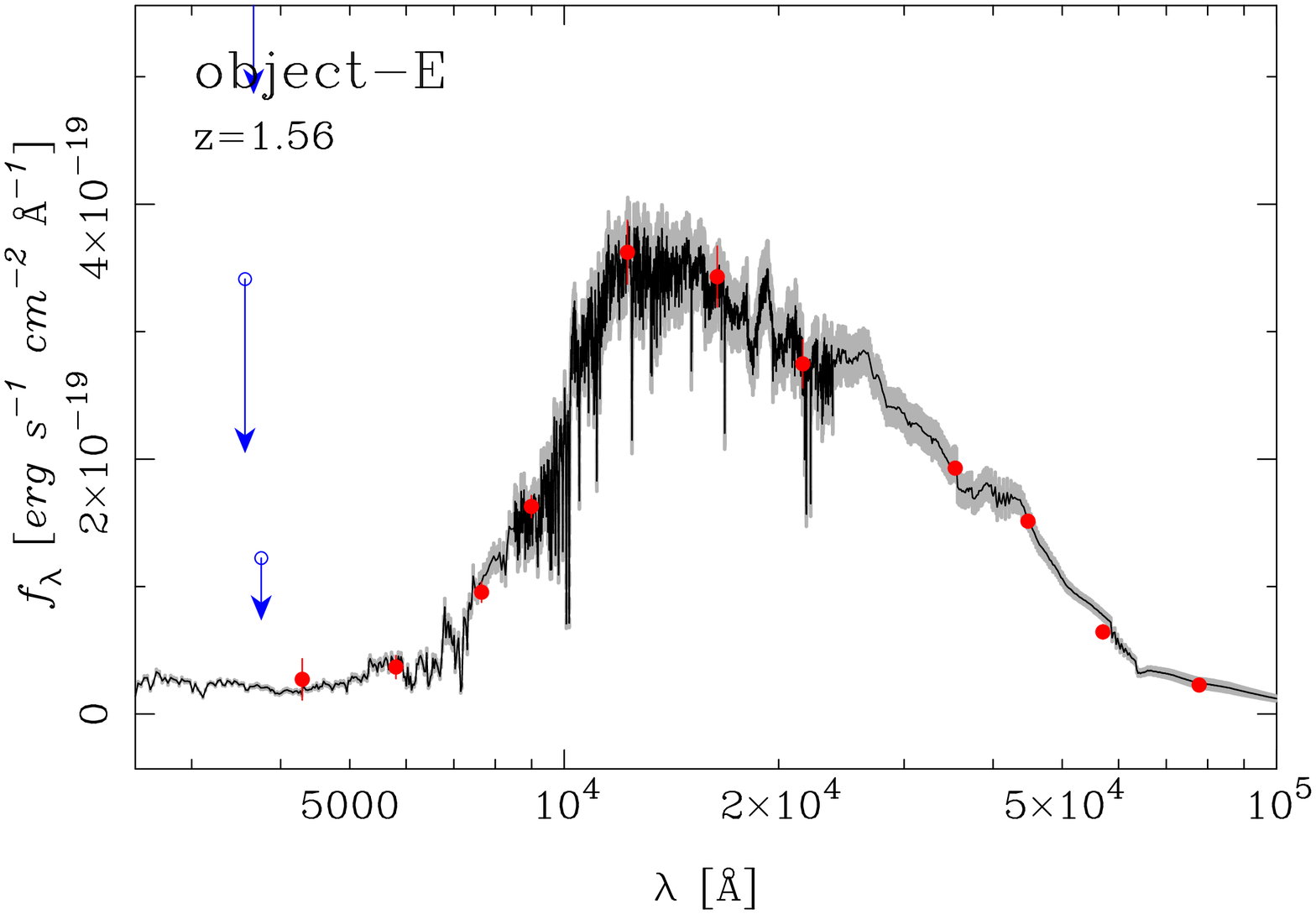}\hspace{0.5cm}
    \FigureFile(80mm,80mm){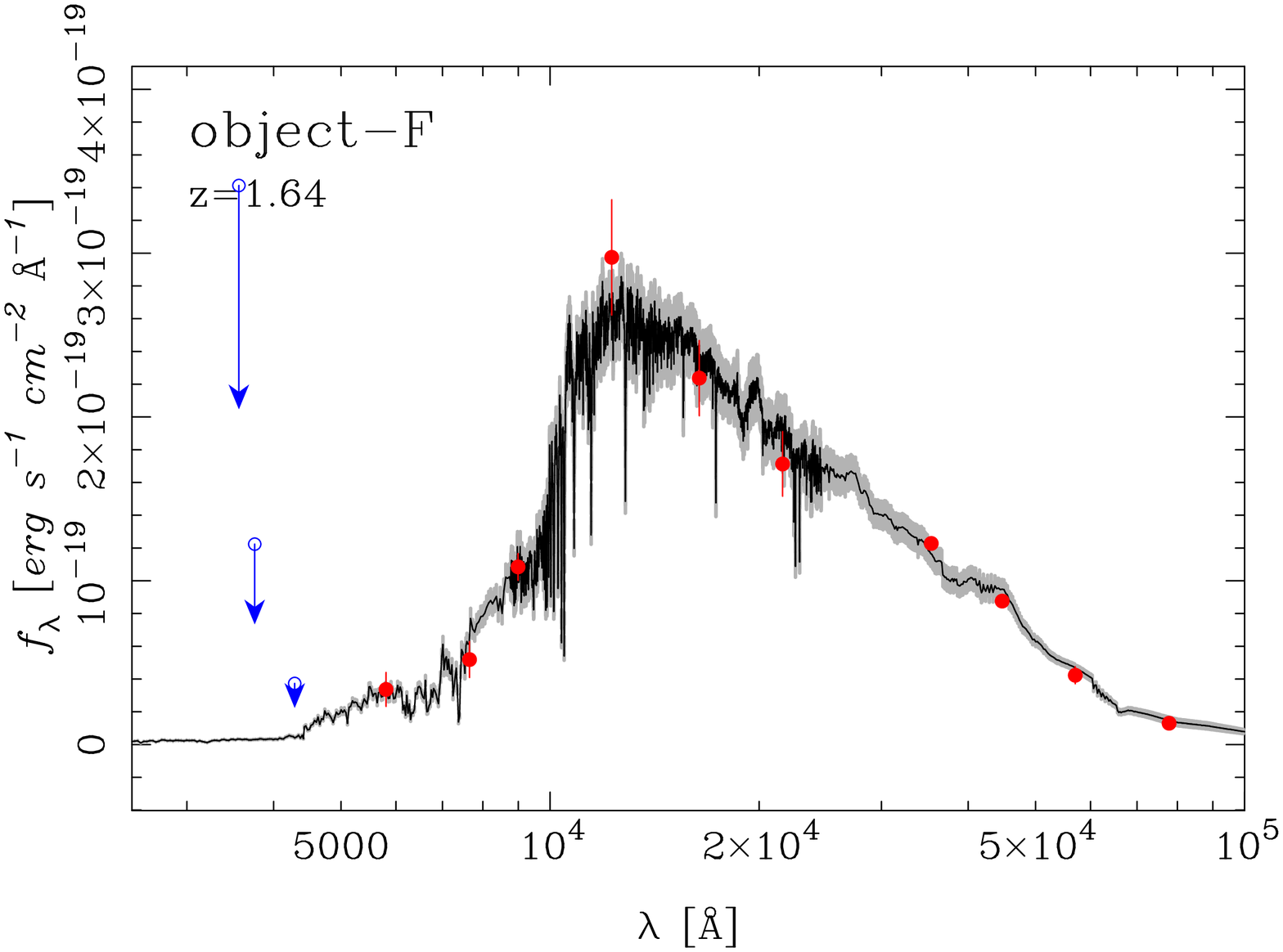}\\
  \end{center}
  \caption{ The best-fit model spectra over-plotted with the observed
    photometry for object A to F.
    The shaded areas around the best-fit spectra show
    the template uncertainties derived in the Appendix 2.
    The best-fit redshift is labeled 
    in each plot.
    For object-A and C, the fits are
    performed at $z_{spec}$. The points with arrows are
    upper limits.  Object-D is close to the BGG and its IRAC
    photometry is over-estimated due to the blending.  We do not use
    the IRAC photometry for the SED fitting for this object.
  }
  \label{fig:sedfits1}
\end{figure*}
\begin{figure*}
  \begin{center}
    \FigureFile(80mm,80mm){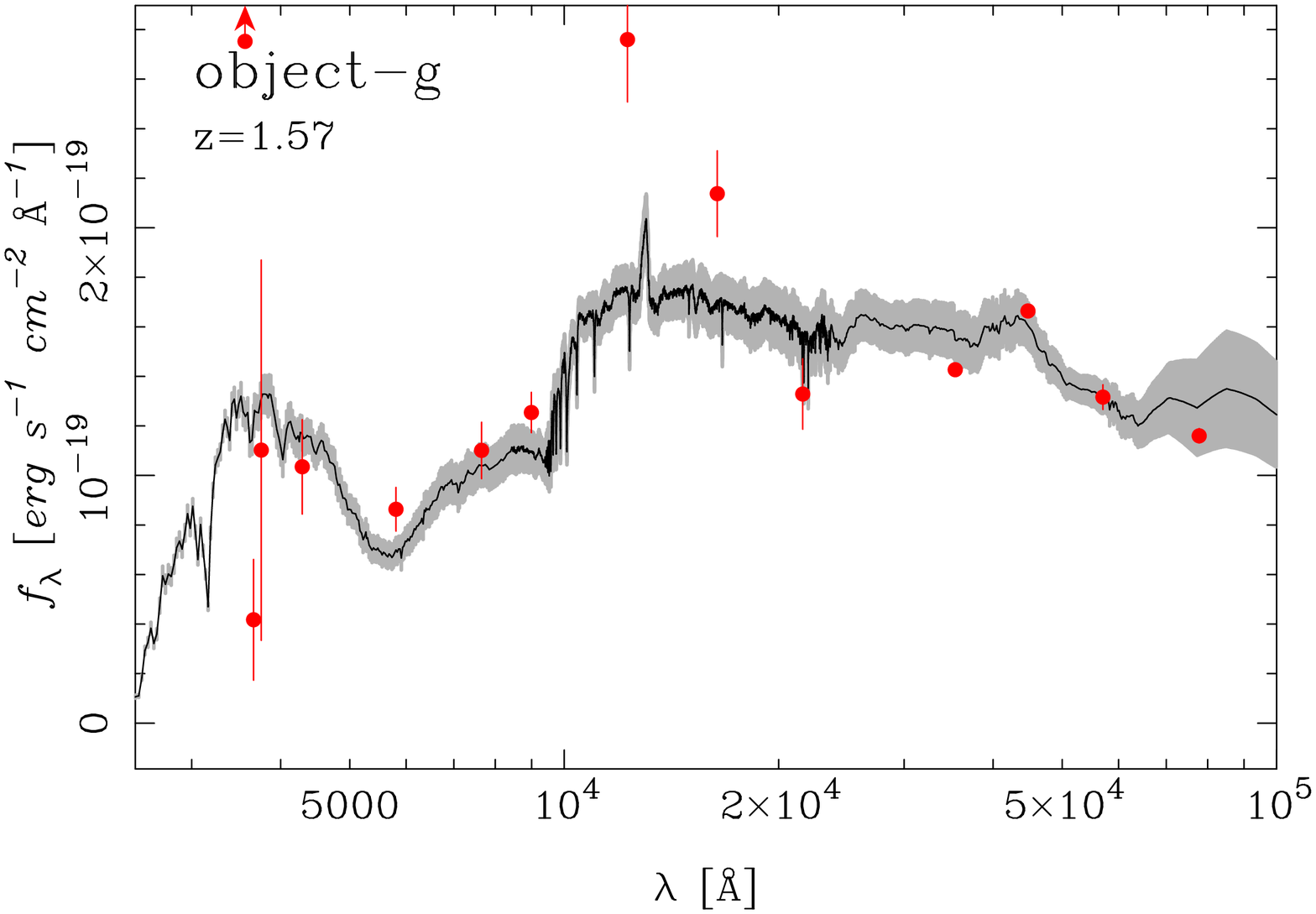}\hspace{0.5cm}
    \FigureFile(80mm,80mm){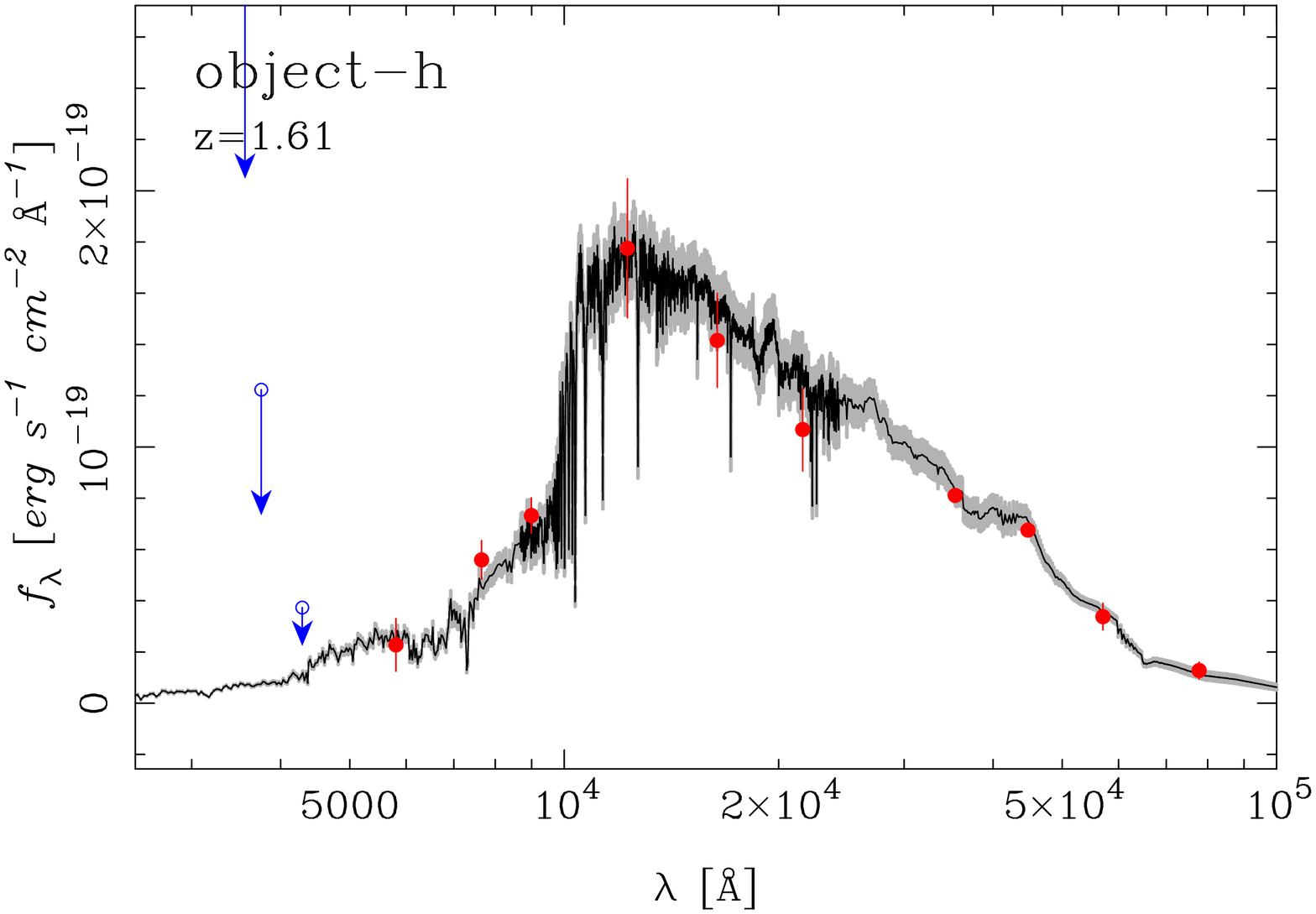}\\
  \end{center}
  \caption{
    As in Fig. \ref{fig:sedfits1}, but for object g and h. 
  }
  \label{fig:sedfits2}
\end{figure*}

The results of the fitting for the six good candidates are shown in Fig. \ref{fig:sedfits1}
and the two likely candidates in Fig. \ref{fig:sedfits2}.
The derived physical parameters of the galaxies are summarized in Table \ref{tab:physical_props}.
For object A and C, we perform the SED fitting with redshifts fixed
to their spectroscopic redshifts.
The fits are generally good
and the objects are consistent with being located at $z\sim1.6$.
Object-D is very close to BGG and its photometry in the IRAC bands
is severely contaminated by the BGG.  Therefore, we remove the IRAC photometry 
from the SED fit.
All the galaxies except for object-g show
a prominent 4000\AA\ break in between the $z$ and $J$-bands.
This strongly suggests that the galaxies have old stellar populations,
as expected from the tight red sequence.

In Section 3, we defined {\it likely} candidates as those with
discrepant photo-$z$'s between ours and \citet{rafferty11} or those
with a poor SED fit.  Let us briefly mention the two likely candidates
in Fig.  \ref{fig:sedfits2}.  The fit for object-g is not very good,
but it is a strong X-ray source and the bad fit might be due to
variability (e.g., $U_{50}$ is much brighter than the other $U$-band
photometry).  Although we have consistent photo-$z$'s both from
$z_{phot,tanaka}$ and $z_{phot,rafferty}$, the fit is not very good
($\chi^2_\nu=2.5$ including the template error).  Due to this bad fit, we define this galaxy as a
likely candidate.  For object-h, we have inconsistent photo-$z$'s:
$z_{phot,tanaka}=1.61$ and $z_{phot,rafferty}=1.18$.  However, our fit
is very good as shown in Fig. \ref{fig:sedfits2}.  Also, the galaxy is
on the red sequence (Fig.  \ref{fig:cmd1}) and is spatially located
close to the center of the group.  For these reasons, we include it in
our analysis as a likely candidate.  We note that the primary
conclusions of this paper do not significantly change if we exclude
these likely candidates.

\subsection{Physical properties}

Now, let us turn our attention to the physical properties of 
the galaxies derived from the SED fits.
We find that the BGG has a stellar mass of about $3^{+0}_{-1}\times10^{11}\rm M_\odot$,
which is $\gtrsim3$ times more massive than the 2nd most massive galaxy (object-B).
If we integrate the stellar mass of the good/likely group members,
we obtain $\sim5^{+0}_{-2}\times10^{11}\rm \ M_\odot$.  Object-g is excluded here
because our stellar mass estimate is not very reliable due to the AGN contamination,
but it probably does not significantly contribute to the overall stellar mass
budget given its faint $H$-band magnitude (it is about 3 magnitudes fainter
than the BGG).
Also, we do not include faint galaxies with $H>24$ in the statistics here, and
thus the number quoted here is a lower limit.
However, the total stellar mass is dominated by the few brightest galaxies
and it is unlikely to change significantly if we included those low-mass galaxies.
The stellar mass to total mass ratio within $r_{200}$
is then $\sim2\%$.
This fraction is nearly a factor of 2 smaller than that observed at
$z<1$ by \citet{giodini09},
but it is consistent with more recent estimates by \citet{leauthaud12}
and Connelly et al. (2012 submitted).

The most striking result here is that the group galaxies have very low SFRs
for $z=1.6$ galaxies ($<0.1\ \rm M_\odot yr^{-1}$).
We recall that SFRs quoted in Table  \ref{tab:physical_props} are subject
to systematic uncertainty due to the assumption of the exponentially
declining SFRs.
Also, SFRs from SED fits may not be very precise at low SFRs \citep{pacifici12}.
However, most of the objects are actually not detected in the very deep MIPS
data at a significant
level as shown in Table \ref{tab:physical_props}, which gives independent
evidence for very low SFRs of the group galaxies.
Only object-D is detected in MIPS at a significant level and its SFRs from
the SED fitting and MIPS are consistent (SFR$\sim20\ \rm M_\odot\ yr^{-1}$).
The BGG is likely forming stars at a low rate of $\sim1\rm M_\odot\ yr^{-1}$,
which is consistent with the upper limit from the MIPS flux.
The only object with active star formation is object-D and it is actually bluer
than the red sequence in Fig. \ref{fig:cmd1}.  

\begin{figure}
  \begin{center}
    \FigureFile(50mm,80mm){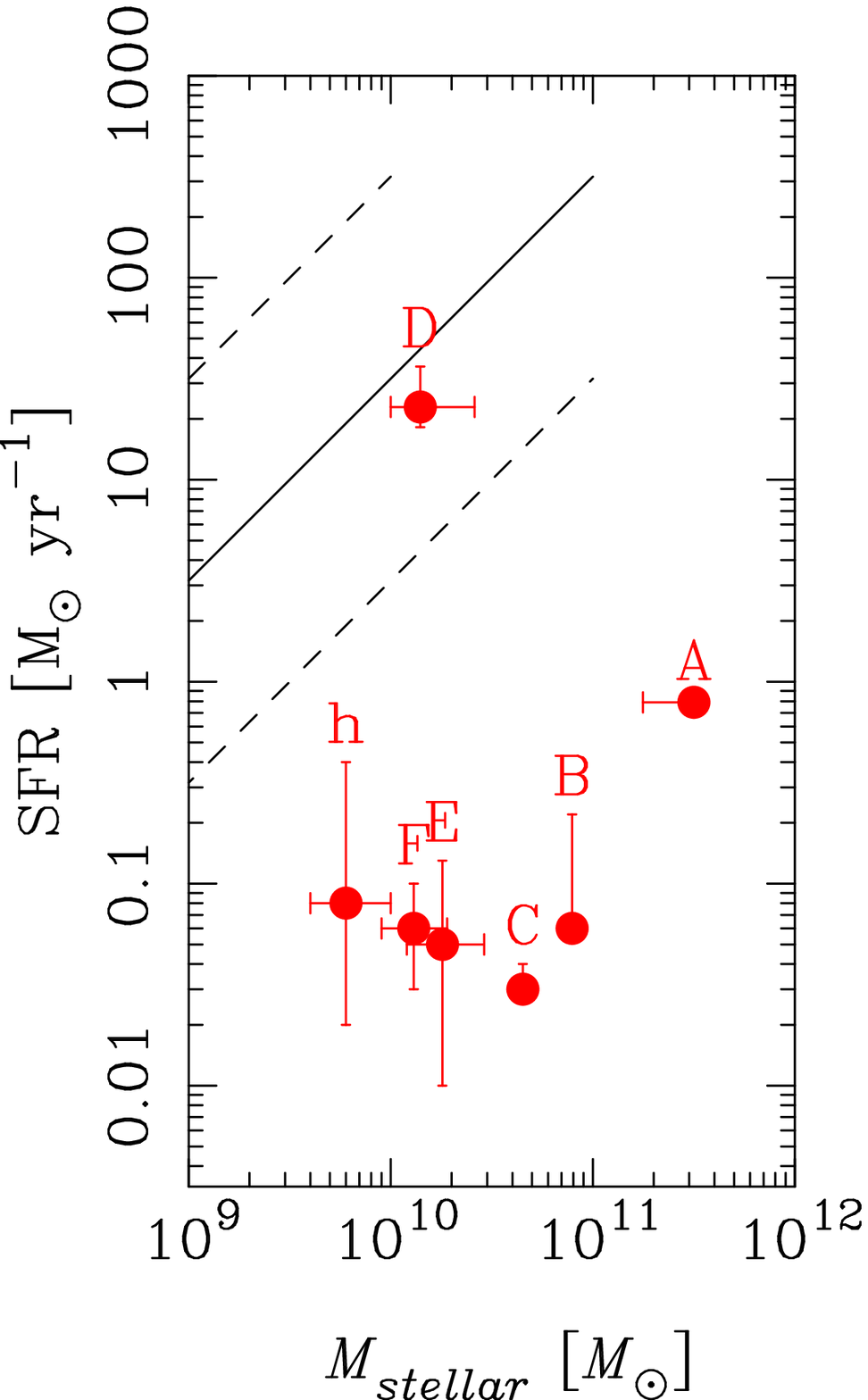}
  \end{center}
  \caption{
    SFR plotted against stellar mass.  The solid line shows a sequence of
    star forming galaxies at $1.5<z<2.5$ \citep{wuyts11}.
    The dashed lines show an approximate width of the sequence ($\pm1$ dex).
    The large symbols are the $z=1.6$ galaxies.
    Object-g is not plotted here because we cannot measure its SFR.
  }
  \label{fig:sfr_vs_smass}
\end{figure}

To quantify how quiescent the group galaxies are, we show a SFR vs stellar
mass diagram in Fig. \ref{fig:sfr_vs_smass}.  Star forming galaxies form
a sequence on this diagram \citep{elbaz07,noeske07} and we show a sequence
for star forming galaxies in the field at $1.5<z<2.5$ from \citet{wuyts11}.
Our group is at the lower bound of the redshift range, but a sequence
at $0.5<z<1.5$ shifts downwards only by $\sim0.3$ dex \citep{wuyts11}
and our conclusion here remains unchanged.
As shown in Fig. \ref{fig:sfr_vs_smass},
most of the $z=1.6$ galaxies are far below the sequence and they have $2-3$
orders of magnitude lower SFRs.
Only object-D is consistent with the star forming sequence.
This clearly shows that the $z=1.6$ group galaxies have suppressed
star formation.  In this case, there is no evidence that the group
environment has led to enhanced star formation.

We note that \citet{quadri12} quoted a quiescent fraction of
$\sim40\%$ in the densest environment at $1.5<z<2.0$ for
galaxies with $>10^{10.2}\rm\ M_\odot$.
If we use object-A to F, which have $\gtrsim10^{10.2}\rm M_\odot$,
the quiescent fraction in the group is $83^{+14}_{-29}\%$.
Also, \citet{popesso12} studied a total SFR within a cluster
normalized by the cluster mass and showed that the normalized SFR increases towards higher
redshifts.  They studied the system found by \citet{kurk09} at $z=1.6$ in CDFS
and measured a normalized total SFR of $\sim500\rm\ M_\odot\ yr^{-1} / 10^{14}M_\odot$.
If we use object-A to h, we obtain $\sim80\rm\ M_\odot\ yr^{-1} / 10^{14}M_\odot$ for our group,
which is a factor of 6 lower than the \citet{kurk09} system.
These results further illustrate that the group is a fairly quiescent place for $z=1.6$.

To summarize, we have shown that the group is dominated by red galaxies with very low SFRs.
Even galaxies with $\sim10^{10}\rm M_\odot$ are not actively forming stars.
These galaxies clearly fall below the sequence of star forming galaxies
(Fig. \ref{fig:sfr_vs_smass}).
This is a surprising result that the environmental dependence of galaxy star formation
is already in place at $z=1.6$ in this group.
It might suggest that at least some groups are dominated by red, quiescent galaxies from early times.
But, let us defer further discussions on this point to Section 6 and focus on 
another important aspect of galaxies -- morphology -- in the following section.

\section{Morphologies of the Group Members}

\subsection{Surface brightness fitting}

For the morphological analysis, we use GALFIT \citep{peng02,peng10}
to measure structural parameters of
the group galaxies. GALFIT performs two-dimensional surface
brightness fitting of a galaxy using an input image, a bad pixel map, a PSF image,
and noise image.  We use the WFC3 F160W image from CANDELS,
which probes the rest frame $r$-band at $z=1.6$,
as an input image of the group.  We use a nearby unsaturated star for
the PSF image.  The noise image is generated from the weight map \citep{koekemoer11}.
We fit the galaxies with a single S\'{e}rsic profile.

Due to the fact that a galaxy group is crowded with galaxies by definition,
one must decide how to determine the boundary of the galaxy of interest
and how to exclude the neighboring objects' light from a fit.
We attempt to solve these problems by  defining elliptical regions around
the galaxy of interest and neighboring objects to specify their boundaries.
The elliptical regions are determined using Source Extractor \citep{bertin96}.
After some experiments, we find that 6 and 2.5 times of half-light radius
are respectively optimum for the galaxy and neighboring objects.
A larger radius for the object of interest is necessary as it must contain
not only light from the outskirts of the galaxy, but also the sky background.
All the initial values for the fitting parameters (position of galaxy
center, integrated magnitude, half-light radius, axis ratio, position
angle, S\'{e}rsic parameter) are set to those derived
by Source Extractor except for the last parameter.  The initial
S\'{e}rsic index is set to 2, which is a boundary of early and late-type galaxies
so that we do not bias our fits.

\begin{figure*}
  \begin{center}
    \FigureFile(30mm,30mm){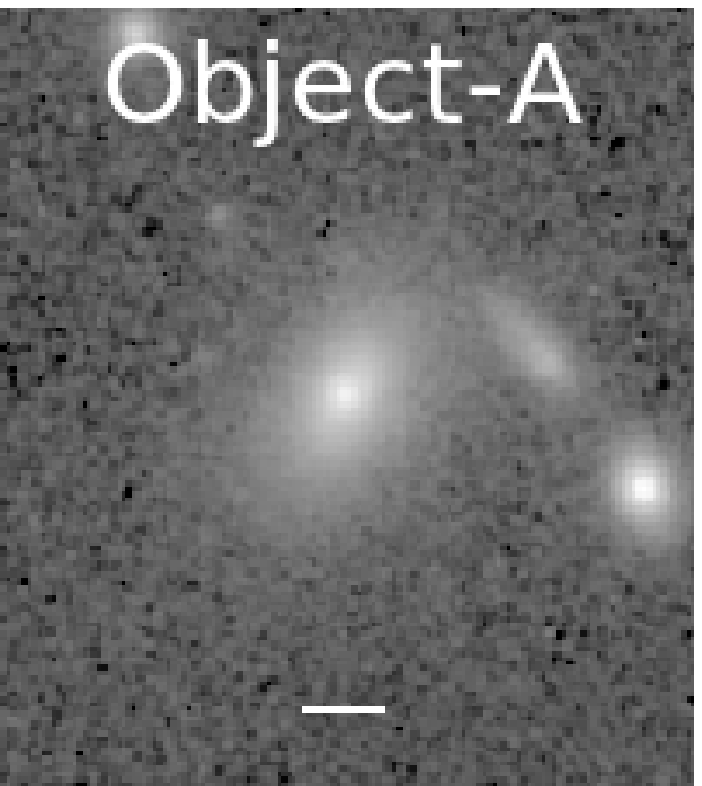}
    \FigureFile(30mm,30mm){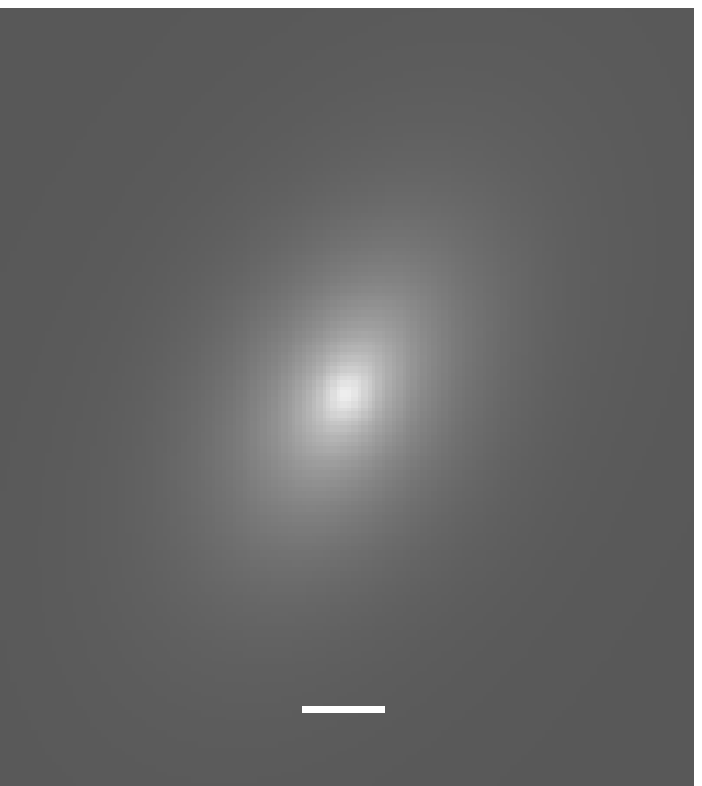}
    \FigureFile(30mm,30mm){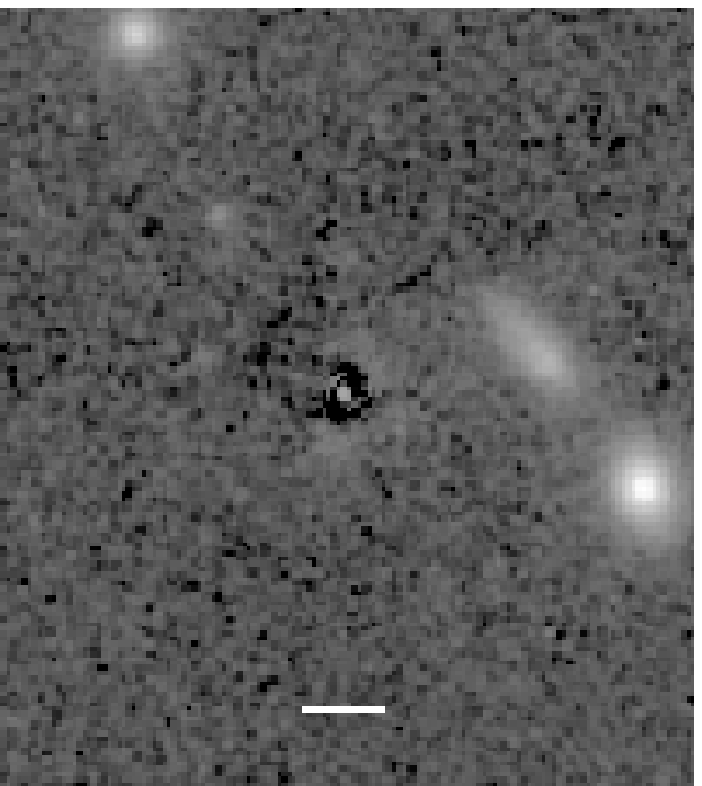}
    \FigureFile(30mm,30mm){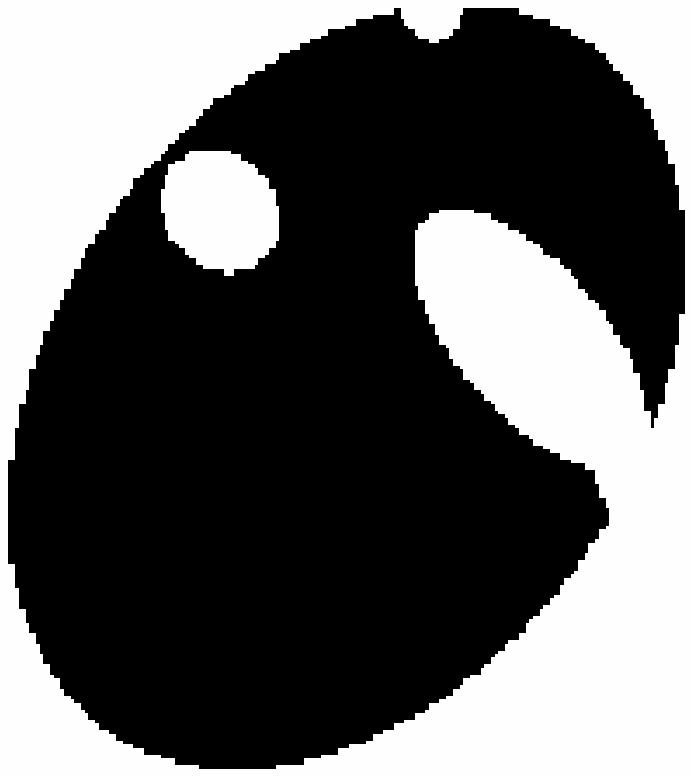}\\\vspace{0.5cm}
    \FigureFile(30mm,30mm){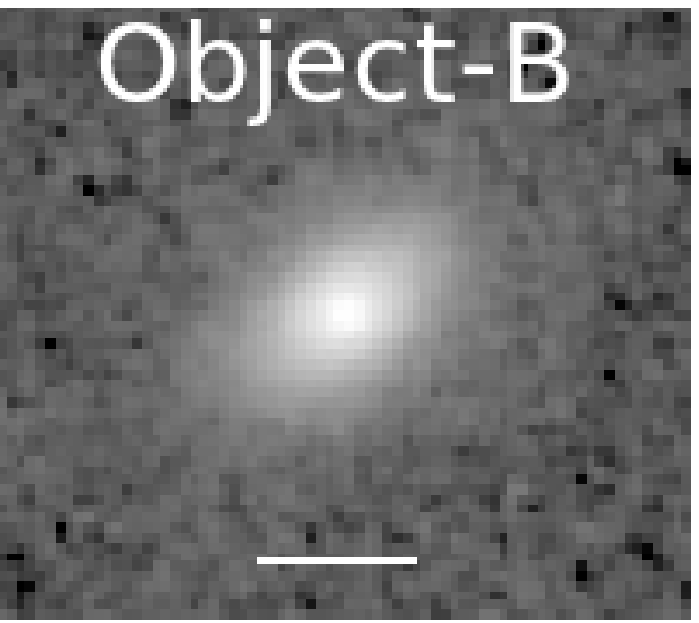}
    \FigureFile(30mm,30mm){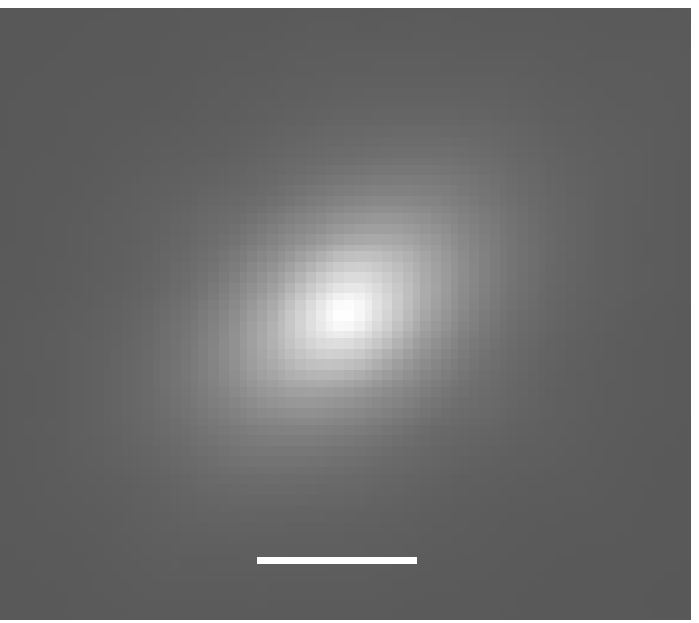}
    \FigureFile(30mm,30mm){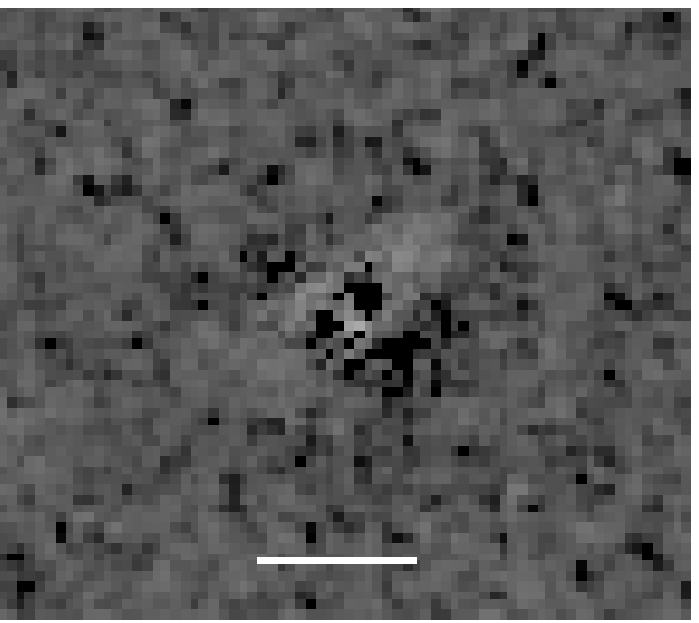}
    \FigureFile(30mm,30mm){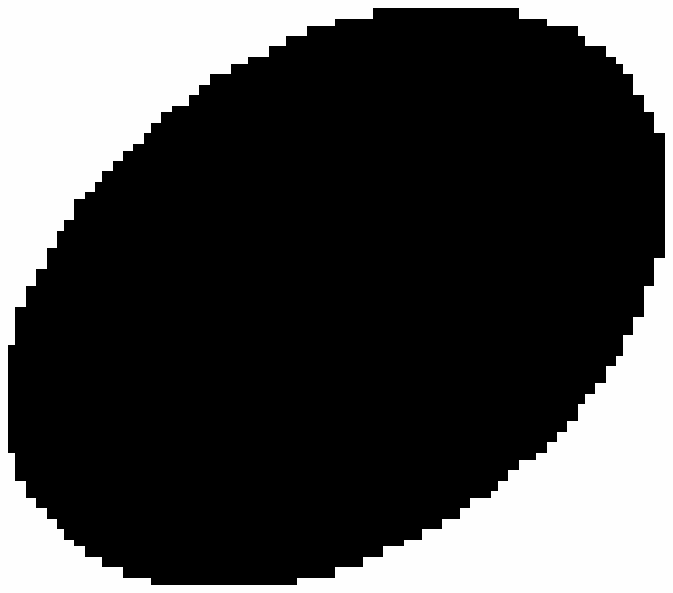}\\\vspace{0.5cm}
    \FigureFile(30mm,30mm){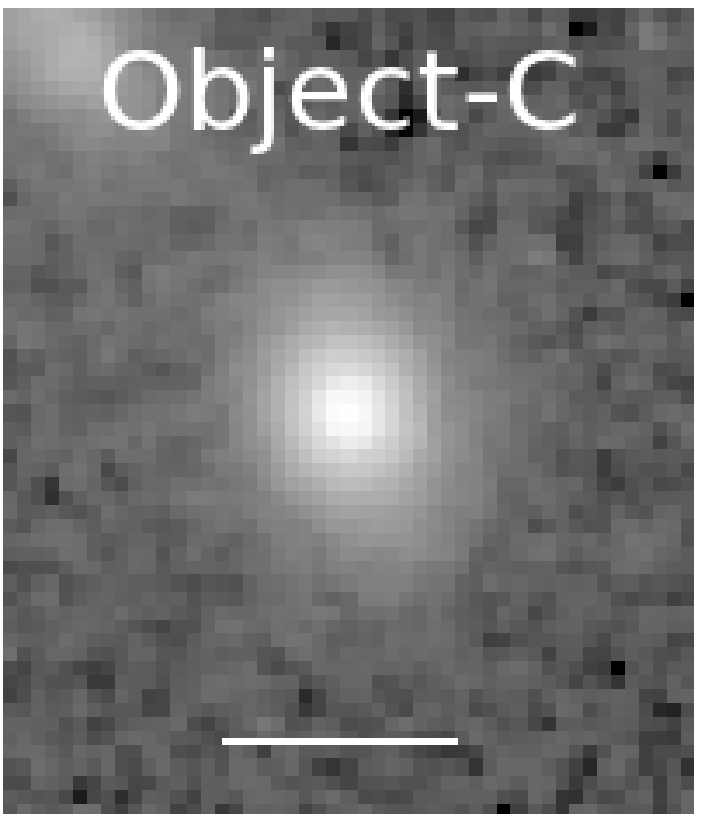}
    \FigureFile(30mm,30mm){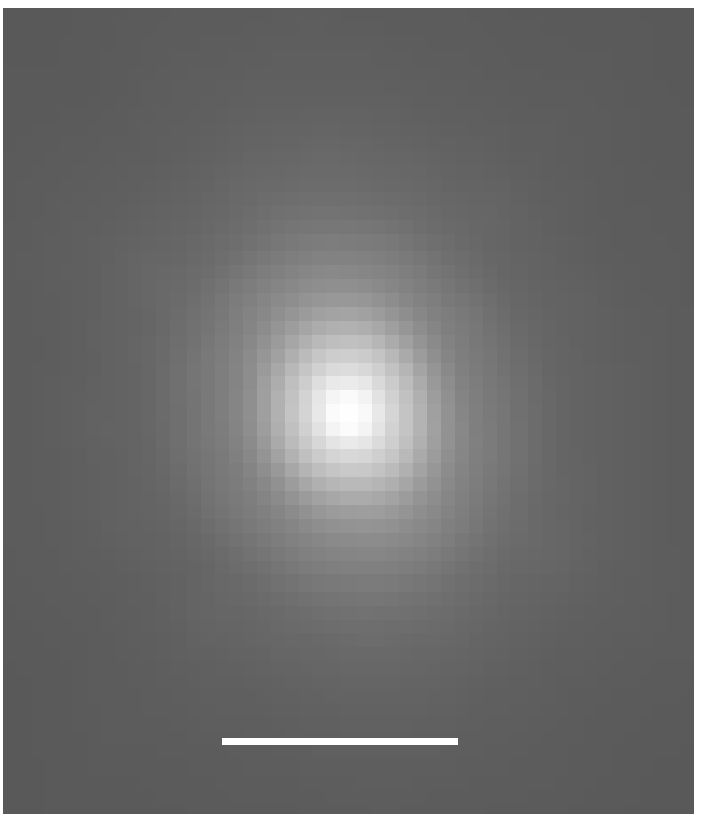}
    \FigureFile(30mm,30mm){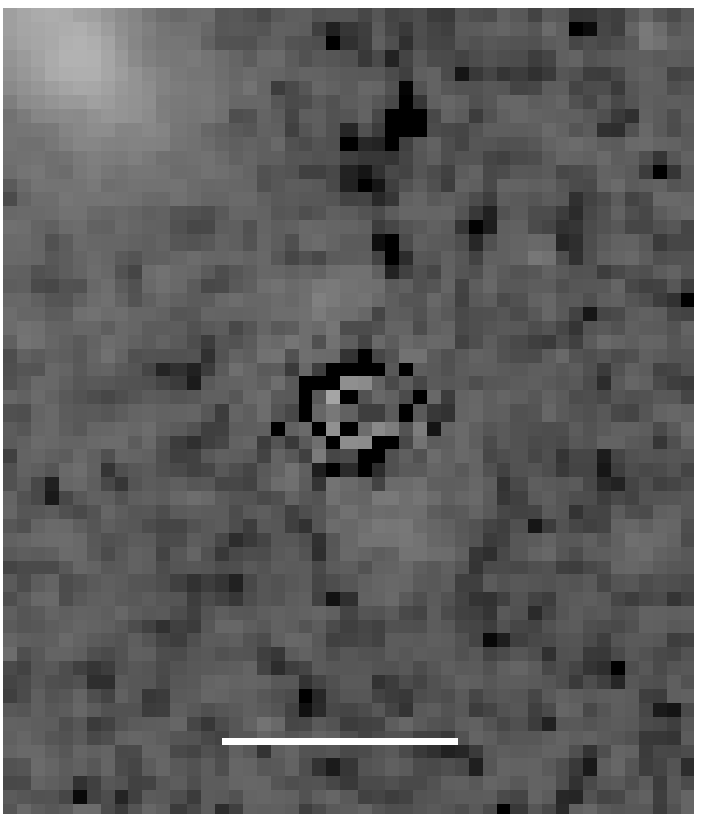}
    \FigureFile(30mm,30mm){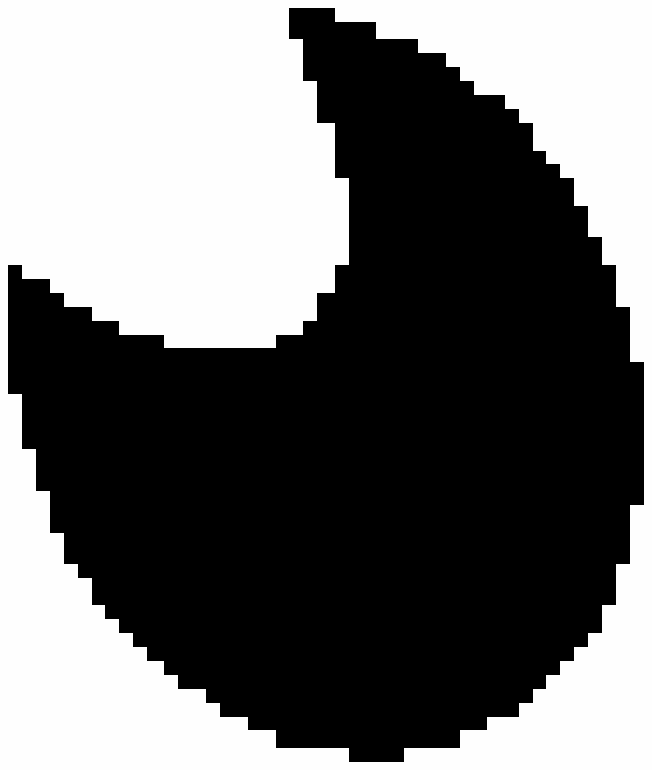}\\\vspace{0.5cm}
    \FigureFile(30mm,30mm){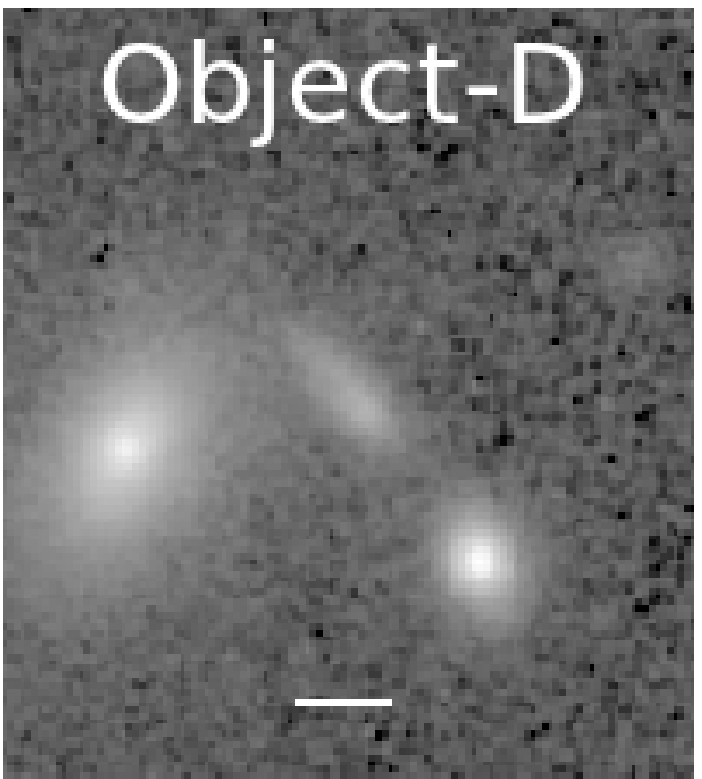}
    \FigureFile(30mm,30mm){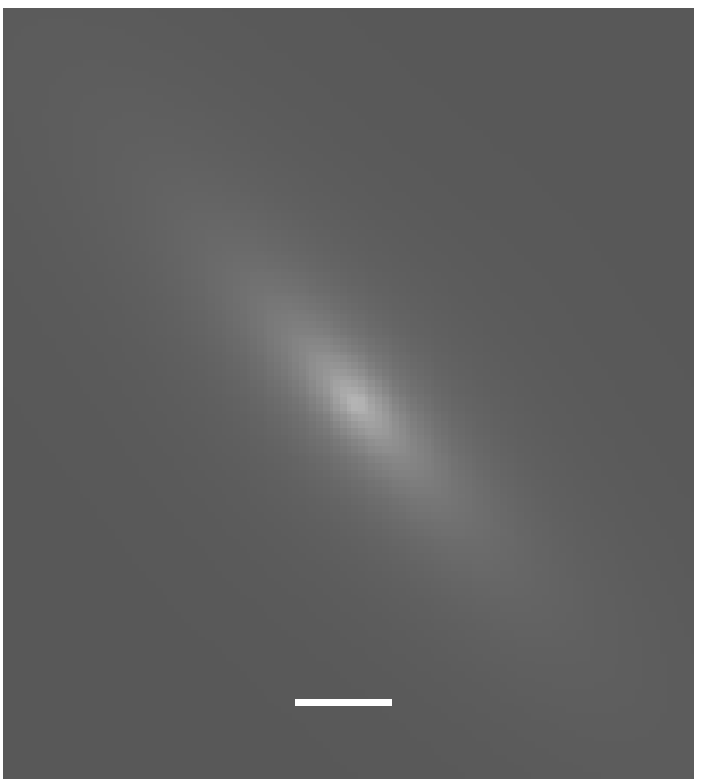}
    \FigureFile(30mm,30mm){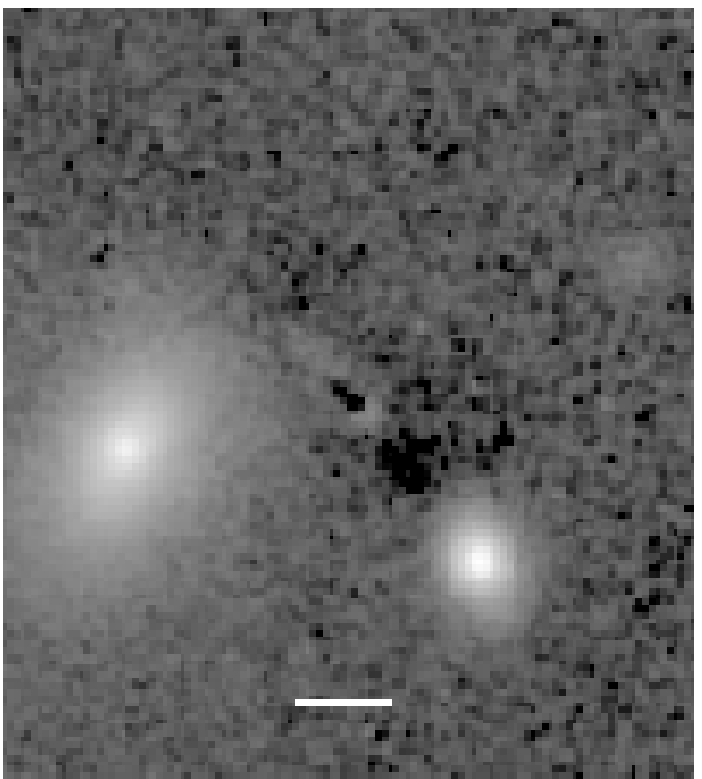}
    \FigureFile(30mm,30mm){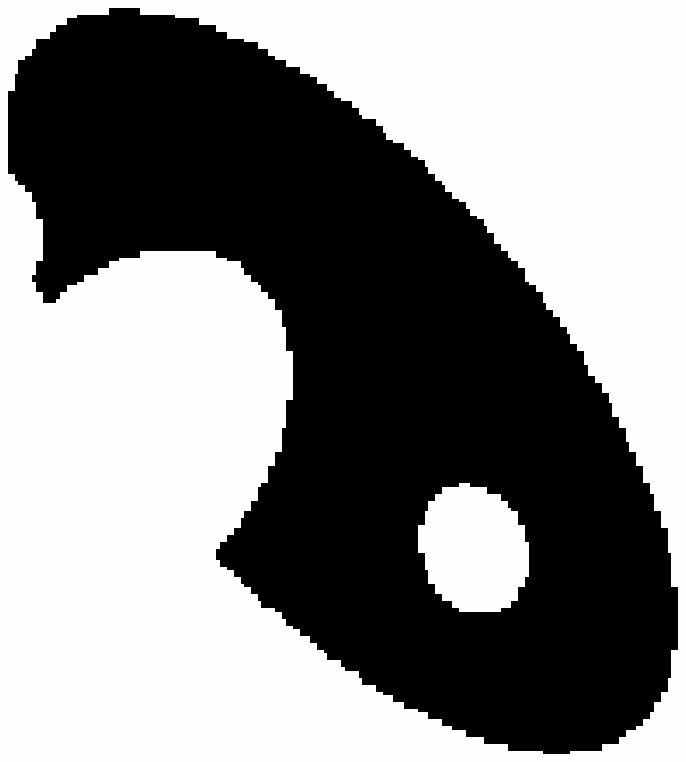}\\\vspace{0.5cm}
    \FigureFile(30mm,30mm){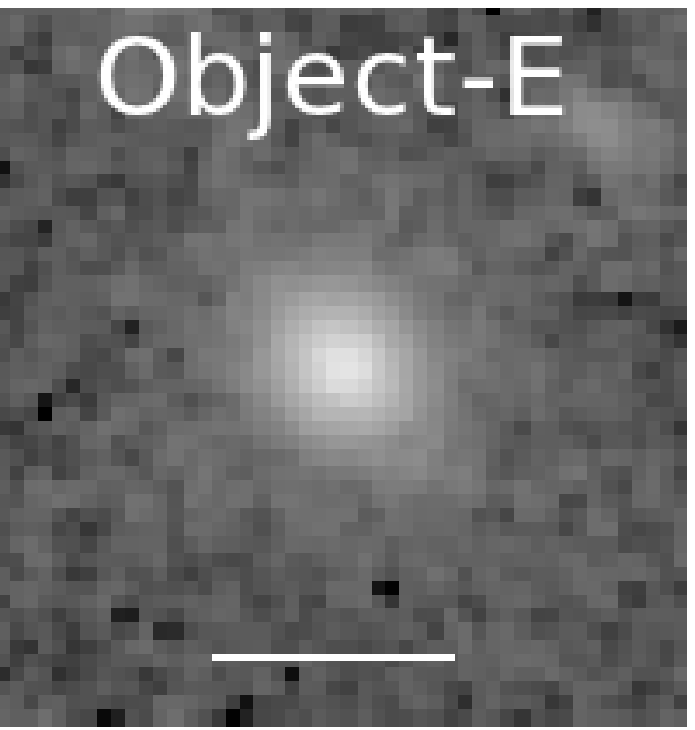}
    \FigureFile(30mm,30mm){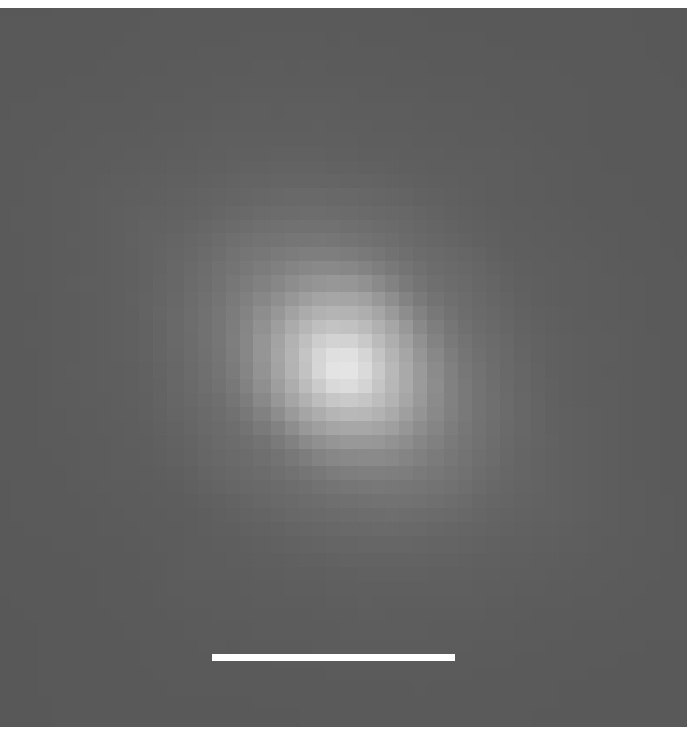}
    \FigureFile(30mm,30mm){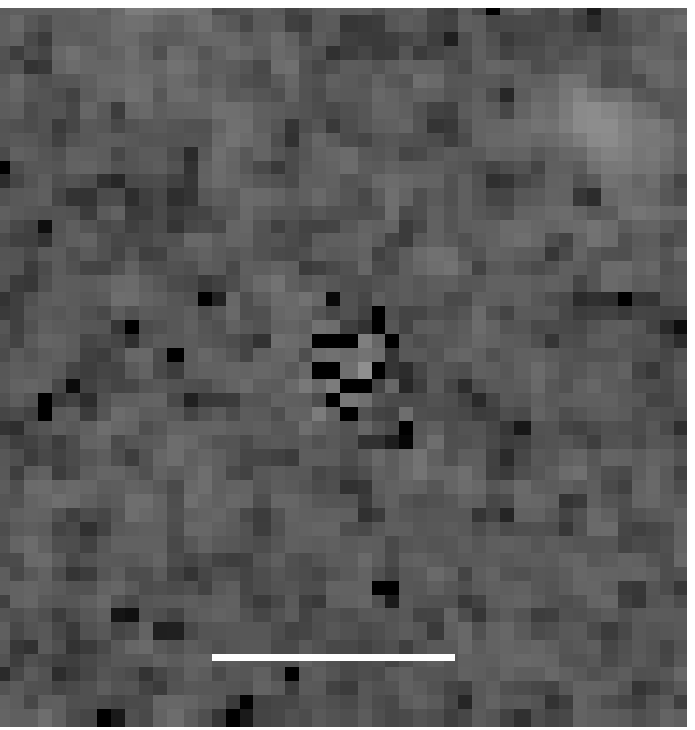}
    \FigureFile(30mm,30mm){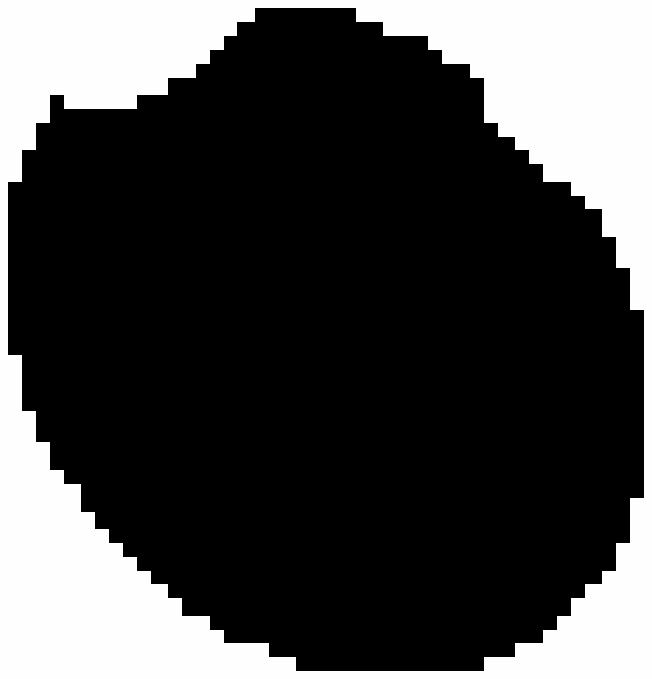}\\\vspace{0.5cm}
    \FigureFile(30mm,30mm){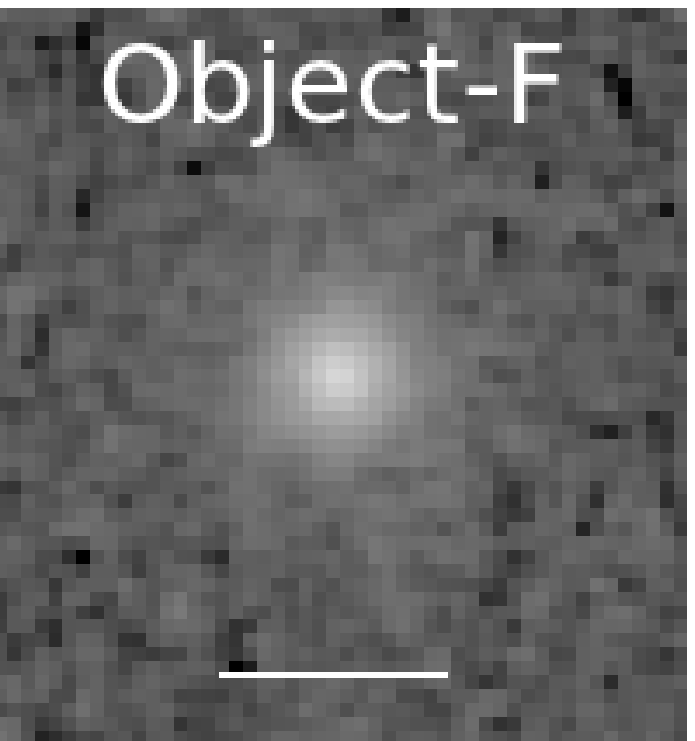}
    \FigureFile(30mm,30mm){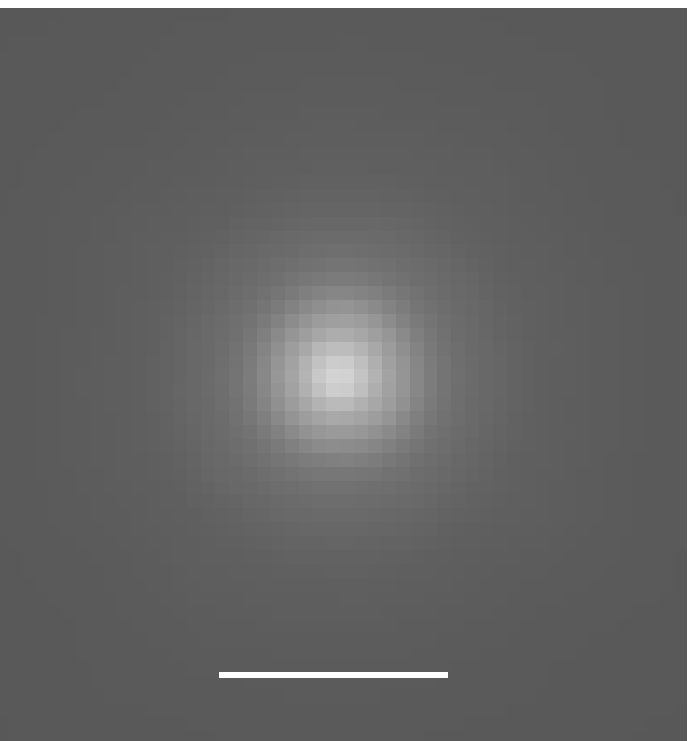}
    \FigureFile(30mm,30mm){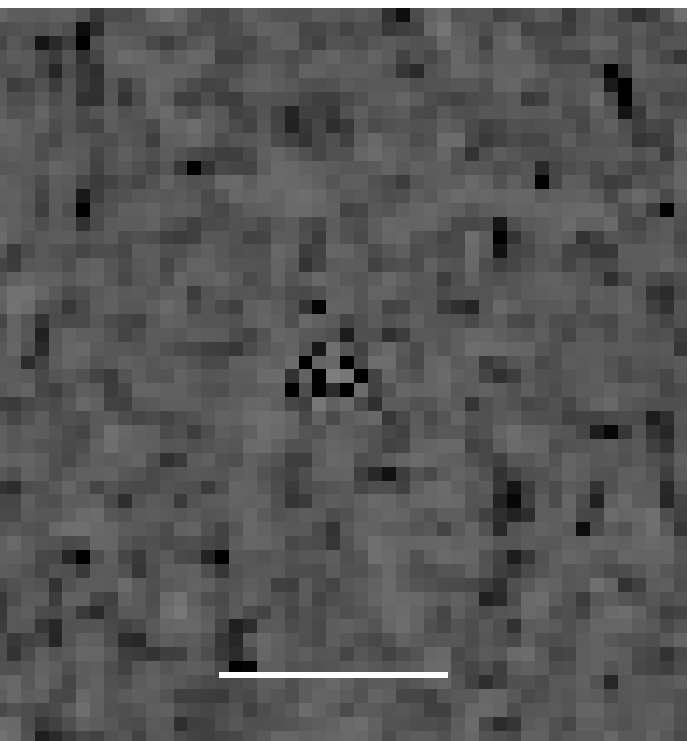}
    \FigureFile(30mm,30mm){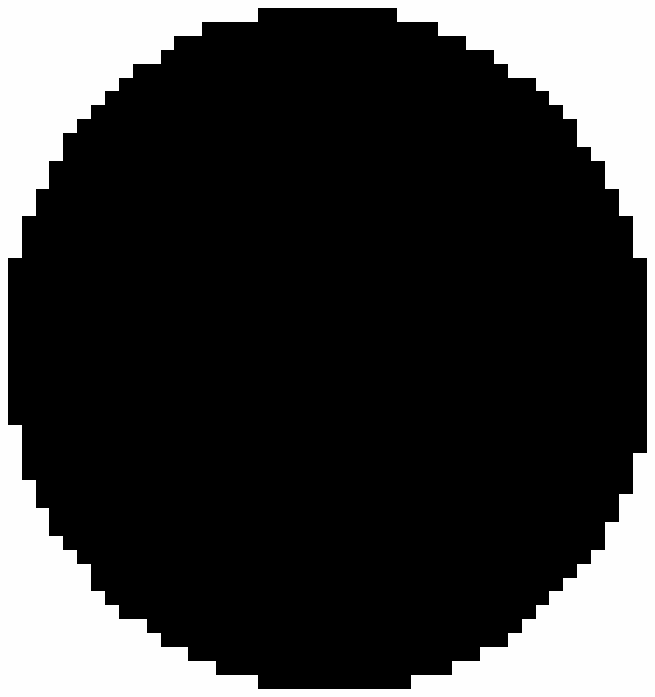}\\\vspace{0.5cm}
  \end{center}
  \caption{
    In each plot, the panels show the input image, model galaxy, residual, and
    mask images from left to right.  The plots are for object-A to F from
    top to bottom, respectively.  The horizontal bar in each panel is 1 arcsec,
    which corresponds to 8.5 kpc (physical) at $z=1.6$.
  }
  \label{fig:galfit}
\end{figure*}
\begin{figure*}
  \begin{center}
    \FigureFile(30mm,30mm){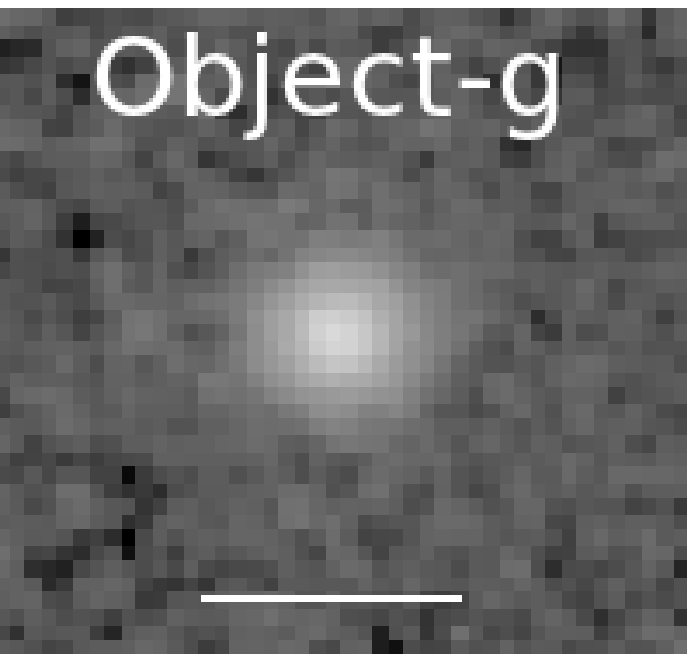}
    \FigureFile(30mm,30mm){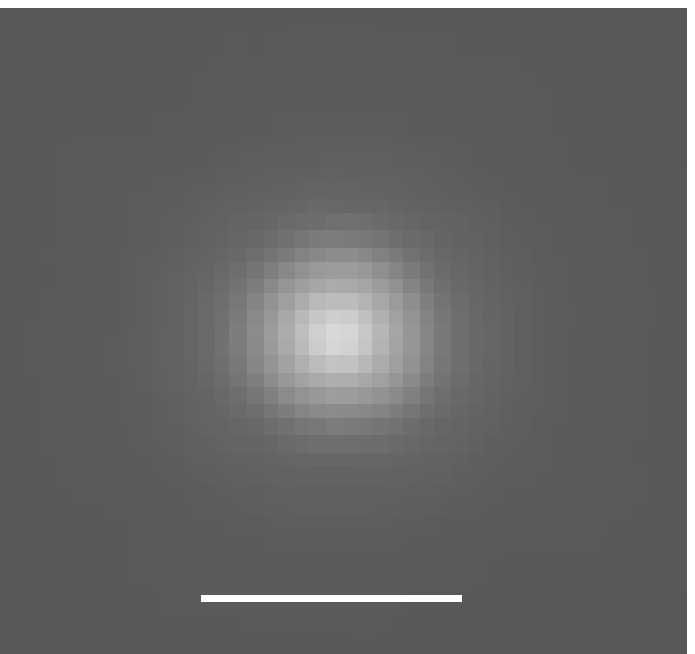}
    \FigureFile(30mm,30mm){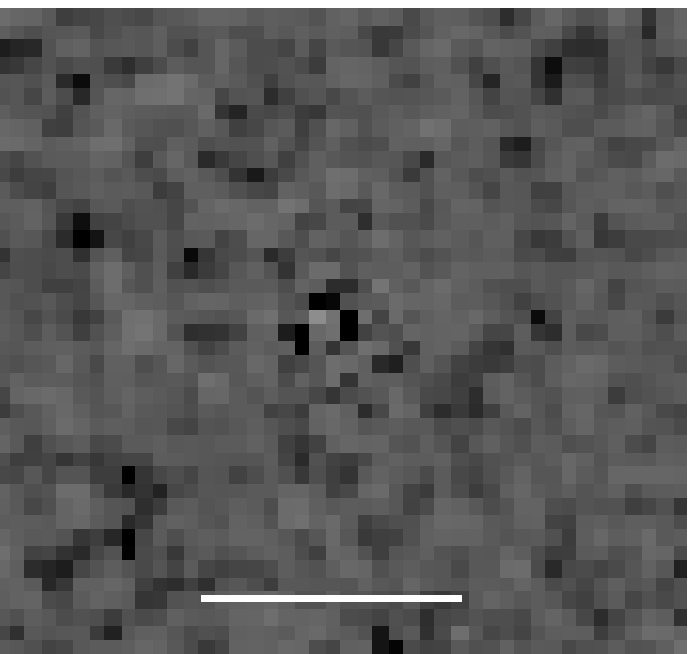}
    \FigureFile(30mm,30mm){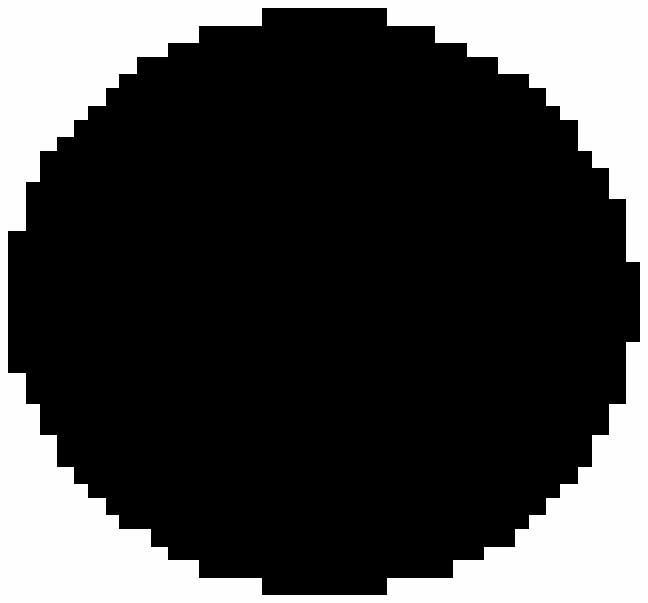}\\\vspace{0.5cm}
    \FigureFile(30mm,30mm){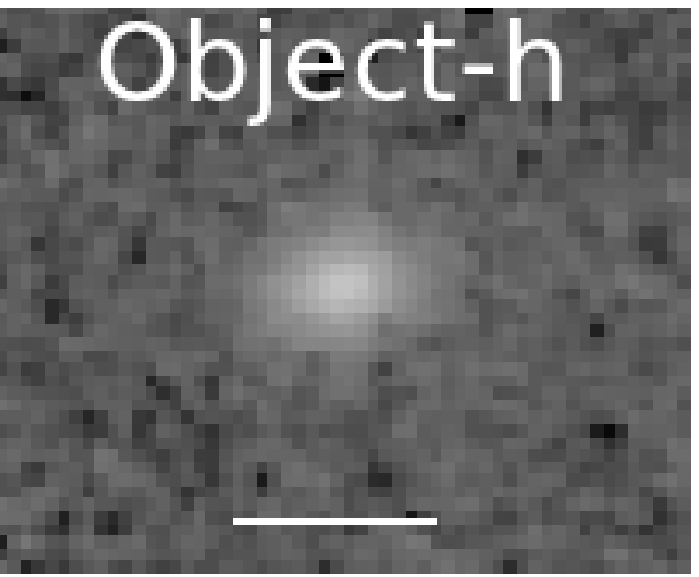}
    \FigureFile(30mm,30mm){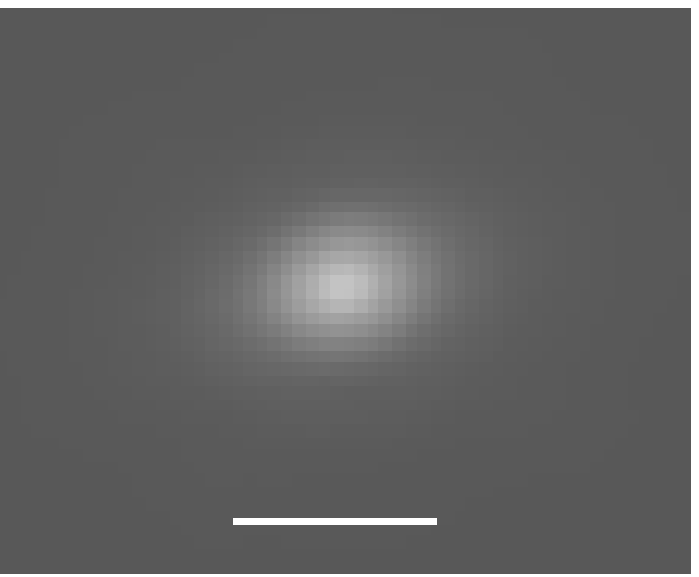}
    \FigureFile(30mm,30mm){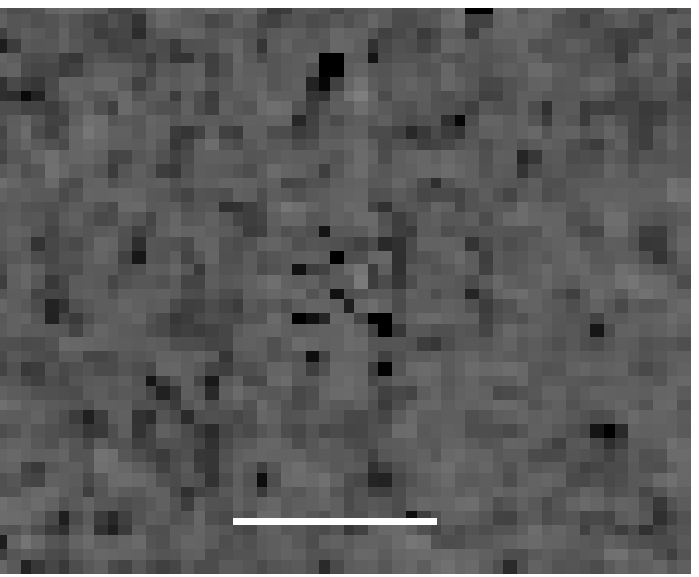}
    \FigureFile(30mm,30mm){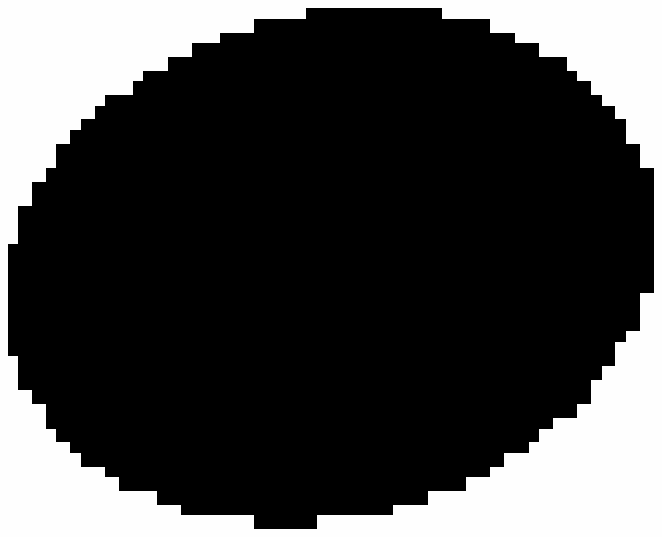}
  \end{center}
  \caption{
    As in Fig. \ref{fig:galfit}, but for object-g and h.
  }
  \label{fig:galfit2}
\end{figure*}

\begin{table*}[t]
  \begin{center}
    \caption{
      Structural properties of galaxies.  The numbers in the brackets are obtained
      by applying a circular mask of 3 pixel radius at the center
      (i.e., $\sim2\times$FWHM region is masked) to avoid contamination
      of central point sources.  Note that object-D is too close to the BGG
      and we fail to fit the object.
      Note as well that the uncertainties quoted here are statistical uncertainties.
      We expect $\sim10\%$ systematics on all the parameters.
    }
    \label{tab:galfit_props}
    \begin{tabular}{clll} 
      ID         &  S\'{e}rsic index & half-light radius (kpc) & axial ratio\\\hline
      object-A & $2.75^{+0.05}_{-0.01}$ ($2.07^{+0.06}_{-0.01}$) & $8.75^{+0.20}_{-0.04}$ ($7.69^{+0.14}_{-0.03}$) & $0.60^{+0.01}_{-0.01}$ ($0.60^{+0.01}_{-0.01}$)\\
      object-B & $1.93^{+0.04}_{-0.01}$ ($1.52^{+0.06}_{-0.01}$) & $2.60^{+0.03}_{-0.01}$ ($2.66^{+0.03}_{-0.01}$) & $0.58^{+0.01}_{-0.01}$ ($0.58^{+0.01}_{-0.01}$)\\
      object-C & $3.42^{+0.18}_{-0.04}$                       & $1.42^{+0.05}_{-0.01}$                      & $0.67^{+0.02}_{-0.01}$\\
      object-D & $3.43$                                   & $21.82$                                  & $0.28$\\
      object-E & $2.99^{+0.29}_{-0.06}$                       & $1.36^{+0.07}_{-0.01}$                      & $0.66^{+0.03}_{-0.01}$\\
      object-F & $5.27^{+1.21}_{-0.23}$                       & $1.53^{+0.37}_{-0.05}$                      & $0.94^{+0.06}_{-0.01}$\\
      object-g & $2.26^{+0.38}_{-0.07}$ ($1.27^{+0.77}_{-0.12}$) & $0.83^{+0.05}_{-0.01}$ ($1.09^{+0.22}_{-0.05}$) & $0.82^{+0.06}_{-0.01}$ ($0.84^{+0.07}_{-0.01}$)\\
      object-h & $2.27^{+0.47}_{-0.07}$                       & $1.59^{+0.14}_{-0.03}$                      & $0.48^{+0.05}_{-0.01}$\\

   \end{tabular}
  \end{center}
\end{table*}

Figs. \ref{fig:galfit} and \ref{fig:galfit2} show the input image, best-fit model image, residuals,
and mask image for the good and likely candidates, respectively.
The fits are generally good and the residuals are relatively small for most objects.
This verifies that the assumption of a single S\'{e}rsic profile is reasonable.
We fail to fit object-D due to its irregular morphology and to its close proximity
to the BGG.

Here we focus on two of the most important structural parameters;
S\'{e}rsic index ($n$) and half-light radius.
We quote half-light radius as $r_{50}\equiv\sqrt{ab}$, where $a$ and $b$
are half-light radius measured along the semi-major and semi-minor axes, respectively.
We estimate an error on each of the S\'{e}rsic index and half-light radius
by Monte-Carlo simulations.
First, we generate the same background level as in the real $H$-band data
assuming Gaussian noise and place a model galaxy with a given brightness,
effective radius and S\'{e}rsic index.  We then run GALFIT and measure
the structural parameters.  We allow the brightness, effective radius,
and S\'{e}rsic index to vary and repeat the analysis above.
A difference between the input and output parameters is fairly small
(a few percent) for bright objects with $H<22$ mag.
For fainter objects, the systematic offset and dispersion increases
to 5\% and 10\% respectively
for objects with $H=23$ mag.  This is roughly the average brightness of
the objects that we study in this section.
For objects with $H=24$, we measure a 10\% systematics and a 20\% scatter.
We note that all of the objects studied here are brighter than $H=24$.

We then perform another set of simulation.
We generate background noise of the same level
as in the real data assuming Gaussian noise and place the group members at
the same relative location with the same S\'{e}rsic
indices and half-light radii as the values measured in the CANDELS image for each galaxy.
We exclude object-D because we fail to fit this object
and therefore cannot use the fitted parameters for the simulation.
We then run GALFIT on a simulated image and store the output parameters.
We repeat this procedure on 2000 simulated images and adopt the 68\% interval of
the output parameters as errors.
In the figures, we show this error or the error estimated in the first simulation,
whichever is larger.  In most cases, the latter error is larger.
For systematic offsets, we find that
the input parameters and the median of the output parameters agree within 10\% for
all the galaxies.  This is consistent with the systematic offsets estimated
in the first simulation and this 10\% is a reasonable estimate of the systematics.

As shown in Fig. \ref{fig:color_pic}, there are a few X-ray point sources among
the group member candidates. This X-ray emission is likely due to nuclear activity. 
To make sure, the presence of AGN does not impact our analysis, 
we repeat the GALFIT analysis with the central
3 pixels in radius masked out. 
The derived structural parameters are summarized in  Table \ref{tab:galfit_props}.
We also show axial ratios ($b/a$) for reference.
Note that the structural parameters derived with the central masking
do not differ significantly from those derived without it.

\subsection{Structural properties}

Let us now compare the structural parameters of the group galaxies
with those measured at $z=0$ to quantify the morphological evolution.
For this comparison, we use data from the Sloan Digital Sky Survey \citep{york00}.
We use galaxies in the Main sample \citep{strauss02} located at
$0.05<z<0.07$.  The F160W filter of WFC3 probes rest-frame $\sim6000\rm\AA$ at $z=1.6$,
which is very close to the effective wavelength of the $r$-band filter in SDSS \citep{doi10}.
Therefore, the morphological $k$-correction is negligible.
The PSF of the stacked F160W image is $0.2$ arcsec, which corresponds to 1.7 kpc
at $z=1.6$.  The average seeing in SDSS is $1.5$ arcsec, which is also 1.7 kpc
at the median redshift of our SDSS sample.  The surface brightness limit
is much deeper for SDSS than for the $z=1.6$ galaxies.  But, we have performed a Monte-Carlo
simulation to quantify uncertainties on our structural parameter estimates at $z=1.6$ as
described above and the shallow limit at $z=1.6$ is folded into the uncertainty.
For the S\'{e}rsic index and half-light radius of $z=0$ galaxies, we use the New-York University
Value Added Catalog \citep{blanton05}.
The stellar mass for the SDSS sample is taken from
\citet{tanaka11a} who fitted the SDSS spectra with \citet{bruzual03} model templates
assuming the Chabrier IMF.  Note that a correction for the fiber loss is applied
in a crude way by assuming that the light within a fiber is representative of
the light from the entire galaxy.

\begin{figure*}
  \begin{center}
    \FigureFile(80mm,80mm){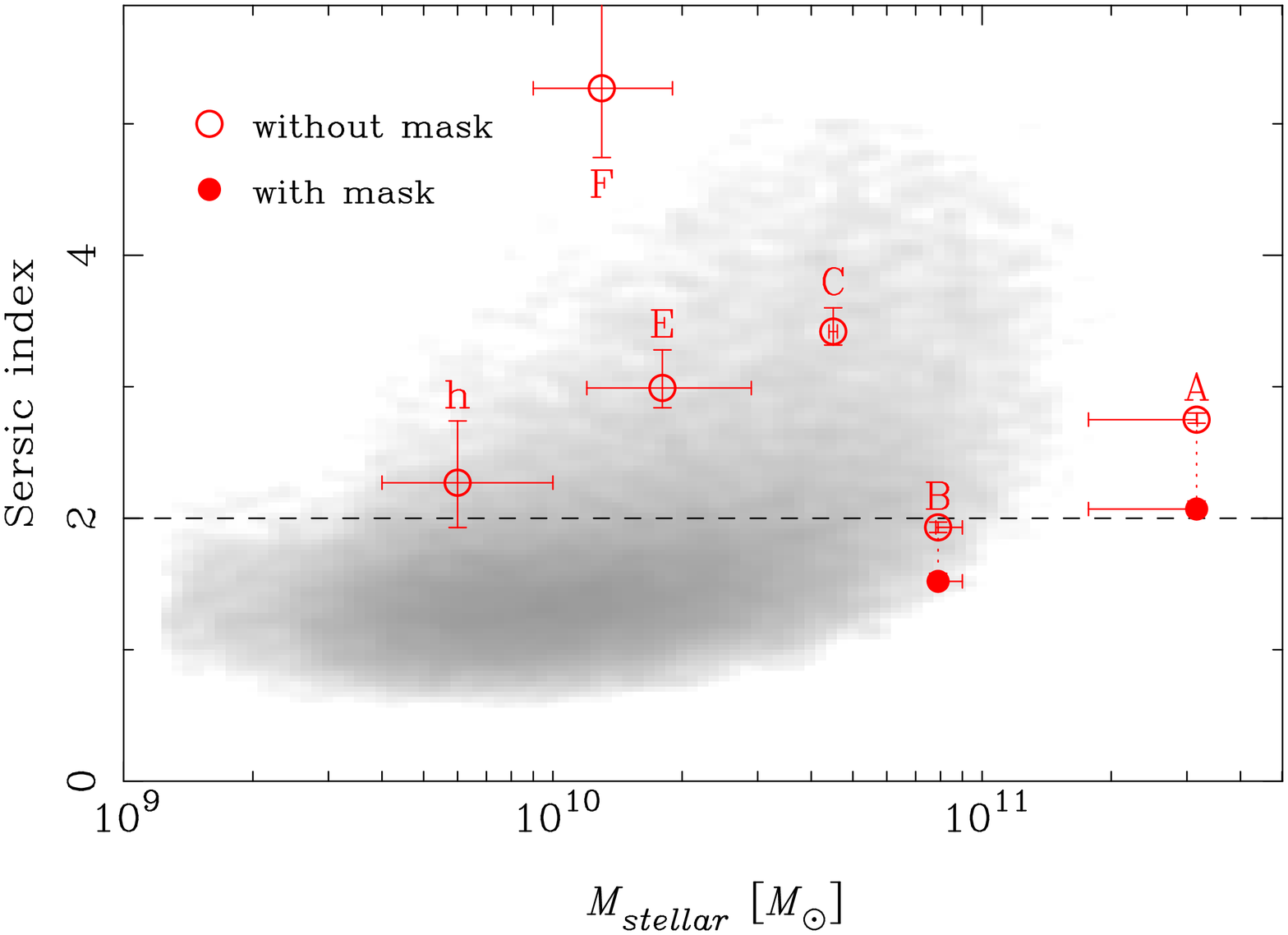}\hspace{0.5cm}
    \FigureFile(80mm,80mm){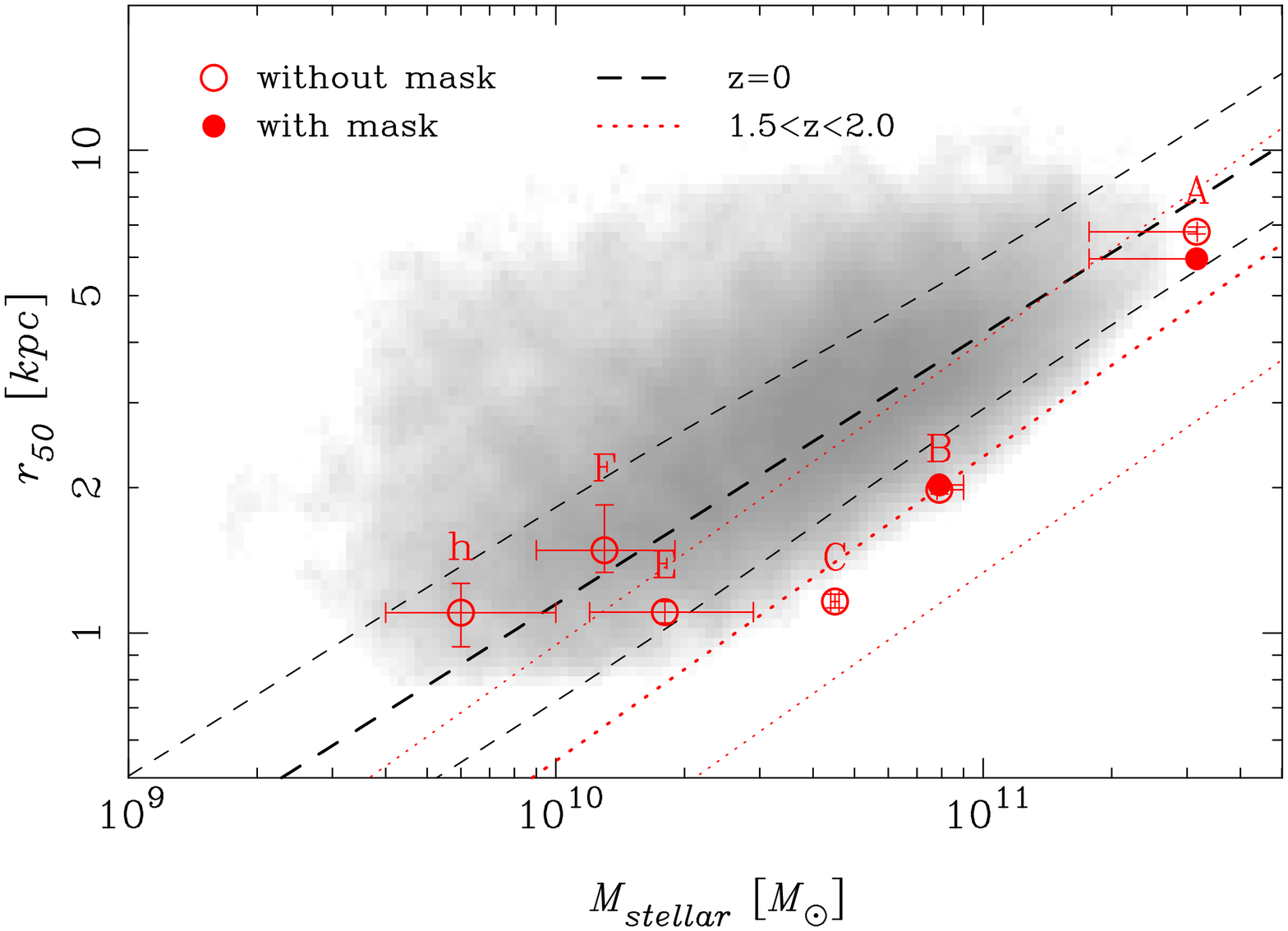}\hspace{0.5cm}
  \end{center}
  \caption{
    {\bf Left:}
    S\'{e}rsic index plotted against stellar mass.
    The gray scales are galaxies at $0.05<z<0.07$ from SDSS.
    For clarify, only 95\% of the galaxies are plotted.
    The large symbols are the $z=1.6$ galaxies and the associated error bars
    show the statistical uncertainty.
    We recall that we have a $\sim10\%$ systematic
    uncertainty on the structural parameters.
    Object-A and B likely host AGNs and we show the S\'{e}rsic index measured both
    with and without the central mask.
    Object-g is a moderately strong AGN and is not plotted here because no reliable
    stellar mass
    estimate is available, while Object-D is not plotted because we fail
    to find an adequate fit to this object.
    The horizontal dashed line shows the rough separation between early-type
    and late-type galaxies.
    {\bf Right:} Half-light radius plotted against stellar mass.
    As in the left panel, the gray scales are for $z=0$ galaxies, but here
    we show only early-type galaxies with $n>2$.
    The thick dashed line is the size-mass relation of early-type galaxies at
    $z=0$ from \citet{shen03} and the thin dashes lines show $1\sigma$ scatter.
    The thick dotted line is for $1.5<z<2.0$ quiescent early-type galaxies
    from \citet{newman12} and the thin dotted lines show a $1\sigma$ scatter
    in the relation.
  }
  \label{fig:sersic_vs_smass}
\end{figure*}

Fig \ref{fig:sersic_vs_smass} presents the S\'{e}rsic index, $n$,
and half-light radius plotted against stellar mass.
As can be seen in the left panel,
most of the $z=0$ galaxies have S\'{e}rsic index
between 1 and 2.  This is the typical range  for late-type galaxies.
The index shows a tail towards a larger index at high mass, which shows
that most massive galaxies tend to be early-type galaxies.
We separate early and late-type galaxies at the S\'{e}rsic index of $n=2$ as shown
by the horizontal dashed line in the left panel.
If we turn our attention to the $z=1.6$ galaxies shown as the large circles in the left panel,
we find that most of them have $n>2$.
In other words, most of the galaxies in the $z=1.6$ group are early-type galaxies.
We know that groups and clusters at $z=0$ are dominated by early-type galaxies
\citep{dressler80,postman84}, but it is surprising that a group at such a high
redshift is also dominated by early-type galaxies.
We have shown in the previous section that these galaxies are mostly quiescent galaxies.
Therefore, the group appears fairly similar to groups and clusters in the local universe.

In addition to the S\'{e}rsic index, another interesting structural parameter
of galaxies is their physical size.
The size evolution is particularly interesting given the recent observations
that distant quiescent galaxies are compact compared to $z=0$ counterparts
(e.g., \cite{daddi05,trujillo06,toft07}).
It would therefore be interesting to study whether the size evolution depends on
environment.  \citet{rettura10} studied a cluster at $z=1.24$ and suggested
that the cluster and field galaxies at the same redshift have similar sizes.
The newly confirmed group is one of the highest redshift systems
discovered so far where high quality WFC3 images are available, allowing us
to investigate galaxy sizes in the group.

We show half-light radius, $r_{50}$, against stellar mass in the right panel of
Fig. \ref{fig:sersic_vs_smass}.
As a $z=0$ reference, we show the size-mass distribution of local early-type
($n>2$) galaxies as well as a size-mass relation of
the local early-type galaxies from \citet{shen03}.  We note that
\citet{shen03} measured sizes in the $z$-band, while we use the $r$-band.
Therefore, care needs to be taken when comparing these two samples.
We also note that the resolution limit of the F160W image is $r_{50}\sim0.5\ \rm kpc$
and thus all the $z=1.6$ galaxies plotted are well resolved.
The Figure also shows the size-mass relation for quiescent galaxies 
from \citet{newman12}.  Their quiescent galaxies typically have early-type
morphology with $n\sim3-4$ and their result is directly comparable to ours.
They did not characterize the environment of the galaxies.  However, most of the galaxies
in their sample are likely field galaxies and we refer to their sample
as a field sample.

The BGG is within the scatter of the local and $1.5<z<2.0$
size-mass relations and it appears that the BGG in the $z=1.6$ group already
has a similar size to local massive galaxies.  The low-mass galaxies with
$\sim10^{10}\rm M_\odot$ are also consistent with both relations.
Only object-B and C are clearly below the local relation, and they are
consistent with the $1.5<z<2.0$ field relation.  Overall, the size-mass
relation of the group galaxies seems to fall in between the $1.5<z<2.0$
and $z=0$ relations.
\citet{papovich11} claimed that galaxies in another $z=1.6$ group in
SXDF exhibit smaller sizes at fixed stellar mass compared to the local
galaxies, but they are larger than field galaxies at the same redshift.
Our result in Fig.  \ref{fig:sersic_vs_smass} may be consistent with
their finding.  \citet{zirm12} reported on a similar trend in a
$z=2.16$ proto-cluster.  However, the statistics in all these studies,
including this paper, are not sufficient to allow a clear conclusion.

\citet{lotz11} reported on an elevated merger/interaction rate in the
$z=1.62$ system in SXDF.
Only one of the 8 group member candidates
(object-D) shows a highly distorted morphology and all the other
galaxies have well-defined early-type morphologies.  We also do not
observe a strong elevated rate of nearby companions (see Fig.
\ref{fig:galfit}).
We may tend to miss interacting galaxies due to
possibly poor photo-$z$'s for such objects.
To be sure, we visually inspect all the bright galaxies with
$H<24$ located within $r_{200}$ without using photo-$z$'s.
Fore-/background galaxies with secure spec-z's are excluded from
this exercise.  We find 3 clear cases for disturbed morphology
in addition to object-D.  We have carefully examined the SEDs
of the 3 objects
and find that only one of them is marginally consistent with
being at the group redshift.  The other SEDs look normal and
they are likely at $z\sim3$. Although the statistics are very poor,
it is unlikely that a large fraction of the group galaxies are
undergoing interaction.
The possible different trend between the two systems might be due to
their different dynamical states.  That is, the newly confirmed $z=1.61$
group in this paper is a more relaxed system than the one in SXDF.  If
the X-ray emission around object-A is partly due to a cool core, it
lends support a relaxed system with no recent merger events.  It would
then not be a surprise that few group members are undergoing
interactions with other galaxies.
On the other hand, the group in SXDS shows a somewhat irregular
distribution of the members and it potentially has a companion group
(but see also the shallow Chandra observations by \cite{pierre11}).
The possibly different dynamical states could explain the different
merger/interaction rates in these two groups at the same redshift.

Finally, we briefly mention the ellipticity of the $z=1.6$ galaxies.
As shown in Table \ref{tab:galfit_props}, most of them have an axis
ratio of $b/a\sim0.6$.  By excluding object-D, which we fail to fit,
we measure an average axis ratio of $0.68$ with a scatter of
$\sigma=0.14$.  This axis ratio is consistent with field galaxies
at similar redshifts ($<b/a>=0.66$; \cite{newman12}).
\citet{holden09} found that cluster early-type galaxies
have the median axis ratio of $0.70$ and this does not strongly evolve at $z<1$.
Our finding here
may extend the result by \citet{holden09} to a redshift of $z=1.6$,
although the cluster mass ranges explored are very different
(our group has a much lower mass than those studied by \cite{holden09}).

To summarize, we find that most of the galaxies in the $z=1.6$ group
are early-type galaxies with S\'{e}rsic index $n>2$.
There are a few galaxies that have smaller physical sizes
than their local counterparts, but the overall size-mass relation of
the group galaxies does not seem to be significantly different from
the local relation.  In all of these aspects,
the group is strikingly similar to local groups and clusters
and the environmental dependence of galaxy properties is clearly 
in place by $z=1.6$.

\section{Discussion}

Our primary finding in this work is that the newly discovered, poor
group of galaxies at $z=1.61$ in the CDFS is dominated by quiescent
early-type galaxies.  This result is possible because of the exquisite,
high quality data available in this field. 
 We started this paper by saying that high-$z$ groups hold a
key to understanding the origin of the environmental dependence of
galaxy properties observed locally because they are the progenitors of
present-day clusters. 
The galaxy populations in the group appear strikingly similar to
those in groups in the local universe. 
A naive interpretation of our results is that
the environmental dependence is established at $z\gg 1.6$.  In fact,
\citet{tanaka10b} found tentative evidence that a fraction of galaxies in a
(proto-)cluster at $z=2.15$ show suppressed star formation activities.
\citet{spitler11} reported on galaxy
concentrations at $z=2.2$ in the COSMOS field and suggested that a red
sequence is present, which might support the view of an environmental
dependence already in place by $z>1.6$.

Groups and clusters form from statistical fluctuations of the
density field and they are thus statistical objects in nature.  They do
not all form at the same time and they do not all evolve in the same
way.  At a given redshift, groups naturally show a diversity in their
properties and we cannot generalize our results in this paper to the
average group properties at $z=1.6$.  However, it is still important
that we find one system that is dominated by quiescent early-type
galaxies.  The environmental dependence of galaxy properties is thought
to be due to both nature and nurture effects.  Most of the group member
candidates are consistent with the formation redshift of $z_f=3$, which
is only $\sim2$ Gyr prior to $z=1.6$  under the assumption of
a single burst model (see Fig. \ref{fig:cmd1}).  Given that galaxies need time
(of order hundred Myr) to settle onto the red sequence even after a
sharp quenching, the galaxies must have grown to $10^{10-11}\rm\
M_\odot$ and then quenched on a time scale of $<2$ Gyr.  What physical
process is responsible for such rapid evolution in this low mass group
at this high redshift?

The $z=1.6$ group shows one significant difference from local groups and
clusters; the high AGN fraction.  Recently, energy feedback from AGNs
has been suggested as a promising quenching mechanism.  Theoretical
work based on a simplified recipe of AGN energy feedback seems to do a
good job of reproducing some observed galaxy properties (e.g.,
\cite{granato04,springel05,bower06,croton06}), although the exact form
of the energy feedback is still highly uncertain.  Our finding that the
group hosts a large fraction of AGNs might indicate some role of AGN
feedback in quenching.  A fraction of AGNs with $L_X>10^{42}\ \rm erg\
s^{-1}$ at $1<z<2$ among galaxies with $>2\times10^{10}\rm M_\odot$ is
about 10-20\% in the field \citep{xue10}.  If we use object-A to E, which
are $>2\times10^{10}\rm\ M_\odot$ within the error, we find an AGN fraction of
$0.40^{+0.30}_{-0.25}$.  The fractions are consistent within the
errors, but the large AGN fraction in groups is not unexpected based on
the increasing AGN fraction with redshift in $z<1.3$ groups and
clusters \citep{eastman07,martini09,tanaka11}.  Even in a proto-cluster
at a higher redshift of $z=3.1$, \citet{lehmer09} observed an enhanced
fraction of AGNs.  However, an enhanced AGN fraction is only
circumstantial hint -- we do not observe any direct evidence that AGNs
quench the galaxies.

Another interesting aspect of the high AGN fraction in high-$z$ groups
is that it may help explain at least partly the apparent diversity of the observed
properties of high-$z$ group galaxies.  \citet{tran10} claimed that the
SFR-density relation reverses in a group at $z=1.62$ in SXDF.  We go
comparably as deep as \citet{tran10} in this study, but we do not find
the group as an active place for star formation (see Table
\ref{tab:physical_props}).  We do not observe an enhanced MIPS
population in our group either.  Not only the group studied by
\citet{tran10}, but another massive cluster at a similar redshift
($z=1.45$) seems to host a large population of emission line objects in
the core \citep{hilton10,hayashi10}.  It is still unclear whether these
MIPS sources and emission line objects are due to enhanced star
formation or AGN activity.  The latter possibility is not unlikely
given the recently observed high AGN fractions in high redshift groups
and clusters as mentioned above and AGNs may contribute at least partly
to the observed fraction of emission line galaxies.

AGN feedback is a possible process to quench galaxies, but are there
any other physical processes that could explain the observed trend?
Physical processes like ram-pressure stripping
\citep{gunn72} and harassment \citep{moore96} may not be very efficient
in groups because these processes are expected to be effective in
rich clusters. 
Strangulation \citep{larson80} may work in groups, and it may be
able to suppress the star formation on a short enough time
scale to bring the galaxies onto the red sequence within $<2$ Gyr.
Galaxy interactions and mergers may play some role in quenching as well,
although we do not observe a hint of frequent interactions in the group.
If interactions are important, they must have happened at higher
redshifts, which would then be closely linked to the nature effects.
The observed large fraction of AGNs makes us speculate their roles,
but as discussed earlier, we do not have direct evidence for the AGN feedback.
In order to further constrain the physical processes, it would be
interesting to perform deep near-IR spectroscopy to measure
the Balmer absorption features.  These absorptions are strongest in
A-type stars and by combining them with a continuum shape
(e.g., a strength of the 4000\AA\ break), one can put a constraint
on the quenching time scale, which will then constrain the physical processes.

Although the discovery of the high-$z$ poor group filled with quiescent early-type
galaxies is interesting, we need a larger sample of groups at high redshifts
to fully address the questions raised in the first section.
We are now in the process
of building such a statistical sample of high-$z$ groups.
Up-coming large imaging surveys such as the one by Hyper Suprime-Cam
will provide a huge data set that allows us to construct a statistical
sample of high redshift groups and study the group evolution
up to $z=2$ and beyond.
If increasing AGN activity in high-$z$ groups is a real trend,
it might imply that future deep X-ray surveys aiming at distant groups
might benefit from moderately high angular resolution.
The XMM-Newton data of the group appears dominated by a collection of
point sources as one can easily imagine from the Chandra data shown in
Fig. \ref{fig:color_pic}.   We are able to subtract such
contamination and detect the extended component in the XMM data
(see Sect. 3.2), but the detection of extended X-ray is
obviously harder where point source contamination is more severe.  
High angular resolution X-ray missions such as WFXT \citep{vikhlinin09}
and SMART-X would be ideal for identifying high redshift
groups and clusters.

\section{Conclusion}

We have discovered a group of galaxies at $z=1.61$ in CDFS
using the deep X-ray, optical, and near-IR data.
This is the lowest mass system discovered so far at such high redshifts
and it provides an interesting system to study galaxy evolution in groups at
high redshift.  Our primary findings can be summarized as follows.

\begin{itemize}
\item We detect extended X-ray emission from the group, which suggests that
  the group is gravitationally bound. 

\item The photo-$z$ selected galaxies at $z=1.6$ exhibit a clear concentration of red galaxies
  around the extended X-ray emission.  A few of them, including the BGG,
  are spectroscopically confirmed at $z\sim1.61$.

\item The group members form a pronounced red sequence, and are mostly
  consistent with a simple model population formed at $z_f=3$. 
  We fit our SEDs with models assuming stellar populations are formed
  with exponentially declining star formation histories. The best fits
  suggest that stellar populations are predominantly old and passively
  evolving, with low star formation rates, SFR$\ll1\rm M_\odot\
  yr^{-1}$. Our galaxies lie clearly below the SFR-mass relation for
  normal star forming galaxies at similar redshifts.

\item We have performed a morphological analysis using the WFC3 data
  and found that most of the group members show early-type morphology with S\'{e}rsic index $n>2$.

\item Although the statistics are poor, the group appears to exhibit a very high
  fraction of AGNs (3 out of 8 good/likely candidates are AGNs), indicating
  an elevated AGN activity in high redshift groups.

\end{itemize}

These findings lead us to conclude that quiescent, early-type galaxies have already
become a dominant population in the group.
The group thus appears similar to present-day groups and clusters with
one possible difference of its high AGN fraction.
This is a surprising result given 
the very low-mass of the system and its very high redshift.
A naive interpretation of our result is that the environmental
dependence of galaxy properties is in place in this group and it must have come in place
at $z>1.6$.  However, the physical process or processes responsible for the galaxy quenching
remain unclear.  Furthermore, there appears to be significant diversity in the observed
properties of group/cluster galaxies at $z\gtrsim1.5$.
To better understand high redshift groups,  statistical samples will be necessary.
Future surveys will yield such samples which will hopefully
allow us to address the long standing issue of the interplay between
structure evolution and galaxy evolution.

\vspace{0.5cm}
We thank the anonymous referees for useful comments,
which helped improve the paper.
This work was supported by World Premier International Research
Center Initiative (WPI Initiative), MEXT, Japan, by KAKENHI No. 23740144,
and by SAO grants SP1-12006B and SP8-9003C to UMBC.
WNB and YQX acknowledge support from Chandra grant SP1-12007A
and JSM acknowledges support from Chandra grant SP1-12006A.
YQX also acknowledges support from the Youth 1000 Plan (QingNianQianRen)
program and the USTC startup funding.
This work is based on observations taken by the CANDELS Multi-Cycle
Treasury Program with the NASA/ESA HST, which is operated by the Association
of Universities for Research in Astronomy, Inc., under NASA contract NAS5-26555.
Funding for the SDSS and SDSS-II has been provided by the Alfred P. Sloan
Foundation, the Participating Institutions, the National Science Foundation,
the U.S. Department of Energy, the National Aeronautics and Space Administration,
the Japanese Monbukagakusho, the Max Planck Society, and the Higher Education
Funding Council for England. The SDSS Web Site is http://www.sdss.org/.

\appendix
\section{Foreground over-density at $z=1$}

There are three objects at $z_{spec}=1$ within the X-ray contours in
Fig. \ref{fig:color_pic}.  In this appendix, we argue that they are
not the primary counterpart of the X-ray emission.

\begin{figure}
  \begin{center}
    \FigureFile(50mm,80mm){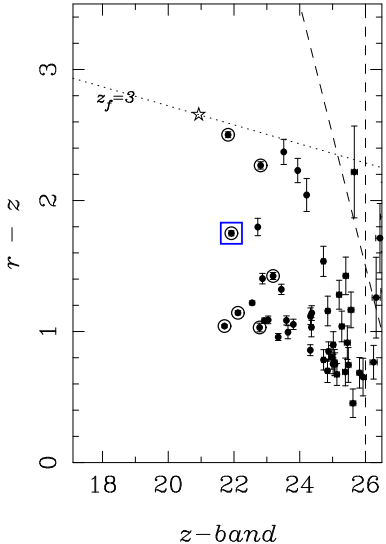}
  \end{center}
  \caption{
    F606W-LP850 plotted against LP850 using the MUSIC catalog.
    Here, we denote F606W and LP850 filters as $r$ and $z$, respectively.
    The large points are galaxies with $P_{gr,z_{gr}=1.03}>0.16$
    within $r_{200}$ of the group.
    The double-circles are spectroscopic
    galaxies at $|z_{spec}-1.03|<0.02$ and the square shows an X-ray point source.
    The slanted dotted line is a model red sequence at $z=1.03$ formed at $z_f=3$ and
    the star indicates $m_z^*$.  The dashed lines are $5\sigma$ limits.
  }
  \label{fig:cmd_z10}
\end{figure}

In Fig. \ref{fig:cmd_z10}, we show  $r-z$ color against $z$-band magnitude
of galaxies at $z\sim1$ around the extended X-ray emission.
Most of the galaxies are blue and there is only a weak red sequence.
In fact, our red sequence finder described in Section 2.3 gives a signal of
only $0.6\sigma$ at $z=1$.
As shown in Fig. \ref{fig:cmd1}, the red sequence is clearly more pronounced
at $z=1.6$.  Furthermore, the brightest galaxy at
$z_{spec}=1.03$ is 1 mag. fainter than $m^*$ and there is no
$m^*$ galaxy around the X-ray emission at $z_{phot}=1$.
The lack of bright galaxies does not favor the interpretation
that they form an X-ray bright group.
Although we cannot completely reject the possibility of
the $z\sim1$ galaxies contributing to the extended X-ray emission,
they are highly unlikely the primary counterpart of the X-ray.

\section{Template error function for SED fitting}

\begin{figure}
  \begin{center}
    \FigureFile(80mm,80mm){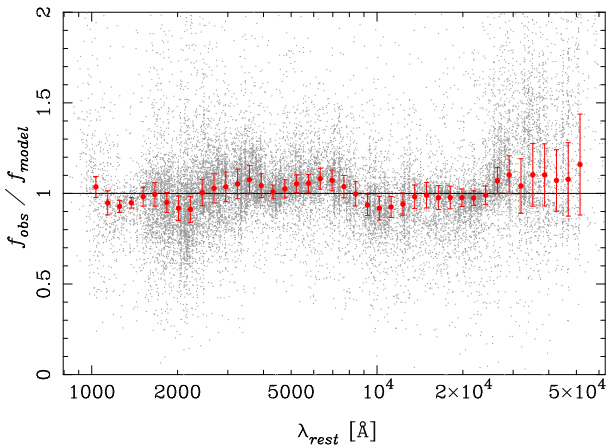}\hspace{0.5cm}
  \end{center}
  \caption{
    Observed to model flux ratio is plotted against rest-frame wavelength.
    The dots are each photometry and the big points show the median
    of the ratio in each wavelength bin.  The associated error bars
    show the $1\sigma$ error range after subtracting the photometric
    uncertainty in the quadrature.
  }
  \label{fig:template_errfn}
\end{figure}

Model templates based on a stellar population synthesis code are
subject to systematic uncertainties.  Such uncertainties include
flux errors in the input stellar spectra used in a code and
errors in the stellar evolutionary track.
These systematics cause a mismatch
between model spectra and observed spectra of galaxies.
A way to crudely reduce such systematics is
to apply flux corrections to model templates as a function of rest-frame wavelength.
In principle, one could apply such correction to templates of a given
spectral type.  However, degeneracies between model parameters and
a limited number of high quality spectra covering a wide enough wavelength
window are major obstacles.  One could instead apply a single
'master' correction to all the templates to crudely correct for
the systematics.  Here we construct a master {\it template error function}
using the photometry of spectroscopically observed objects in CDFS
and apply the error function to all the templates used in the SED fits.
As noted in the main body of the paper, our conclusions do not change
at all if we do not include the template error function in our analysis.

We first select galaxies with secure spectroscopic redshifts 
from the public spectroscopic redshift catalogs in the literature.
We then fit SEDs of these galaxies with redshifts fixed to their
spectroscopic redshifts using the photometry from the MUSIC catalog.
The dots in Fig. \ref{fig:template_errfn} show the ratio between
observed fluxes and best-fit model templates. 
The large points are the median of the ratio in each wavelength bin and
the associated error bars are the $1\sigma$ scatter in the ratio
after subtracting the photometric error.
The template error function defined here is a set of two quantities:
the flux stretch and dispersion.
We use both information in the SED fitting performed in the main body
of the paper.  We first apply flux stretches to the model templates
and then take into account uncertainties in model fluxes in a given
band in the fits.
The latter procedure is important as pointed out by \citet{brammer08}
because the accuracy of model templates is dependent on wavelength.
Our model templates are most reliable (i.e., the dispersion is smallest)
in the rest-frame optical,
where population synthesis models are calibrated to real data.
They become less reliable at shorter and longer wavelengths.
SED fits can be improved by taking such model uncertainties into account.

The template flux uncertainties can be larger
than photometric uncertainties especially for bright objects
and this makes a fit too good -- reduced $\chi^2$ of a fit for a bright object
is typically below 1.  This does not mean that observed photometry
has pessimistic errors, but it simply means that model templates
are uncertain.  In fact, the template error function increases
the error ranges of all the physical parameters such as SFR and
stellar mass, but the increased uncertainties likely represent more realistic
uncertainties on the physical parameters.  Note that the template
error function does not completely remove the systematics.  For example,
as mentioned in the main body of the paper, we assume $\tau$-models,
but no real galaxies would exactly follow the exponential decay.
We force them to fit with $\tau$-models, which introduces systematic
uncertainties.  There is no straightforward way to quantify such
systematics and it remains one of the major uncertainties in our analysis.


\end{document}